\documentclass[aps,prd,twocolumn,showpacs,floatfix,preprintnumbers,amsmath,amssymb,nofootinbib,groupedaddress,superscriptaddress]{revtex4-1}

\input epsf
\usepackage{graphicx}
\usepackage{color}
\usepackage{mathtools}

\newcommand{\yacine}[1]{#1}

\newcommand{\beq}{\begin{equation}}
\newcommand{\eeq}{\end{equation}}
\newcommand{\barr}{\begin{eqnarray}}
\newcommand{\earr}{\end{eqnarray}}

\newcommand{\rme}{\textrm{e}}

\newcommand{\bs}{\boldsymbol}

\newcommand{\apjl}{Astrophys. J. Lett.}
\newcommand{\aap}{Astron. Astrophys.}
\newcommand{\apjs}{Astrophys. J. Suppl. Ser.}

\newcommand{\aj}{Astron. J.}

\newcommand{\mnras}{Mon. Not. R. Astron. Soc.}
\newcommand{\jcap}{J. Cosm. Astropart. Phys.}

\newcommand{\physrep}{Phys. Rep.}

\newcommand{\lsim}{\mathrel{\hbox{\rlap{\lower.55ex\hbox{$\sim$}} \kern-.3em \raise.4ex \hbox{$<$}}}}
\newcommand{\gsim}{\mathrel{\hbox{\rlap{\lower.55ex\hbox{$\sim$}} \kern-.3em \raise.4ex \hbox{$>$}}}}

\begin{document}
\title{A new light on 21 cm intensity fluctuations from the dark ages}
\author{Yacine Ali-Ha\"imoud}
\email{yacine@ias.edu}
\affiliation{Institute for Advanced Study, Einstein Drive, Princeton, New Jersey 08540}
\author{P.~Daniel Meerburg}
\email{meerburg@princeton.edu}
\author{Sihan Yuan}
\email{sihany@princeton.edu}
\affiliation{Department of Astrophysical Sciences, Princeton
  University, Princeton, New Jersey 08540}
\date{\today}

\begin{abstract}
Fluctuations of the 21~cm brightness temperature before the
  formation of the first stars hold the promise of becoming a
  high-precision cosmological probe in the future. The growth of overdensities is
  very well described by perturbation theory at that epoch and the
  signal can in principle be predicted to arbitrary accuracy for given
  cosmological parameters. Recently, Tseliakhovich and Hirata pointed out
  a previously neglected and important physical effect, due to the fact that baryons and
  cold dark matter (CDM) have supersonic relative velocities after
  recombination. This relative velocity
  suppresses the growth of matter fluctuations on scales $k \sim
  10-10^3$ Mpc$^{-1}$. In addition, the amplitude of the small-scale
  power spectrum is modulated on the large scales over which the
  relative velocity varies, corresponding to $k \sim 0.005-1$
  Mpc$^{-1}$. In this paper, the effect of the relative velocity on
  21~cm brightness temperature fluctuations from redshifts $z \geq 30$ is computed. We show that
  the 21~cm power spectrum is affected on {\it most} scales. On small
  scales, the signal is typically suppressed several tens of percent, except for extremely small scales ($k
  \gtrsim 2000$ Mpc$^{-1}$) for which the fluctuations are boosted by
  resonant excitation of acoustic waves. On large scales, 21~cm fluctuations
  are enhanced due to the non-linear dependence of the brightness temperature on the underlying gas density
  and temperature. The enhancement of the 21 cm power
  spectrum is of a few percent at $k \sim 0.1$ Mpc$^{-1}$ and up to tens of percent at $k \lesssim 0.005$
Mpc$^{-1}$, for standard $\Lambda$CDM cosmology. In principle this
effect allows to probe the small-scale matter power spectrum not only
through a measurement of small angular scales but also through its effect on large angular scales. 
\end{abstract}

\maketitle

\section{Introduction}

One of the exciting frontiers of cosmology in the post-WMAP\footnote{http://map.gsfc.nasa.gov/} and
\emph{Planck}\footnote{http://sci.esa.int/planck/} era is the observation of the high-redshift 21 cm
spin-flip transition of neutral hydrogen. Observations of the sub-mK fluctuations of the brightness temperature in this line are challenging
but can potentially provide unprecedented information about the early
universe \cite{Fuetal2006, Pritchard_2008,Fur2009proc}.
They are the only direct probe of large-scale structure during the
cosmic ``dark ages'', which follow the last scattering of cosmic
microwave background (CMB) photons and precede the formation of the
first luminous objects\footnote{The term ``cosmic dark ages'' is
  somewhat loosely used in the literature; here we mean it in the
  strict sense, i.e.~we refer to the epoch before the formation of the first
  stars, at $z \gtrsim 30$.} \cite{Loeb_2004}. 
21 cm intensity fluctuations contain in principle much more information than CMB
anisotropies: firstly, they can be used to probe a fully
  three-dimensional volume rather than a thin shell near the last
  scattering surface \cite{Mao:2008ug}, and secondly, they are limited only by the
  baryonic Jeans scale, $k_{\rm J} \sim 300$ Mpc$^{-1}$, whereas CMB fluctuations are
  damped for scales smaller than the Silk diffusion scale, $k_{\rm Silk} \sim 0.15$ Mpc$^{-1}$. In addition, overdensities remain small during the dark ages and their growth is very well described by perturbation theory. Linear
perturbation theory is sufficient to describe redshifts $z \gtrsim
50$, whereas non-linear corrections can become important at later
times \cite{Lewis_2007}; however, contrary to the present-day density
field which reaches order unity fluctuations on scales $k \gtrsim k_{\rm NL} \sim
0.1$ Mpc$^{-1}$, for $z \gtrsim 30$ non-linear corrections remain perturbative on all scales of
interest and the dark-ages 21 cm power
spectrum can in principle be computed accurately with
analytic methods. 

\yacine{Loeb and Zaldarriaga \cite{Loeb_2004} were the first to
  computate of the angular power spectrum of 21 cm fluctuations from the dark ages, and show its
  potential as a cosmological probe. Their computation did not account for the fluctuations of the local velocity gradient
  or of the gas temperature, shown to be important in Ref.~\cite{Bharadwaj_2004}. Since
  then Lewis and Challinor \cite{Lewis_2007} (hereafter LC07), have
  provided the most detailed calculation, including relativistic and
velocity corrections, as well as approximate non-linear
corrections. If 21 cm observations are to fulfill their promise
of an unprecedented high-precision cosmological probe, one must be able to predict the
signal to very high accuracy.} The goal of the present paper is to account for an important
physical effect previously overlooked and recently unveiled by Tseliakhovich
and Hirata \cite{Tsel_2010} (hereafter TH10): the fact that the baryons and the cold dark matter (CDM) have supersonic
relative velocities after primordial recombination. In this paper we will show that this physical effect modifies the theoretical 21~cm power spectrum on {\it all} scales. 

The relative velocity effect is present in standard $\Lambda$CDM cosmology with Gaussian
adiabatic initial conditions but was previously overlooked because it
is non-perturbative, even at redshift $z \sim 1000$. The basic idea is as follows. Prior to
recombination (or more accurately, kinematic decoupling), the tightly coupled photon-baryon fluid resist
gravitational growth due to its high pressure, resulting in acoustic oscillations. Meanwhile,
the CDM is oblivious to photons and its perturbations grow under their
own gravitational pull. At recombination, CDM and baryons have
therefore very different density and velocity fields; in particular,
their \emph{relative} velocity is of order $30$ km/s at recombination, a factor of $\sim5$ times larger
than the post-recombination baryonic sound speed.

TH10 pointed out two
consequences of these supersonic motions. First, the growth of
structure is hampered on scales smaller than the characteristic
advection scale over a Hubble time, and the matter density fluctuations are
suppressed by $\sim$15\% around $k \sim 200$ Mpc$^{-1}$. Second,
the small-scale power is modulated on the large scales over which the
relative velocity field varies, corresponding to $k \sim 0.005 - 1$
Mpc$^{-1}$.

 \begin{figure}
 \includegraphics[width = 89mm]{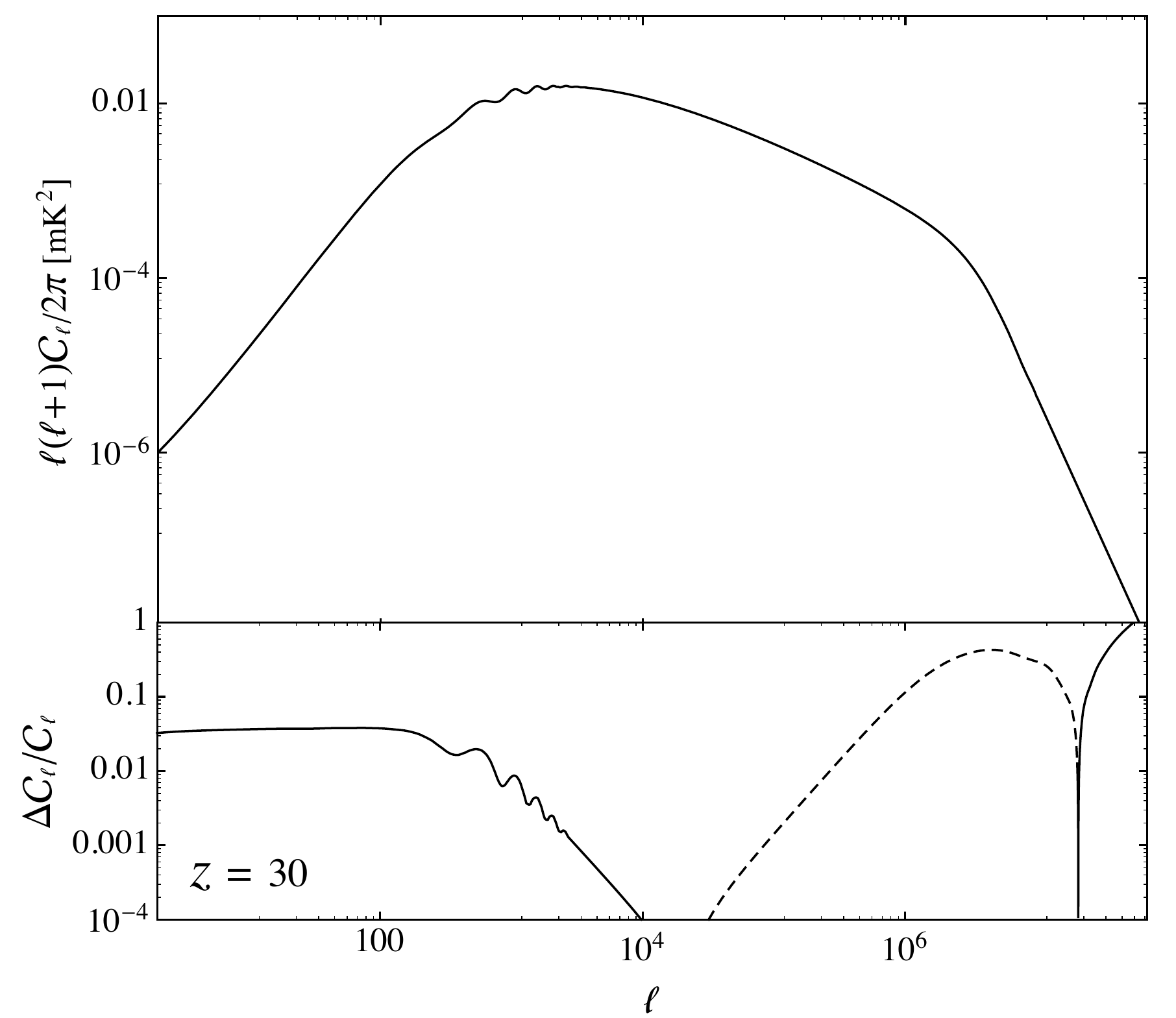}
 \caption{21 cm angular power spectrum at redshift $30$ for a window
   function of width $\Delta\nu=1$ MHz without relative velocity
   corrections (obtained using \textsc{camb} sources\footnote{http://camb.info/sources/}). The bottom panel shows the
   relative correction when accounting for the relative velocity
   effect: solid lines represent an enhancement and dashed lines a suppression.} \label{fig:intro} 
 \end{figure}
As we shall demonstrate in this paper, the relative velocity affects
the 21 cm fluctuations in three different ways. First, on small scales, $k \sim
200$ Mpc$^{-1}$, the perturbations are suppressed by several tens of percent; this is
because the 21 cm brightness temperature depends on the \emph{baryonic} density
and temperature fluctuations, which is more dramatically affected by
the relative velocity than the CDM \cite{Tsel_2011}. Second, on extremely
small scales ($k \gtrsim 2000$ Mpc$^{-1}$), we actually find an
\emph{enhancement} of baryonic density and temperature fluctuations,
hence of 21 cm fluctuations. This comes from the quasi-resonant
excitation of baryon acoustic oscillations as the baryonic fluid is advected across
CDM density perturbations, an effect which was not pointed out
previously. \yacine{Third, and most importantly, we also find enhanced 21cm fluctuations on large scales, $k\sim 0.005 - 1$ Mpc$^{-1}$.} This effect is less intuitive but can be
summarized as follows. The relation between the 21 cm intensity and
the underlying baryonic fluctuations $\delta$ is fundamentally non-linear, and
we may formally write $\delta T_{21} \approx \alpha \delta + \beta
\delta^2$\yacine{, where $\alpha$ and $\beta$ are of comparable
  magnitude}. When considering large-scale fluctuations of the
brightness temperature, we therefore have $\delta T_{21}|_l \approx \alpha
\delta_l+ \beta (\delta^2)_l$. In the absence of relative
velocities, the second term would be negligible for Gaussian initial
conditions and as long as perturbations are in the linear
regime. However, relative velocities lead to a large-scale, order
unity modulation of the amplitude of small-scale fluctuations $\delta_s$, and as a consequence, $(\delta^2)_l \sim \langle
\delta_s^2\rangle$. The small-scale fluctuations are much larger that
the large-scale ones, $\delta_l \ll \delta_s \ll 1$; for $z
\lesssim 100$, we even have $\delta_s^2 \sim \delta_l$, and the
quadratic term usually neglected in 21 cm fluctuations is actually
comparable to the linear term, leading to an order unity enhancement of the
large-scale 21 cm power spectrum. The effect on the \emph{angular}
power spectrum is not so dramatic, since power on large angular scales
is dominated by the rapidly rising small-scale power spectrum due to
standard terms. We find that the angular power spectrum is enhanced by
a few percent at $z = 30$ for $\ell \lesssim 1000$. \yacine{We emphasize that the large-scale
  enhancement is formally a non-linear effect, even if the
  perturbations remain small. The change to the large-scale power
  spectrum of 21cm fluctuations is indeed of order $(\delta_s^2/\delta_l)^2 \sim 1$, even
  though $\langle \delta^2 \rangle \ll 1$. The latter condition allows
  us to neglect ``standard'' non-linear terms which are not affected
  by the relative velocity.} 

Our results are summarized in Figure \ref{fig:intro}, where we show the standard theoretical 21~cm angular power spectrum
at redshift 30 and the corrections resulting from including the
relative velocity effect. 

We note that several previous works have already computed the consequences of
the relative velocity on the 21 cm signal in the
pre-reionization era, \emph{after the first stars have formed}, at redshifts $z
\lesssim 30$ \cite{Bittner_2011, Fialkov_2012, Visbal_2012, Fialkov_2013,
  Fialkov_2013b}. At that epoch the relevant physical ingredients are
very different than during the dark ages. On the one hand, the 21 cm spin
temperature is determined by the strength of the ambient stellar ultraviolet radiation field through
resonant scattering of Lyman-$\alpha$ photons (the Wouthuysen-Field
effect \cite{Wouthuysen_1952, Field_1958, Hirata_2006}). On the other hand,
the gas temperature, which sets the color temperature in the
Lyman-$\alpha$ line, and hence the spin temperature, is determined by
the rate of X-ray heating. Because the physics involved is complex, modeling the 21 cm emission from $z \lesssim 30$ requires numerical simulations, is model-dependent, and observing this signal is more likely to inform us
about the details of the formation of the first luminous sources than about
fundamental physics. Our work is therefore complementary to these
studies, extending the physical analysis of relative velocities to higher redshifts. The 21 cm signal from the dark ages is
even more challenging to observe due to ionospheric opacity and other
complications \cite{Carilli_2007}, but can be modeled exactly, with
relatively simple tools, and can potentially be a very clean probe
of the very early Universe. 

This paper is organized as follows. In Section \ref{sec:small-scale} we
compute the evolution of small-scale fluctuations accounting for the
relative velocity of baryons and CDM. We closely follow previous works
\cite{Tsel_2010, Tsel_2011} while consistently accounting for fluctuations of
the free-electron fraction as in LC07. Section \ref{sec:large-scale} describes the computation of large-scale fluctuations of quantities
which depend non-linearly on the underlying density field. Finally, we
apply our results to the 21 cm power spectrum from the dark ages in
Section \ref{sec:21cm}. We conclude in Section
\ref{sec:conclusions}. Appendix \ref{app:correlation} details our
method for computing autocorrelation functions of quadratic
quantities, and Appendix \ref{app:Cl-Limber} gives some analytic results
for the angular power spectrum. All our numerical results are obtained assuming a minimal flat
$\Lambda$CDM cosmology with parameters derived from \emph{Planck}
observations \cite{Planck_params} $T_{\rm cmb, 0} = 2.726$ K, $H_0 =
67.8$ km~s$^{-1}$Mpc$^{-1}$,
$\Omega_b = 0.0456$, $\Omega_c = 0.227$, $Y_{\rm He} = 0.24$, $N_{\rm
  eff} = 3.046$, $\tau_{\rm reion}=0.089$ $A_s = 2.196 \times10^{-9}$, $n_s = 0.96$, $k_{\rm pivot} =
0.05$ Mpc$^{-1}$.



\section{Effect of the relative velocity on small-scale fluctuations} \label{sec:small-scale}

\subsection{Statistical properties of the relative velocity field} \label{sec:vbc-stats}

In this section we briefly summarize the statistical properties of the
relative velocity field and the characteristic scales associated with
the problem (see also TH10).

\begin{figure*}
\includegraphics[width = 180mm]{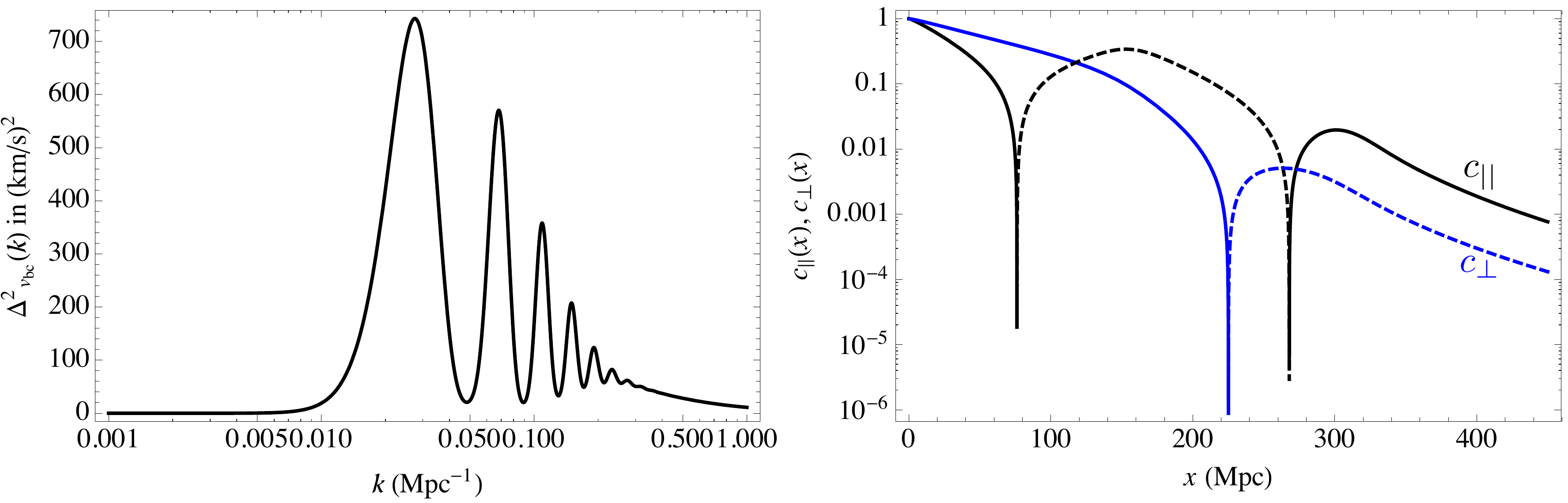}
\caption{Statistical properties of the relative velocity
  field. \emph{Left}: power per logarithmic $k$-interval
  $\Delta^2_{v_{\rm bc}}(k) \equiv k^3 P_{v_{\rm bc}}(k)/(2
  \pi^2)$ at redshift $z = 1010$. \emph{Right}: absolute value of the dimensionless auto-correlation coefficients
  for the relative velocity as a function of separation $x$ (solid
  lines for $c >0$ and dashed lines for $c < 0$).} \label{fig:corr} 
\end{figure*}

While the cold dark matter density perturbations grow unimpeded under
the influence of their own gravity, baryonic matter is kinematically
coupled to the photon gas by Thomson scattering until the abundance of
free electrons is low enough. Using the fitting formulae of
Ref.~\cite{Eisenstein_1998} with the current best-fit cosmological
parameters, the redshift of kinematic decoupling is $z_{\rm dec}
\approx 1117$. Later on, baryons and CDM evolve as
pressureless fluids on all scales greater than the baryonic Jeans
scale $k_{\rm J}\sim 300$ Mpc$^{-1}$. However they have notably different
initial conditions at $z_{\rm dec}$, in particular, for their peculiar velocities. In the absence of
vorticity perturbations, the Fourier transform of the gauge-invariant \emph{relative}
velocity field takes the form
\beq
\bs{v}_{\rm bc}(\bs k) \equiv \bs{v}_{\rm b}(\bs k)- \bs{v}_{\rm c}(\bs k) = \hat{k} \mathcal{V}(\bs{k}),
\eeq
where from the continuity equations for baryons and CDM we have
\beq
\mathcal{V}(\bs{k}) \equiv
-\frac{1}{i k(1+z)}\frac{d}{dt}\left(\delta_b(\bs{k}) - \delta_c(\bs{k})\right).
\eeq
We define the relative velocity power spectrum $P_{v_{\rm bc}}(k)$ such
that 
\beq
\langle \mathcal{V}(\bs{k}) \mathcal{V}(\bs{k'})^* \rangle = (2 \pi)^3\delta_{\rm D}(\bs{k'} - \bs{k}) P_{v_{\rm bc}}(k),
\eeq
where $\delta_{\rm D}$ is the Dirac delta function. The variance of
the relative velocity along any fixed axis is denoted by
$\sigma_{1d}^2$. It is one third of the variance of the magnitude of
the three-dimensional relative velocity vector, which we denote by
$\sigma_{3d}^2$. They are given by
\beq
\sigma_{1d}^2 \equiv \frac13 \sigma_{3d}^2 \equiv \frac13 \int \frac{d^3
  k}{(2 \pi)^3} P_{v_{\rm bc}}(k).
\eeq
From symmetry considerations, the autocorrelation function of the relative velocity takes the form
\beq
\frac{\langle v_{\rm bc}^i(\bs{0}) v_{\rm bc}^j(\bs{x})
  \rangle}{\sigma_{1d}^2} = c_{\parallel}(x) \hat{x}^i \hat{x}^j +
c_{\bot}(x)(\delta^{ij} - \hat{x}^i \hat{x}^j), \label{eq:cpara-cperp}
\eeq
where the dimensionless coefficients $c_{\parallel}$ and $c_{\bot}$ give the
correlation of the velocity components parallel and perpendicular to
the separation vector, respectively. They are given by \cite{Dalal_2010}
\yacine{
\barr
c_{\parallel}(x) &=& \frac{1}{\sigma_{3d}^2} \int \frac{d^3
  k}{(2 \pi)^3} P_{v_{\rm bc}}(k)\Big{(}j_0(kx) - 2 j_2(kx)\Big{)},\\
c_{\bot}(x) &=& \frac{1}{\sigma_{3d}^2} \int \frac{d^3
  k}{(2 \pi)^3} P_{v_{\rm bc}}(k)\Big{(}j_0(kx) +  j_2(kx)\Big{)},
\earr}
where $j_i$ is the $i$-th spherical Bessel functions of the first
kind. We have extracted the baryon and CDM power spectra and their
derivatives at $z_i = 1010$ from \textsc{camb} \cite{Lewis:1999bs},
and computed $P_{v_{\rm 
 bc}}(k)$. We obtain $\sigma_{1d} \approx 17$ km/s and $\sigma_{3d}
\approx 29$ km/s at $z_i$. We
show the power per logarithmic interval
$\Delta^2_{v_{\rm bc}}(k) \equiv k^3/(2 \pi^2) P_{v_{\rm bc}}(k)$ and
the correlation
coefficients of the relative velocity field in Fig.~\ref{fig:corr}. After kinematic decoupling, the
relative velocity decreases proportionally to $1/a$ on all scales
larger than the baryonic Jeans scale since dark
matter and baryons are subjected to the same acceleration on these
scales \cite{Tsel_2010}.

The correlation coefficients $c_{\parallel}(x), c_{\bot}(x)$ are greater than
95\% for $x \lesssim 3$ Mpc and $x \lesssim 6$ Mpc, respectively,
which means that the relative velocity is very nearly homogeneous on
scales of a few Mpc. This defines a coherence scale for the relative
velocity, $x_{\rm coh} \approx 3$ Mpc, corresponding to a wavenumber
$k_{\rm coh} = (x_{\rm coh})^{-1} \approx 0.3$ Mpc$^{-1}$, which can also be
inferred directly by considering the power spectrum $P_{v_{\rm bc}}(k)$.

On the other hand, starting from kinematic decoupling at time $t_{\rm
  dec}$, the relative velocity displaces baryons with respect to CDM perturbations by a characteristic
comoving distance 
\barr
x_{v_{\rm bc}} = \int_{t_{\rm dec}}^t \sigma_{1d}(t') \frac{dt'}{a(t')} \approx
\frac{2 \sigma_{1d}(a_{\rm dec}) a_{\rm dec}^{1/2}}{H_0
  \Omega_m^{1/2}} \approx 30 \textrm{ kpc},~~ \label{eq:xvbc}
\earr
where in the second equality we have taken the limit $t \gg t_{\rm dec}$, assumed a matter dominated
universe and used a characteristic velocity $\sigma_{1d}$ (instead
of $\sigma_{3d}$) as only the component of the relative velocity along
the wavevector is relevant. Baryonic fluctuations with wavenumbers $k \gtrsim 2 \pi x_{v_{\rm bc}}^{-1}
\approx 200$ Mpc$^{-1}$ are therefore advected across several peaks and troughs of the
gravitational potential, sourced mostly by the CDM overdensity. The
net acceleration partially cancels out, which slows down the growth of baryonic perturbations, and, in
turn, that of the CDM. The effect is most pronounced for $k \gtrsim 200$
Mpc$^{-1}$ but it is still important at slightly larger scales, and we define $k_{v_{\rm bc}} \equiv 30$ Mpc$^{-1}$ as the
typical scale at which the suppression is of the order of a percent (as we shall
confirm a posteriori).

Throughout this paper, unless otherwise stated, we shall use ``small scales'' (and use the
subscript $s$ in relation to them) to refer to scales with a wavenumber $k_s
\gtrsim k_{v_{\rm bc}} \approx 30$ Mpc$^{-1}$, and use ``large scales''
(subscript $l$) for those with a wavenumber $k_l
\lesssim k_{\rm coh} \approx 0.3$ Mpc$^{-1}$.

\subsection{Basic equations}

\subsubsection{Moving background perturbation theory}

As first pointed out in TH10 and brought to mind in the previous section, the
scales at which the relative velocity affect the growth of structure
are about two orders of magnitude smaller than the coherence scale
of the relative velocity field. This makes it possible to use
moving-background perturbation theory, i.e.~compute the evolution of
small-scale fluctuations given a local background value of the
relative velocity. This approximation is equivalent to the eikonal
approximation recently introduced in the context of cosmological
perturbations \cite{Bernardeau_2012, Bernardeau_2013}. As a result,
the small-scale fluctuations $\delta(\bs{k}_s; \bs{v}_{\rm
  bc}(\bs{x}))$ are functions of the small-scale
wavevector $\bs{k}_s$ \emph{and} of the local relative velocity
$\bs{v}_{\rm bc}(\bs{x})$. Let the reader not be confused by this
mixture of Fourier-space and real-space dependence: it is justified
because the relative velocity field only fluctuates significantly on
large scales $k_l \lesssim k_{\rm coh} \ll k_{v_{\rm bc}} \lesssim k_s$. Moving-background perturbation theory allows us to account non-perturbatively
for a fundamentally non-linear term that is active as early as $z
\approx 1000$. Other non-linearities become important in the evolution
of the small-scale fluctuations at lower redshifts. In this paper, we shall not concern ourselves with the
latter, which can in principle be treated with standard perturbation
theory methods. One should keep in mind that they do become important for the computation of
21~cm fluctuations from $z \lesssim 50$ \cite{Lewis_2007}, and should
eventually be consistently included for a high-precision computation
of the 21~cm signal.

Following TH10, we place ourselves in the local baryon rest-frame
(defined such that the baryon velocity averaged over a few Mpc patch
vanishes). We consider the evolution of small-scale modes with $k_s
\gtrsim 30$ Mpc$^{-1}$ and can therefore neglect relativistic
corrections since the scales of interest are much
smaller than the horizon scale $k_{\rm hor} = a H \approx 0.001
\left(\frac{1+z}{101}\right)^{1/2}$ Mpc$^{-1}$. The relative velocity
is locally uniform and decreases proportionally to the inverse of the
scale factor, $v_{\rm bc} \propto 1/a$.

\subsubsection{Fluid equations}

The linear evolution of small-scale perturbations in Fourier
space is given by the usual fluid equations in an expanding universe, with an additional
advection term:
\barr
&&\dot{\delta}_{c} - i a^{-1}(\bs{v}_{\rm bc} \cdot \bs{k}) \delta_c +
\theta_c = 0, \label{eq:fluid1}\\
&&\dot{\theta}_c - i a^{-1}(\bs{v}_{\rm bc} \cdot \bs{k}) \theta_c + 2 H
\theta_c - k^2 \phi = 0,~~~~\\
&&\dot{\delta}_{b}  + \theta_b = 0,\\
&&\dot{\theta}_b + 2 H \theta_b - \frac{k^2}{a^2} \phi - \frac{\overline{c}_s^2}{a^2}
  k^2\left(\delta_b + \delta_{T_{\rm gas}}\right) =0, \label{eq:baryon-momentum}\\
&&\frac{k^2}{a^2} \phi = -\frac32 \frac{H_0^2}{a^{3}}\left(\Omega_{b}^0 \delta_b +
  \Omega_{c}^0 \delta_c\right), \label{eq:fluid5}
\earr
where the subscripts $b$ and $c$ refer to baryons and CDM,
respectively, $\theta$ is the velocity divergence with respect to
proper space\footnote{Here we use the notation of
  Ref.~\cite{Tsel_2010}, which differs from the more commonly
  used definition of $\theta$ given in Ref.~\cite{Ma_1995} by a factor
of $a$.}, overdots denoted differentiation with respect to proper
time, and $\phi$ is the Newtonian gravitational potential. In
Eq.~(\ref{eq:baryon-momentum}) $\overline{c}_s$ is the average baryon isothermal
sound speed, given by
\beq
\overline{c}_s^2 \equiv \frac{\overline{T}_{\rm gas}}{\mu ~m_{\rm H}}.
\eeq
Here  $\mu$ is the mean molecular weight given by
\beq
\mu \equiv \frac{1 + \frac{m_{\rm He}}{m_{\rm H}} x_{\rm He}}{1 +
  x_{\rm He} + x_e(z)},
\eeq
where $x_{\rm He} \equiv n_{\rm He}/n_{\rm H}$ is the constant ratio of helium
to hydrogen by number and $x_e(z) \equiv n_e/n_{\rm H}$ is the free
electron fraction. For a helium mass fraction $Y_{\rm He} = 0.24$ and
for an essentially neutral plasma, $\mu \approx 1.22$. 

Following Refs.~\cite{Naoz_2005, Tsel_2011}, we have included matter temperature
fluctuations $\delta_{T_{\rm gas}} \equiv \delta T_{\rm gas}/T_{\rm
  gas}$ in the baryon momentum equation (\ref{eq:baryon-momentum}). We
do not include fluctuations of the mean molecular weight due to
fluctuations of the free electron fraction as the latter is very
small at the redshifts of interest, with $x_e \approx 5\%$ at $z =
1000$ and falling below 0.1\% for $z < 600$. 

\subsubsection{Temperature fluctuations} \label{sec:Tgas}

To complete the system we need an evolution equation for $\delta_{T_{\rm
    gas}}$. Because some mistakes exist in the literature we rederive
this equation here, following Ref.~\cite{Switzer_2008}. We start by writing down the first law of thermodynamics
in a small volume $V$ containing a fixed number of hydrogen nuclei (i.e.~a
fixed total amount of protons \emph{and} neutral hydrogen atoms), so
that $n_{\rm H} V$ is constant: 
\beq
\frac{d}{d t}\left(\frac32 n_{\rm tot} V T_{\rm gas}\right) + n_{\rm tot} T_{\rm
  gas} \frac{dV}{dt}= \dot{Q},
\eeq
where $n_{\rm tot} \equiv n_{\rm HI} + n_p + n_e + n_{\rm He} = n_{\rm
  H}(1 + x_{\rm He} + x_e)$ is the
total number density of all free particles (neutral hydrogen, free protons, free electrons, and Helium), $n_{\rm H} \equiv
n_{\rm HI} + n_p$, and $\dot{Q}$ is the rate of
energy injection in the volume $V$. In the absence of any non-standard heating sources such as dark matter
annihilation or decay, two sources contribute to $\dot{Q}$:
photoionization / recombination and heating by CMB photons scattering
off free electrons which then rapidly redistribute their energy to the
rest of the gas through Coulomb scattering.

Let us first consider recombinations and photoionizations. We denote by $d \dot{x}_e/dE_e$
the differential net photoionization rate (i.e. the rate of photoionizations minus the rate of recombinations) per total abundance of
hydrogen, and per interval of energy of the electron, whether it is the initial,
recombining electron or the final free electron after photoionization. The
source term due to recombinations and photoionizations can be written
as
\beq
\dot{Q}_{\rm rec} = \int dE_e ~E_e\frac{d \dot{x}_e}{dE_e} n_{\rm H} V,
\eeq
where we used the fact that $n_{\rm H}V$ is constant. Without loss of
generality we may rewrite this
quantity as 
\barr
\dot{Q}_{\rm rec} &=& \frac32 T_{\rm gas} \dot{x}_e  n_{\rm H} V +
\Delta \dot{Q}_{\rm rec}, \label{eq:heating-rec}\\
\Delta \dot{Q}_{\rm rec} &\equiv& \int dE_e ~\left(E_e - \frac32 T_{\rm gas}\right)\frac{d \dot{x}_e}{dE_e} n_{\rm H} V.
\earr
The first term in Eq.~(\ref{eq:heating-rec}) contains the bulk of
$\dot{Q}_{\rm rec}$, and corresponds to the rate of energy injection
if every net recombination event removed on average exactly $\frac32
T_{\rm gas}$ of kinetic energy from the gas. This is nearly exact
since almost all of the kinetic energy of recombining electrons goes
into the emitted photon (with a very small fraction going into the recoil of the
formed nucleus), and the term
$\Delta \dot{Q}_{\rm rec}$ accounts for small corrections to this
relation. This term is completely negligible in comparison to
Compton heating and adiabatic cooling (in fact, even the much
bigger term $\dot{Q}_{\rm rec}$ which was not properly included in
Refs.~\cite{Seager_2000, Senatore_2009} is negligible). Neglecting the small
correction term $\Delta \dot{Q}_{\rm rec}$, after simplification we get the evolution equation for
the gas temperature
\beq
\dot{T}_{\rm gas} - \frac23 \frac{\dot{n}_{\rm H}}{n_{\rm H}} T_{\rm
  gas} = \frac23 \dot{q}_{\rm C}, \label{eq:first-law}
\eeq
where $\dot{q}_{\rm C}$ is the Compton heating rate per particle:
\barr
\dot{q}_{\rm C} &=& \frac{4 \sigma_{\rm T} a_r T_{\rm cmb}^4}{(1 + x_{\rm He} + x_e) m_e}  x_e(T_{\rm cmb} -
  T_{\rm gas})\nonumber\\
 &\equiv& \frac32
\Gamma_{\rm C}\frac{x_e}{\overline{x}_e}(T_{\rm cmb} -
  T_{\rm gas}). 
\earr
Here $\sigma_{\rm T}$ is the Thomson cross-section, $a_r$ is the
radiation constant, $m_e$ is the electron mass, and we have defined
the rate
\beq
\Gamma_{\rm C} \equiv \frac{8 \sigma_{\rm T} a_r T_{\rm cmb}^4}{3(1 +
  x_{\rm He} + x_e) m_e}  \overline{x}_e,
\eeq
which we shall assume to be homogeneous as it only depends on the
local free electron fraction through the term $1 + x_{\rm He} + x_e
\approx 1 + x_{\rm He}$. The homogeneous part of Eq.~(\ref{eq:first-law}) gives the
evolution of the average matter temperature:
\beq
\dot{\overline{T}}_{\rm gas} + 2 H \overline{T}_{\rm gas} =
\Gamma_{\rm C} (T_{\rm cmb} -
  \overline{T}_{\rm gas}).
\eeq
We now turn to the perturbations. Assuming the helium to hydrogen
ratio is uniform, and up to very small corrections of order $x_e \times(m_e/m_p)$,
we have $\delta n_{\rm H}/ n_{\rm H} = \delta_b$. Since we are considering scales deep inside
the Horizon, photon temperature perturbations are negligible
compared to any other perturbations, and we set $T_{\rm cmb} =
\overline{T}_{\rm cmb}$. The non-perturbative evolution equation for the
gas temperature fluctuation therefore reads:
\barr
&& \dot{\delta}_{T_{\rm gas}} - \frac23 \dot{\delta}_b \frac{1 +
  \delta_{T_{\rm gas}}}{1 + \delta_b} =\nonumber\\
&& \Gamma_{\rm C} \left[\frac{\overline{T}_{\rm cmb} - \overline{T}_{\rm
      gas}}{\overline{T}_{\rm gas}} \delta_{x_e} -
 \left( \frac{\overline{T}_{\rm cmb}}{\overline{T}_{\rm gas}} + \delta_{x_e}\right)\delta_{T_{\rm gas}} \right],\label{eq:Tgas-nonpert}
\earr
which corresponds to Eq.~(16) of Ref.~\cite{Pillepich_2007} if
$\delta_{x_e} \equiv \delta x_e/\overline{x}_e= 0$. To first order, the evolution equation for the temperature perturbations is therefore
\barr
\dot{\delta}_{T_{\rm gas}} - \frac23 \dot{\delta}_b = \Gamma_{\rm C} \left[\frac{\overline{T}_{\rm cmb} - \overline{T}_{\rm
      gas}}{\overline{T}_{\rm gas}} \delta_{x_e}  - \frac{\overline{T}_{\rm cmb}}{\overline{T}_{\rm gas}}\delta_{T_{\rm gas}} \right]. ~~~~~\label{eq:Tgas-pert}
\earr
Refs.~\cite{Naoz_2005, Tsel_2011} did not account for the fluctuations
of the free-electron fraction. This is justified at high redshifts at which the matter
temperature is very close to the radiation temperature and the
prefactor of $\delta_{x_e}$ in Eq.~(\ref{eq:Tgas-pert}) is small; it
is also justified at $z \ll 200$ when $\Gamma_{\rm C} \ll H$ and
the gas simply cools adiabatically. However, at intermediate stages this term cannot be neglected, at least
formally. Besides our neglect of photon temperature perturbations and
relativistic corrections (of order $\sim a^ 2H^2/k^2
\delta_m \ll \delta_m$ in the deep sub-horizon regime), our equation
(\ref{eq:Tgas-pert}) is identical to Eq.~(B12) of LC07, and does not include spurious molecular weight
terms as in Ref.~\cite{Senatore_2009}, where the term $\dot{Q}_{\rm
  rec}$ was not accounted for. 

In order to account for other potential heating sources such as dark matter
annihilation \cite{Chen_2004,
  Giesen_2012}, one would simply have to add the corresponding heating rate to the right-hand side of
Eq.~(\ref{eq:first-law}), and perturb the equation consistently \cite{Dvorkin_2013}.

\subsubsection{Free-electron fraction fluctuations} \label{sec:xe}

To close our system of equations we require an evolution equation for
the fluctuations in the ionization fraction of the gas. Because the
pre-factor of $\delta_{x_e}$ in Eq.~(\ref{eq:Tgas-pert}) is less than
1\% for $z \gtrsim 500$ \cite{hyrec}, we only need to have an accurate
equation at late times and we do not need to worry about details of the
radiative transfer in the Lyman-$\alpha$ line, which affect the
recombination history near the peak of the CMB visibility function (see
for example Refs.~\cite{Hirata_2009, Chluba_2010} and references
therein). We compute the background recombination history exactly with
\textsc{hyrec}\footnote{http://www.sns.ias.edu/$\sim$yacine/hyrec/hyrec.html}
\cite{hyrec} but when computing the perturbations, we simply adopt an effective 3-level atom model
\cite{Peebles_1968, Zeldovich_1969}, for which the recombination rate is given by 
\beq
\dot{x}_e = - C \left(\mathcal{A}_{\rm B} n_{\rm H}
  x_e^2 - 4 (1 - x_e) \mathcal{B}_{\rm B} \rme^{- E_{21}/T_{\rm cmb}} \right), \label{eq:xedot}
\eeq
where $E_{21} = 10.2$ eV is the energy of the Lyman-$\alpha$
transition, $\mathcal{A}_{\rm B} (T_{\rm cmb} ,T_{\rm gas})$ is the
effective case-B recombination coefficient, $\mathcal{B}_{\rm
  B}(T_{\rm cmb})$ is the
corresponding effective photoionization rate, and $C$ is the Peebles
C-coefficient \cite{Peebles_1968}, which gives the ratio of the net rate
of downward transitions from the first excited states to their total
effective lifetime:
\barr
C &\equiv& \frac{3 R_{\rm Ly \alpha} + \Lambda_{2s, 1s}}{3 R_{\rm Ly
    \alpha} + \Lambda_{2s, 1s} + 4 \mathcal{B}_{\rm B}}, \label{eq:Peebles-C}\\
R_{\rm Ly \alpha} &\equiv& \frac{8 \pi (H + \frac13 \theta_b)}{3 \lambda_{\rm Ly \alpha}^3 (1
- x_e) n_{\rm H}}. \label{eq:RLya}
\earr
Equation~(\ref{eq:xedot}) is identical in spirit to that of Peebles
\cite{Peebles_1968} and of Ref.~\cite{Seager_2000}, with however two
technical differences. First, following LC07 and Ref.~\cite{Senatore_2009}, we have replaced the Hubble rate in the
Lyman-$\alpha$ escape rate (\ref{eq:RLya}) by the local expansion
rate, which is enhanced by one third of the baryon peculiar velocity
divergence. This simple replacement relies on the implicit assumption
that the recombination process is local, in the sense that the Lyman-$\alpha$
radiation field is determined by the density and temperature within a
distance much smaller than the wavelength of the scales
considered. Checking this assumption quantitatively is non trivial, however at the low redshifts of interest the net recombination rate
is independent of the details of the Lyman-$\alpha$ radiative transfer
($C \rightarrow 1$ for $z \lesssim 900$), and the detailed value of the
perturbed $C$-factor is not critical.

Second, instead of using the case-B recombination
coefficient $\alpha_{\rm B}(T_{\rm gas})$ of
Ref.~\cite{Pequignot_1991} or a fudged version of it as in Ref.~\cite{Seager_2000}, we use the \emph{effective} recombination
coefficient $\mathcal{A}_{\rm B}(T_{\rm gas}, T_{\rm cmb})$, which
accounts \emph{exactly} for stimulated recombinations to, ionizations from, and transitions between,
the highly-excited states of hydrogen during the cascading process \cite{emla}. These coefficients are
related through $\alpha_{\rm B} = \mathcal{A}_{\rm B}(T_{\rm cmb} =
0)$. The temperature dependence of $\alpha_{\rm B}$ (even rescaled by
a fudge factor) differs from the correct one given by
$\mathcal{A}_{\rm B}$ at the level of $\sim 10-20$\%.

For $z < 1010$ the free-electron fraction is already much larger than
its value in Saha equilibrium and the second term in
Eq.~(\ref{eq:xedot}) is less than $10^{-4}$ times the first term. We
therefore have, to an excellent accuracy,
\beq
\dot{x}_e \approx - C \mathcal{A}_{\rm B} n_{\rm H} x_e^2. \label{eq:dotxe-final}
\eeq
This allows us to get a simple expressions for the evolution of
$\delta_{x_e}$, to first order:
\barr
\dot{\delta}_{x_e} &=&
\frac{\dot{\overline{x}}_e}{\overline{x}_e}\Bigg{[}\delta_{x_e} +
    \delta_b + \frac{\partial \ln \mathcal{A}_{\rm B}}{\partial \ln T_{\rm
        gas}} \delta_{T_{\rm gas}}  \nonumber\\
&&~~~~~+ \frac{\partial
  \ln C}{\partial \ln R_{\rm Ly \alpha}}\left(\frac{\theta_b}{3 H} - \delta_b\right) \Bigg{]}, \label{eq:xe-pert} 
\earr
where we have used the fact that $C$ depends on the baryon
density and velocity divergence $\theta_b$ through the Lyman$-\alpha$ escape
probability, and we have neglected fluctuations of the free electron
fraction in the Lyman-$\alpha$ escape rate since $x_e \ll 1$ at the
times of interest.

Here again, one can easily include additional ionization sources, for example resulting from dark matter annihilation \cite{Chen_2004,
  Giesen_2012, Dvorkin_2013}. 

\subsubsection{Initial conditions} \label{sec:init}

The initial conditions for
$\delta_b, \theta_b, \delta_c$ and $\theta_c$ are
extracted from \textsc{camb} at $z_{\rm ini} = 1010$. The initial condition for
$\delta_{T_{\rm gas}}$ is obtained from noticing that at $z_{\rm
  ini}$, $H/\Gamma_{\rm C} \approx 3 \times 10^{-5} \ll 1$, and $T_{\rm gas} \approx
T_{\rm cmb}$ to an excellent accuracy. Up to
corrections of order $\dot{\delta}_b / \Gamma_{\rm C}
\ll \delta_b$ and $\delta_{T_{\rm cmb}}$, we therefore have $\delta_{T_{\rm gas}}(z_{\rm init}) = 0$.

In principle one should start computing the evolution of ionization
fraction perturbations from an earlier time in order to get the proper
initial conditions at $z_{\rm ini} = 1010$. However, since
the perturbations of $\delta_{x_e}$ only affect the 21 cm signal at
late times through their coupling to $\delta_{T_{\rm gas}}$,
and since the entire system is driven by $\delta_c \gg \delta_{x_e} \sim \delta_b$
initially, the value of $\delta_{x_e}(z_{\rm ini})$ is quickly forgotten
and has virtually no effect on the observables of interest here\footnote{It is
however important to compute $\delta_{x_e}$ accurately if one is
interested in the effect of perturbations on CMB anisotropies}. We may
therefore safely set $\delta_{x_e}(z_{\rm init}) = 0$.

\subsection{Results: evolution of small-scale fluctuations} \label{sec:deltab-ss}

We have numerically solved the coupled differential equations
(\ref{eq:fluid1})-(\ref{eq:fluid5}), (\ref{eq:Tgas-pert}) and
(\ref{eq:xe-pert}) for $\delta_b, \delta_c, \delta_{T_{\rm
    gas}}$ and $\delta_{x_e}$, as a function of $k$ and $\bs{v}_{\rm
  bc}(z_{\rm dec}) \cdot \hat{k}$, starting at $z_{\rm ini} = 1010$
with initial conditions described above, down to $z = 20$. The evolution of the background free-electron fraction and matter
temperature is computed with the recombination code \textsc{hyrec}. 

\begin{figure*}
\includegraphics[width = 180mm]{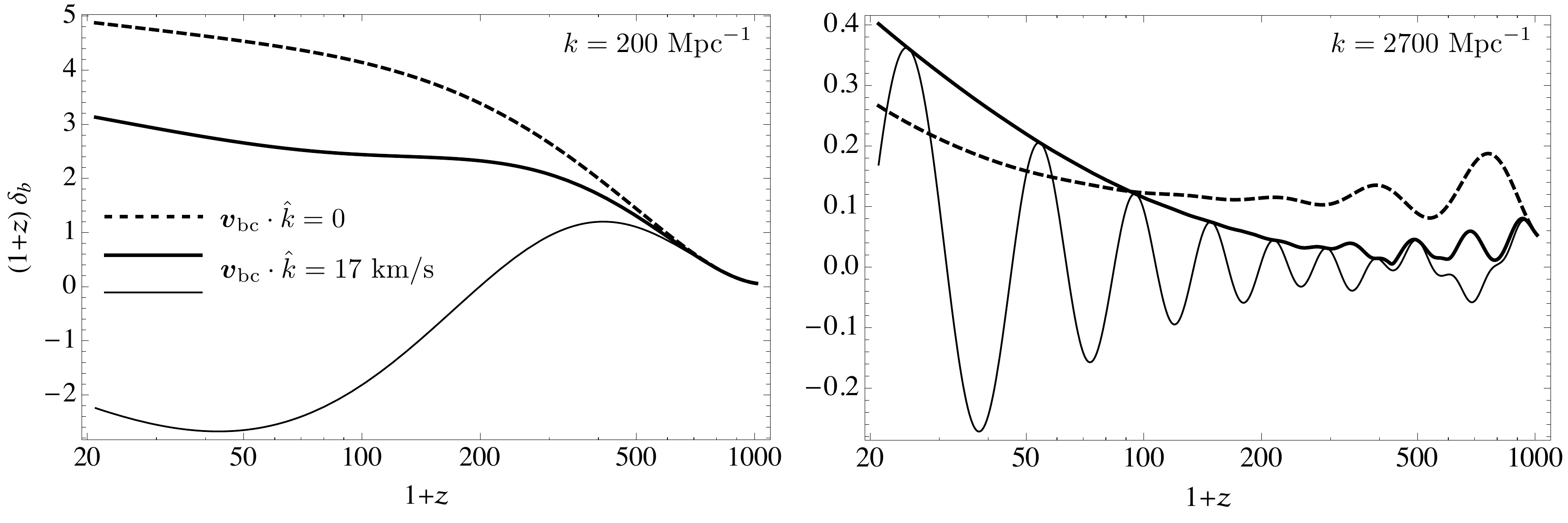}
\caption{Evolution of $\delta_b (k, z)$ for two small-scale modes with $k = 200$ Mpc$^{-1}$ and $k
  = 2700$ Mpc$^{-1}$ (specifically, what is plotted is $(1+z) T_{\delta_b}(k, z)
  [A_s (k/k_{\rm pivot})^{n_s -1}]^{1/2}$, where $T_{\delta_b}(k, z)$ is the
  transfer function). The values of the relative velocity are $\bs{v}_{\rm bc} \cdot
  \hat{k} = 0$ (dotted) and $\bs{v}_{\rm bc} \cdot \hat{k} = 17$ km/s
  (solid), the latter corresponding to the rms relative velocity along
  a given axis. Thick lines represent the
  absolute value of $\delta_b$ and thin lines show its real
  part (the two quantities are equal for $\bs{v}_{\rm bc} \cdot \hat{k} = 0$). For $k = 200$ Mpc$^{-1}$ the relative velocity leads to a suppression of
  fluctuations, whereas for $k
  = 2700$ Mpc$^{-1}$ the streaming of baryons
  relative to the dark matter leads to a resonant amplification of
  baryonic acoustic oscillations.} \label{fig:deltab} 
\end{figure*}

\begin{figure}
\includegraphics[width = 86mm]{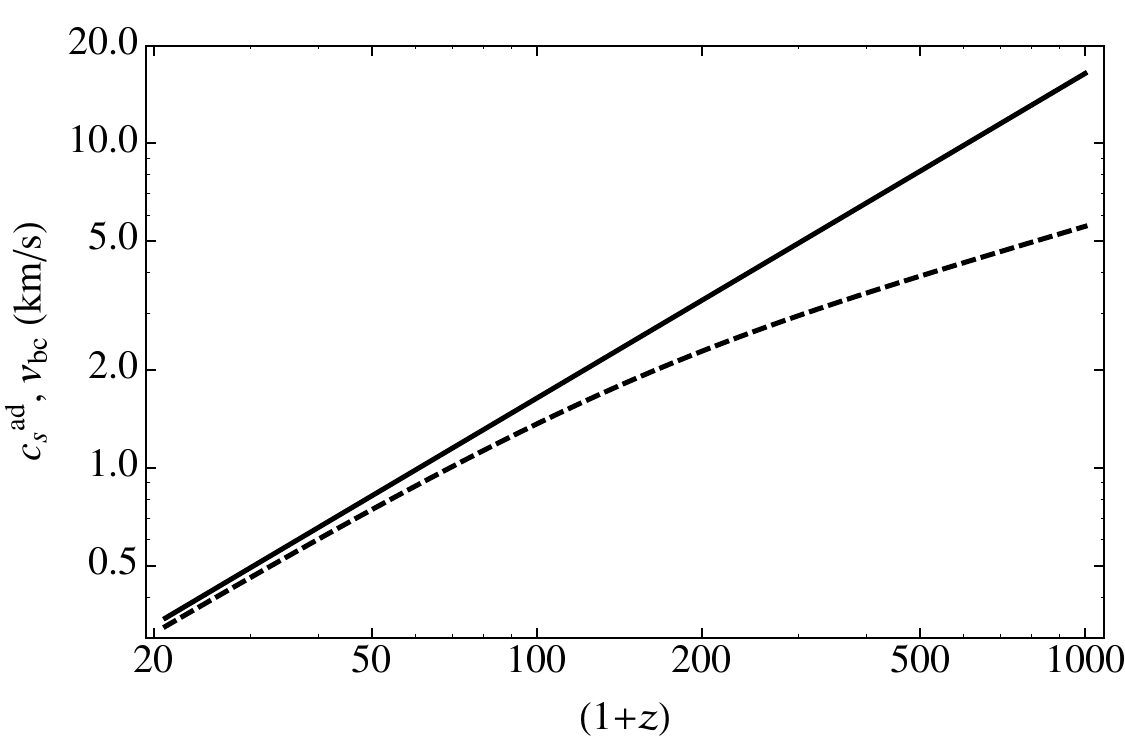}
\caption{Rms value of the relative velocity along a given axis (solid
  line) compared to the
adiabatic sound speed $\overline{c}_s^{\rm ad} = \sqrt{5/3} ~
\overline{c}_s^{\rm iso}$ (dashed line), as a function of redshift.} \label{fig:cs-vbc} 
\end{figure}

We show the evolution of the baryon density fluctuations $\delta_b$ for
two modes in Fig.~\ref{fig:deltab}. For a scale $k = 200$ Mpc$^{-1}$ of the order of the
advection scale but somewhat larger than the Jeans
scale ($k_{\rm Jeans} \approx 300$ Mpc$^{-1}$), the relative velocity destroys the phase coherence between
baryons and dark matter by advecting their perturbations across more
than a wavelength in a Hubble time. The result is to suppress the growth of structure,
as illustrated in the left panel of
Fig.~\ref{fig:deltab}. On the other hand, for scales much smaller than the
Jeans scale, we find that a typical value of the relative velocity actually
leads to a \emph{resonant amplification} of baryon density and temperature fluctuations (see
evolution of the mode $k = 2700$ Mpc$^{-1}$ in Fig.~\ref{fig:deltab}). This can be
understood as follows. On sub-Jeans scales, baryonic fluctuations are suppressed due to their
pressure support, and $\delta_b \ll \delta_c$. One can solve explicitly
for the evolution of the growing mode of CDM perturbation in the limit
$\delta_b = 0$ and obtain, during matter domination:
\barr
\delta_c &\propto& \exp\left[i \bs{k} \cdot \int^t  \frac{\bs{v}_{\rm
      bc}}{a}  dt\right] a^{\alpha}, \label{eq:forcing}\\
\alpha &\equiv& 1 - \frac54\left(1 - \sqrt{1 - \frac{24}{25}
    f_b}\right)\approx 1 - \frac35 f_b,\label{eq:alpha}\\
f_b &\equiv& \frac{\Omega_b}{\Omega_c + \Omega_c},
\earr
where the approximate value of the growth rate in Eq.~(\ref{eq:alpha})
is valid in the limit $f_b \ll 1$. With our fiducial cosmology $f_b
\approx 0.17$ and $\alpha \approx 0.90$. 
Baryonic perturbations undergo acoustic
oscillations forced by the gravitational attraction from the dark matter and damped by the Hubble expansion:
\beq
\ddot{\delta}_b+ 2 H \dot{\delta}_b + \frac{\overline{c}_s^2}{a^2} k^2
\left(1 + \frac{\delta_{T_{\rm gas}}}{\delta_b}\right)
\delta_b = \frac32 H^2 (1 - f_b) \delta_c. \label{eq:forced-sound}
\eeq
Figure~\ref{fig:cs-vbc} shows that the characteristic relative velocity
along a given axis is very close to the adiabatic sound speed for $z
\lesssim 200$. For typical
relative velocities, the forcing term in Eq.~(\ref{eq:forced-sound}) therefore
oscillates with a frequency close to that of acoustic oscillations,
which leads to a resonant amplification of acoustic waves. 

\begin{figure*}
\includegraphics[width = 180mm]{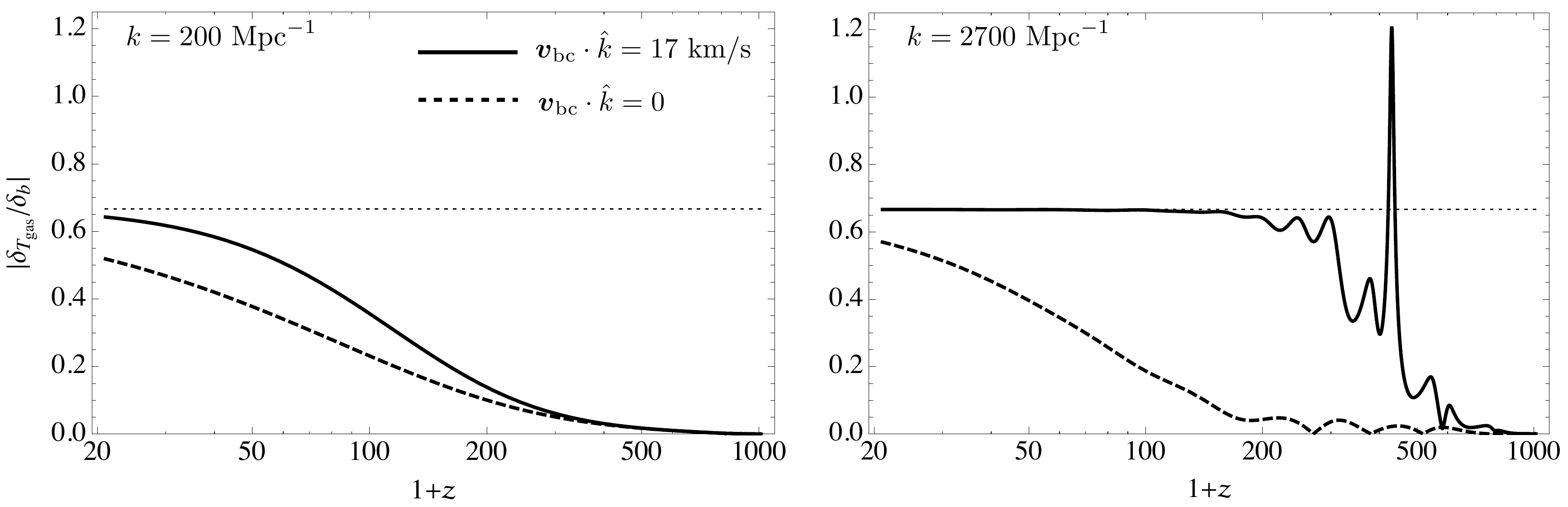}
\caption{Evolution of the ratio $|\delta_{T_{\rm gas}}/\delta_b|$ as a
function of redshift, for $k = 200$ Mpc$^{-1}$ and $k
  = 2700$ Mpc$^{-1}$, as a function of the local relative velocity. In
both cases the relative velocity speeds up the convergence towards the
adiabatic limit $\delta_{T_{\rm gas}} = \frac23 \delta_b$ (indicated with
a dotted line). The effect is much more pronounced for very small scales.} \label{fig:deltaT} 
\end{figure*}

Figure \ref{fig:deltaT} shows the evolution of the ratio $|\delta_{T_{\rm
  gas}}/\delta_b|$ for $k = 200$ Mpc$^{-1}$ and $k = 2700$
Mpc$^{-1}$. We see that the relative velocity leads to a faster
convergence to the adiabatic regime $\delta_{T_{\rm gas}} \rightarrow
\frac23 \delta_b$, with a very pronounced effect for scales much
smaller than the Jeans scale. This can be understood by considering
Eq.~(\ref{eq:Tgas-pert}), neglecting fluctuations of the free electron
fraction for simplicity. In this equation, the term $\frac23
\dot{\delta}_b$ can be seen as a forcing term; physically, it arises
from the work done by the compression and expansion of the baryonic
fluid. The term linear in $\delta_{T_{\rm gas}}$ is a
friction term, which translates the tendency for the gas temperature
to equilibrate with the (nearly) homogeneous CMB temperature through
Thomson scattering. In the
deep sub-Jeans regime the baryonic overdensity oscillates in time like
its own forcing term (\ref{eq:forcing}), so that $\dot{\delta}_b \sim (\bs{k} \cdot
\bs{v}_{\rm bc}/a) \delta_b$, which increases with wavenumber. For
very small scales, this term can be much larger than the friction term, in which case the gas
temperature fluctuation rapidly equilibrates to 2/3 of the baryon
density fluctuations.

\begin{figure}
\includegraphics[width = 87mm]{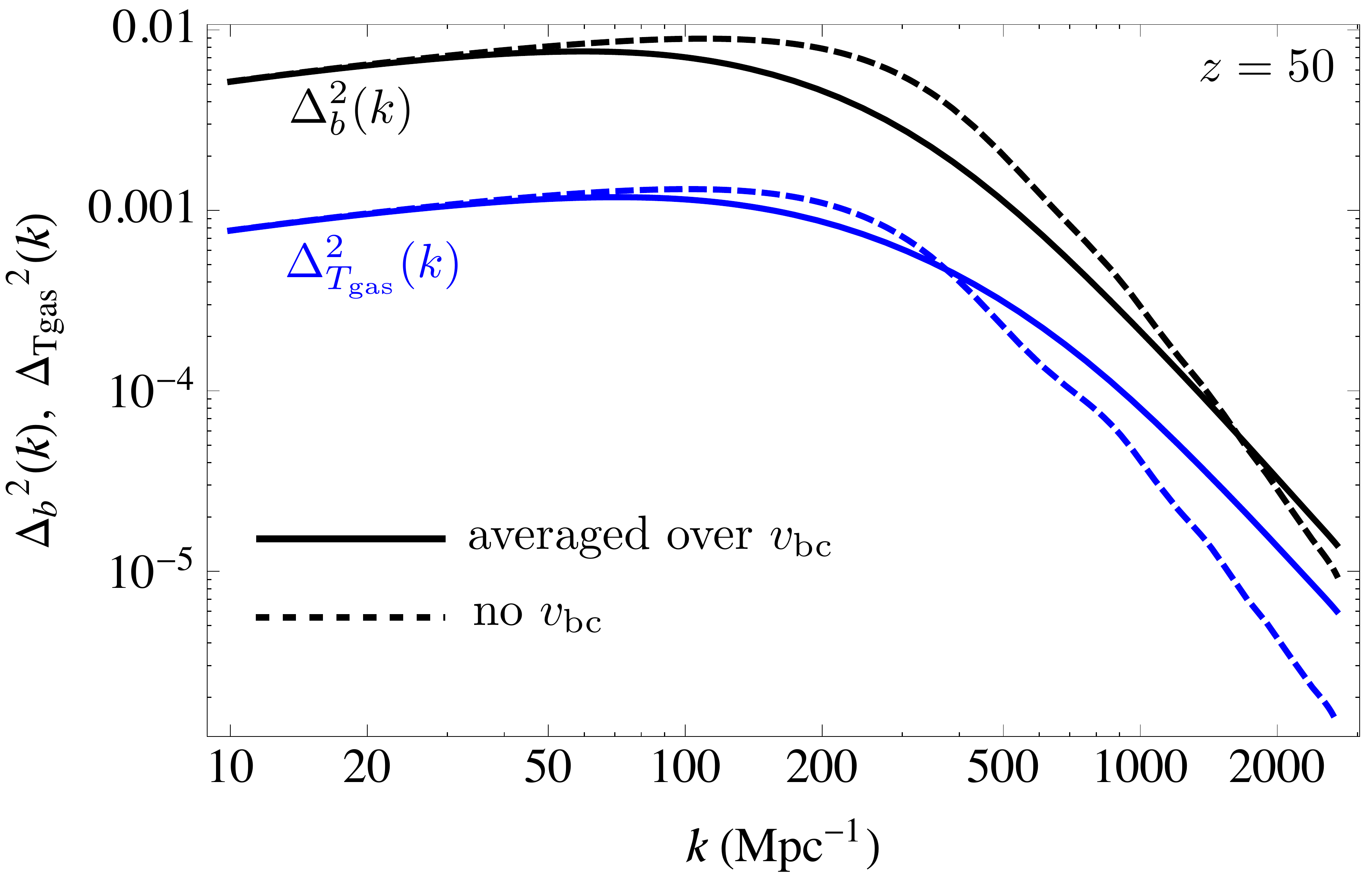}
\caption{Power per logarithmic $k$-interval for baryon density and
  temperature fluctuations at $z = 50$, neglecting the relative
  velocities (dashed lines), and averaging over their probability
  distribution (solid lines). Accounting for relative velocities leads
to a suppression of power around $k \sim 200$ Mpc$^{-1}$ and an
enhancement at smaller scales due to resonant excitation of acoustic
waves. The enhancement is more pronounced for temperature
fluctuations, which are driven towards the adiabatic regime
$\delta_{T_{\rm gas}} \rightarrow \frac23 \delta_b$ earlier on when
relative velocities are present.} \label{fig:Pdelta} 
\end{figure}

In Fig.~\ref{fig:Pdelta} we show the small-scale power
spectra of the baryon density and temperature fluctuations at $z = 50$, both in the
standard case (setting $\bs{v}_{\rm bc} = 0$), and averaged over the
Gaussian distribution of the relative velocity vector. The latter is most
efficiently computed by averaging over the one-dimensional distribution of
$\bs{v}_{\rm bc} \cdot \hat{k}$. We have checked that our result for
the total matter power spectrum agrees with that of TH10. We have also
checked that our results are in good agreement with those of
\textsc{camb} when setting $\bs{v}_{\rm bc} = 0$. The main effect of the
relative velocity is to suppress power by several tens of percent on scales $k
\sim 100-300$ Mpc$^{-1}$, and enhance it on very small scales for
which baryon acoustic oscillations get resonantly forced. The transition from suppression to enhancement occurs at
larger scales for the temperature fluctuations, due to the faster
convergence to the adiabatic regime described above.

\section{Modulation of non-linear quantities on large scales} \label{sec:large-scale}

\subsection{Motivation} \label{subsec:motiv}

The 21~cm brightness
temperature is a non-linear function of the baryon
density and temperature (see Section \ref{sec:21cm} for details). In addition, as can be seen from
Eq.~(\ref{eq:first-law}), the gas temperature itself depends
non-linearly on the gas density. The goal of this section is to show
how the large-scale fluctuations of the relative velocity between baryons and CDM leads to a
large-scale modulation of non-linear quantities, which can be
comparable to the large-scale fluctuations of linear
perturbations. 

\begin{figure}
\includegraphics[width = 87 mm]{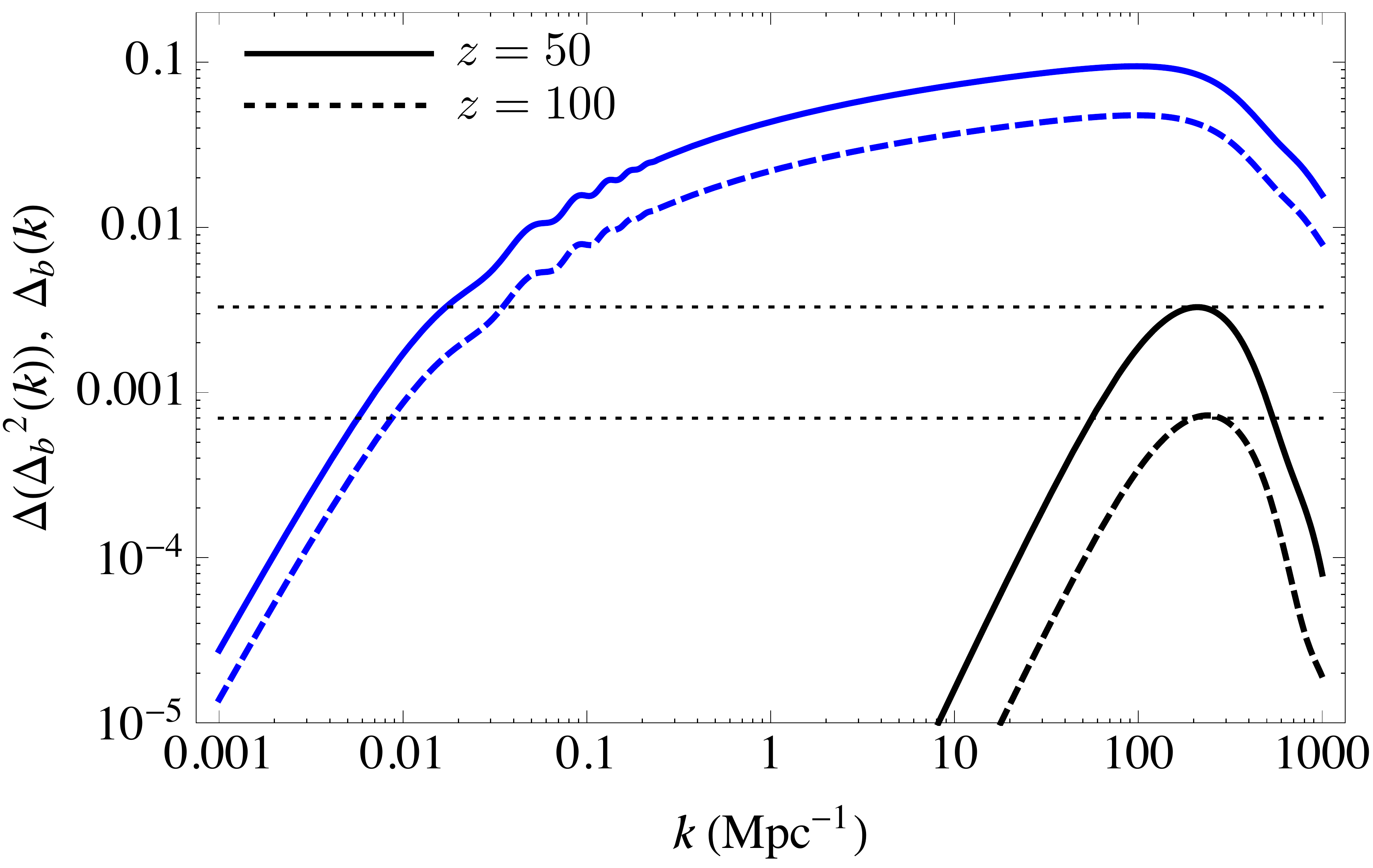}
\caption{Characteristic change in the small-scale baryon power
  $\Delta(\Delta_b^2(k)) \equiv |\langle \Delta_b^2(k)\rangle -
  \Delta_b^2(k, v_{\rm bc} = 0)|$ (black, lower two curves) and
  characteristic baryon overdensity $\Delta_b(k) \equiv \left[k^3
    P_b(k)/(2 \pi^2)\right]^{1/2}$ (blue, upper two curves), as a function
  of wavenumber, and at redshifts 100 and 50. The dotted
  lines illustrate that the long-wavelength modulation of the
  small-scale quadratic fluctuations is of the same order as the
  long-wavelength fluctuations of the linear overdensity: $(\delta_s^2)_l = \Delta \delta_s^2 \sim \delta_l$.} \label{fig:db-db2} 
\end{figure}

Let us consider a quantity $X(\rho_b)$ that depends non-linearly on the
local baryon density $\rho_b(\bs{x}) \equiv \overline{\rho}_b(1 +
\delta(\bs{x}))$. The following argument can be immediately generalized to a dependence on
multiple perturbations, such as density, temperature or ionization
fraction. Since during the dark ages $\delta \ll 1$ on all scales, we may write $X$ as a
Taylor expansion:
\barr
X\left(\rho_b(\bs{x})\right) = \chi_0 + \chi_1 \delta(\bs{x}) +
\chi_2\delta(\bs{x})^2 + \mathcal{O}(X \delta^3),
\earr
where the coefficients $\chi_0, \chi_1, \chi_2$ are functions of redshift only and are
in general of comparable magnitude. We now decompose the density fluctuation in a long-wavelength part and a short-wavelength part:
\beq
\delta(\bs{x}) = \delta_l(\bs{x}) + \delta_s(\bs{x}).
\eeq
Both $\delta_l$ and $\delta_s$ are small quantities; however, there
exists a hierarchy between them: 
\beq
\delta_l \ll \delta_s \ll1. \label{eq:hierarchy}
\eeq
In fact, for $z \lesssim 100$, taking $k_s \sim 100$ Mpc$^{-1}$ and $k_l
\sim 0.01$ Mpc$^{-1}$, the hierarchy between long- and
short-wavelength fluctuations is such that
\beq
\delta_s^2 \sim \delta_l.
\eeq
We therefore ought to write a two-parameter Taylor expansion of
$X$. To first order in $\delta_l$ and second order in $\delta_s$, we
have
\beq
X(\rho) = \chi_0 + \chi_1 (\delta_l + \delta_s) +
\chi_2 \delta_s^2 + \mathcal{O}(X \delta_s \delta_l).
\eeq
If we consider the small-scale fluctuations of $X$, we see that, to
lowest order,
\beq
X_s = \chi_1 \delta_s + \mathcal{O}(X \delta_s^2), \label{eq:Xs}
\eeq
i.e.~at small scales we only need to account for the linear term, up to corrections of
relative order $\delta_s$. However, when
computing the long-wavelength fluctuations of $X$, the quadratic term
does become important and can be comparable to the linear term, provided it
is significantly modulated on large scales:
\beq
X_l = \chi_1 \delta_l + \chi_2 (\delta_s^2)_l + \mathcal{O}(X
\delta_s^3, X \delta_l^2). \label{eq:X_l}
\eeq
In the absence of relative velocities, $\delta_s^2$ does vary 
stochastically, but mostly on small scales. On the other hand, fluctuations of the
relative velocity over large scales lead to order unity fluctuations
of the small-scale power spectrum, and therefore $(\delta_s^2)_l \sim
\delta_s^2 \sim \delta_l$. This is illustrated in Fig.~\ref{fig:db-db2}.

\begin{figure}
\includegraphics[width = 85 mm]{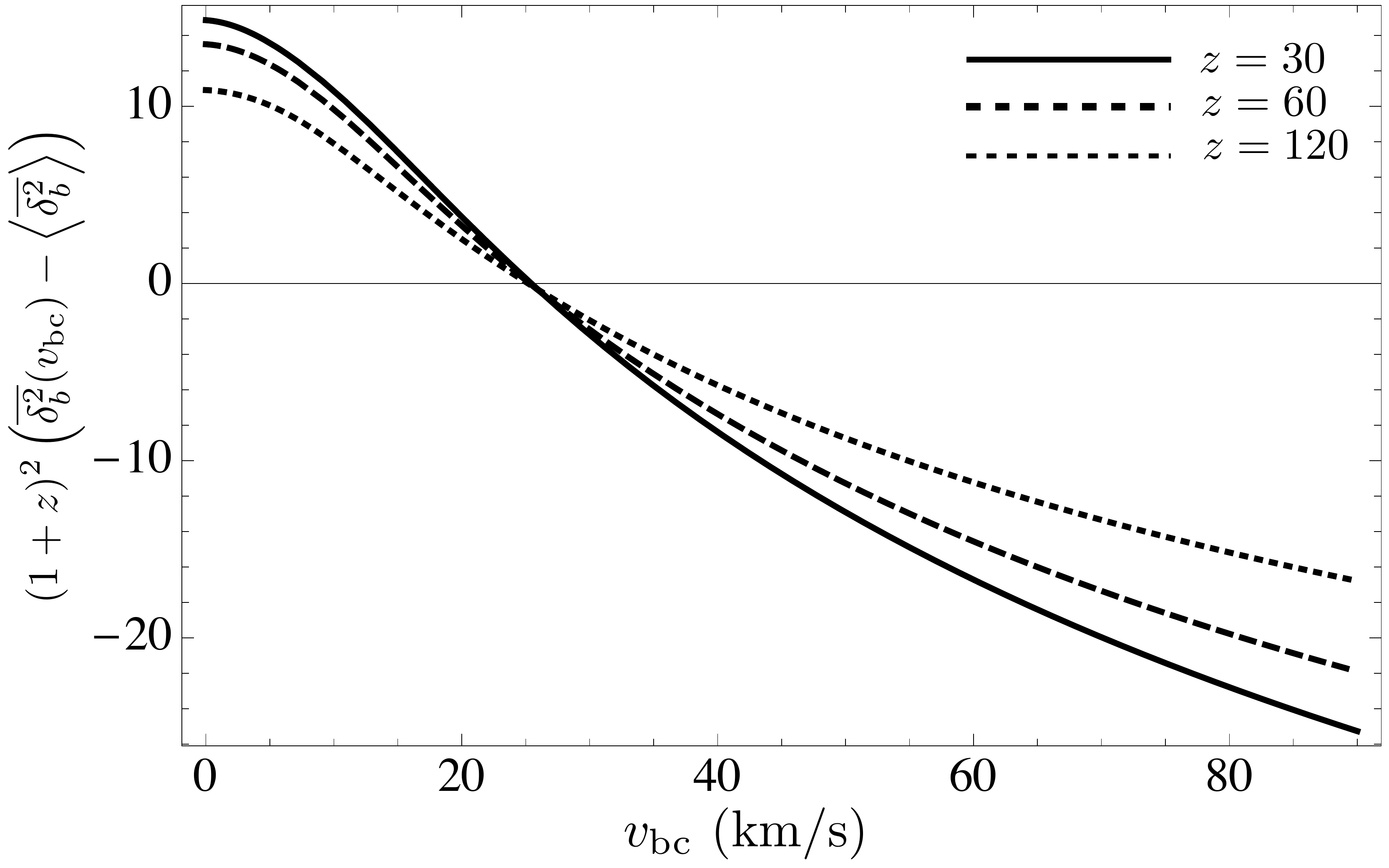}
\caption{Variation of the variance of the small-scale baryon overdensity
  as a function of the magnitude of the local relative velocity, at $z
= 30, 60$ and 120. We have multiplied $\delta_b^2$ by $(1+z)^2$ in
order to factor out the approximately linear growth of overdensities
with the scale factor during matter domination (in practice the growth
rate is slightly faster than linear with the scale factor as baryons
fall in the pre-existing dark matter potential wells).} \label{fig:db2_vbc} 
\end{figure}

In order to compute the long-wavelength fluctuation of
$\delta_s^2$, we may first smooth it over an intermediate scale of a few tenths of
Mpc, such that the smoothing scale satisfies
\beq
k_{\rm coh} \ll k_{\rm smooth} \ll k_{v_{\rm bc}}.
\eeq
The first inequality ensures that the
long-wavelength fluctuations of the field are unaffected by smoothing:
denoting the smoothed field by $\widetilde{\delta_s^2}$, we have $(\delta_s^2)_l \approx (\widetilde{\delta_s^2})_l$, up to corrections
of order $(k_l/k_{\rm smooth})^2$ with a Gaussian smoothing
kernel. The second inequality allows us to
replace the spatial averaging involved in the smoothing by a
\emph{statistical} averaging:
\beq
\widetilde{\delta_s^2} \approx \overline{\delta_s^2}(v_{\rm bc}) \equiv
\int \frac{d^3 k_s}{(2 \pi)^3} P_{\delta}(\bs{k}_s, \bs{v}_{\rm bc}).
\eeq
Finally, the fluctuating part is obtained by
subtracting the average over the Gaussian distribution of relative
velocities:
\beq
(\delta_s^2)_l (v_{\rm bc}) = \Delta \delta_s^2  \equiv
\overline{\delta_s^2} (v_{\rm bc}) - \big{\langle}
\overline{\delta_s^2} \big{\rangle}. \label{eq:ds2_l}
\eeq
As an illustration, we show the
fluctuation of the variance $\overline{\delta_s^2}(v_{\rm bc})$ as a
function of the relative velocity, and at several redshifts in
Fig.~\ref{fig:db2_vbc}.

\subsection{Correlation functions and power spectra}

In this section we give a more detailed and quantitative description
of the method to compute statistical
properties of non-linear quantities, accounting for the relative
velocity effect. A summary of this section can be found in paragraph \ref{subsubsec:summary}.

\subsubsection{Probability distribution for the overdensity}

We first need to determine the joint probability distribution for the
overdensity pair $(\delta_0, \delta_x)$ at two points with separation
$\bs{x}$. We start by describing the constrained distribution
$\mathcal{P}(\delta_0, \delta_x | \bs{v}_0, \bs{v}_x)$: \emph{the probability of the pair $(\delta_0, \delta_x)$
given fixed values of the relative velocities $\bs{v}_0 \equiv \bs{v}_{\rm bc}(\bs{0})$
and $\bs{v}_x \equiv \bs{v}_{\rm bc}(\bs{x})$}. From there the full
distribution $\mathcal{P}(\delta_0, \delta_x)$ is obtained by convolving with the six-dimensional joint
Gaussian probability distribution for $(\bs{v}_0, \bs{v}_x)$, which we
denote by $\mathcal{P}(\bs{v}_0, \bs{v}_x)$, i.e. 
\beq
\mathcal{P}(\delta_0, \delta_x) = \int d^3 v_0 d^3
v_x \mathcal{P}(\bs{v}_0, \bs{v}_x) \mathcal{P}(\delta_0, \delta_x |\bs{v}_0, \bs{v}_x). 
\label{eq:fullprob}
\eeq
Throughout this section an overline $\overline{X}$ denotes the
averaging with respect to the distribution of overdensities \emph{at fixed values of the relative velocities}
and brackets $\langle . \rangle$ denote the subsequent averaging over the
distribution of relative velocities. 

We decompose the density field into its small-scale
contribution $\delta_s$, which only contains modes with $k \geq
k_{v_{\rm bc}}$ and its long-wavelength contribution $\delta_l
\equiv \delta - \delta_s$ (here $\delta_l$ includes not only large-scale modes but all modes
with $k \leq k_{v_{\rm bc}}$). 

The distribution of the small-scale modes $\mathcal{P}_s$ is a two-dimensional Gaussian with vanishing
means and variances $\overline{\delta_{0s}^2}(v_0),
\overline{\delta_{xs}^2}(v_x)$ obtained from
\beq
\overline{\delta_{s}^2}(v_{\rm
  bc}) \equiv \int_{k \geq k_{v_{\rm bc}}} \frac{d^3k}{(2 \pi)^3} P_{\delta}(\bs{k},
 \bs{v}_{\rm bc}). \label{eq:<delta^2>}
 \eeq
Since $\delta_s$ has support
only on $k \geq k_{v_{\rm bc}}$, the covariance $\overline{\delta_{0 s}
\delta_{x s}}$ rapidly vanishes for $x \gtrsim$ few $k_{v_{\rm
    bc}}^{-1} \ll x_{\rm coh}$. It is therefore only
significant for separations well within the coherence scale of the relative velocity, for which
$\bs{v}_0 = \bs{v}_x$. It can be computed at all separations by
Fourier transforming either $P(\bs{k}, \bs{v}_0)$ or $P(\bs{k},
\bs{v}_x)$:
\beq
\overline{\delta_{0s} \delta_{xs}}(\bs{x}, \bs{v}_0) = \int_{k \geq k_{v_{\rm bc}}}
\frac{d^3k}{(2 \pi)^3} \rme^{i \bs{k} \cdot \bs{x}} P_{\delta}(\bs{k},
 \bs{v}_0). \label{eq:d0s_dxs}
\eeq
It will be useful in what follows to understand the symmetries of this
function. First, consideration of the system
(\ref{eq:fluid1})-(\ref{eq:fluid5}) shows that the transfer function
of the overdensity is a function of $k$ and $\hat{k} \cdot \bs{v}_{\rm
  bc}$ only, and so will be the power spectrum. Moreover, the complex conjugate $\delta^*(k, \hat{k} \cdot \bs{v}_{\rm
  bc}) = \delta(k, -\hat{k} \cdot \bs{v}_{\rm bc})$, which implies that the power spectrum depends on $k$ and the
absolute value $|\bs{k} \cdot \bs{v}_{\rm bc}|$, i.e. is symmetric in
$\bs{v}_{\rm bc}$. This implies that the correlation function
$\overline{\delta_{0s} \delta_{xs}}$ is a function of $x, v_{\rm bc}$
and $|\bs{x} \cdot \bs{v}_{\rm bc}|$ only, and is also an even function of
$\bs{v}_{\rm bc}$.

The large-scale pieces $(\delta_{0l}, \delta_{xl})$ have a priori
non-zero correlations with the relative velocity field. Specifically,
symmetry considerations show that the non-vanishing
correlations are $\langle \delta_{0l} v_{x,||} \rangle = - \langle
\delta_{xl} v_{0,||} \rangle$, where $v_{i||}
\equiv \bs{v}_i \cdot \hat{x}$ is the projection of the relative
velocity along the separation vector. For given values of the relative
velocity, the distribution $\mathcal{P}_l$ is therefore a constrained Gaussian, with means
\barr
\overline{\delta_{0}} = \frac{\langle \delta_0 v_{x ||}\rangle }{(1 - c_{||}^2) \sigma_{1d}^2}\left(v_{x||}
  - c_{||} v_{0||}\right), \label{eq:<d0>}\\
 \overline{\delta_{x}} = \frac{\langle \delta_x v_{0 ||}\rangle }{(1 - c_{||}^2) \sigma_{1d}^2}\left(v_{0||}
  - c_{||} v_{x||}\right), \label{eq:<dx>}
\earr
where we have dropped the subscripts ``$l$'' since these expressions are
also valid for the total overdensity. The covariance matrix has
elements
\barr
\overline{\delta_{0l}^2} - \left(\overline{\delta_0}\right)^2&=& \langle \delta_l^2\rangle - \frac{\langle
  \delta_0 v_{x||}\rangle^2}{(1- c_{||}^2)\sigma_{1d}^2},\\
\overline{\delta_{xl}^2} - \left(\overline{\delta_x}\right)^2&=& \langle \delta_l^2\rangle - \frac{\langle
  \delta_x v_{0||}\rangle^2}{(1- c_{||}^2)\sigma_{1d}^2},\\
\overline{\delta_{0l}\delta_{xl}} - \overline{\delta_0} \times \overline{\delta_x} &=& \langle \delta_{0l} \delta_{xl}
\rangle + c_{||} \frac{\langle \delta_0 v_{x||} \rangle \langle \delta_x v_{0||} \rangle }{(1- c_{||}^2)\sigma_{1d}^2}, 
\earr
where the right-hand sides are independent of the relative velocities
$(\bs{v}_0,\bs{v}_x)$. 

For a given pair of relative velocities $(\bs{v}_0, \bs{v}_x)$, the small-scale parts
$(\delta_{0s}, \delta_{0x})$ and the large-scale parts $(\delta_{0l},
\delta_{xl})$ are independent pairs of variables, so that we may
rewrite the probability distribution for $(\delta_0, \delta_x)$ given
$(\bs{v}_0, \bs{v}_x)$ as
\barr
\mathcal{P}(\delta_0, \delta_x |\bs{v}_0, \bs{v}_x) = \int d
\delta_{0s} d \delta_{x s} \mathcal{P}_s(\delta_{0s}, \delta_{xs}
|\bs{v}_0, \bs{v}_x)\nonumber\\
\times \mathcal{P}_l(\delta_0 -\delta_{0s}, \delta_x - \delta_{xs} |\bs{v}_0, \bs{v}_x).
\earr
As a consequence, at fixed relative velocities, the sums
$\delta_0 = \delta_{0s} + \delta_{0l}$, $\delta_x = \delta_{xs} +
\delta_{xl}$ also have a two dimensional Gaussian distribution, whose
first and second order moments are just the sums of those of
$\mathcal{P}_s$ and $\mathcal{P}_l$.

The independence of small-scale and large scale modes is
only valid \emph{at fixed relative velocities} and no longer
holds after convolution with the probability distribution of relative
velocities to obtained the full probability distribution of
$(\delta_0, \delta_x)$ through Eq.~\eqref{eq:fullprob}.

When computing the cosmic average $\langle \overline{F} \rangle$ of a function $F(\delta_0, \delta_x)$, we must evaluate the integral
\beq
\Big{\langle}\overline{ F(\delta_0, \delta_x)} \Big{\rangle} \equiv \int d \delta_0 d
\delta_x  \mathcal{P}(\delta_0, \delta_x) F(\delta_0, \delta_x).
\eeq
After a change of variables we arrive at
\beq
\Big{\langle}\overline{F(\delta_0, \delta_x)} \Big{\rangle}= \Big{\langle}
\overline{F(\delta_{0s} + \delta_{0l}, \delta_{xs} + \delta_{xl})} \Big{\rangle},
\eeq
where the first averaging, denoted by an overline, is to be performed
over the independent distributions of $(\delta_{0s}, \delta_{xs})$ and
$(\delta_{0l}, \delta_{xl})$ at fixed relative velocities, and is
followed by averaging over the distribution of velocities, denoted by brackets.
With this probability distribution at hand, we may compute various
correlation functions. This will allow us to compute the
autocorrelation function and power spectrum of 21~cm fluctuations in
the next section.

\subsubsection{Autocorrelation of the density field}
Let us start by computing the autocorrelation of the density field:
\barr
\xi_{\delta}(x) \equiv \Big{\langle} \overline{\delta_0 \delta_x}
\Big{\rangle} &=& \Big{\langle} \overline{(\delta_{0s} + \delta_{0l})
  (\delta_{xs} + \delta_{xl})} \Big{\rangle}\nonumber\\
 &=& \Big{\langle}
\overline{\delta_{0s} \delta_{xs}} \Big{\rangle}  + \Big{\langle}
\overline{\delta_{0l} \delta_{xl}} \Big{\rangle}, 
\earr
where we have used the independence of small-scale and large-scale
modes at fixed relative velocity. The second avergage is just $\langle
\delta_{0l} \delta_{xl} \rangle$, obtained from Fourier transforming
$P(k < k_{v_{\rm bc}})$, which is independent of the relative velocity. The average of the small-scale correlation
function is obtained from averaging Eq.~(\ref{eq:d0s_dxs}) over the
distribution of $\bs{v}_0$, which amounts to taking the Fourier
transform of the velocity-averaged small-scale power spectrum. We
therefore arrive at
\beq
\xi_{\delta}(x) = \int
\frac{d^3k}{(2 \pi)^3} \rme^{i \bs{k} \cdot \bs{x}} \langle P_{\delta}(\bs{k},
 \bs{v}_0) \rangle. 
\eeq
By taking the Fourier transform, we see that the full-sky power
spectrum is simply obtained by averaging the local power spectrum over
the distribution of relative velocities, as one may expect intuitively.

\subsubsection{Autocorrelation of the density field squared}

We now compute the autocorrelation
function of $\delta^2$:
\beq
\xi_{\delta^2}(x) \equiv \Big{\langle} \overline{\delta_0^2
    \delta_x^2} \Big{\rangle} - \Big{\langle} \overline{\delta^2} \Big{\rangle}^2.
\eeq
Using Wick's theorem for the Gaussian variables $(\delta_0, \delta_x)$ at fixed relative
velocities (and accounting for the non-zero means), we arrive at
\barr
\xi_{\delta^2}(x) &=&    2 \bigg{\langle}
\Big{(}\overline{\delta_0 \delta_x}\Big{)}^2 -
\Big{(}\overline{\delta_0}\times  \overline{\delta_x} \Big{)}^2
\bigg{\rangle}\nonumber\\
&+& \bigg{\langle}\Big{(}\overline{\delta_0^2} - \big{\langle}
\overline{\delta^2}\big{\rangle}\Big{)} \Big{(}\overline{\delta_x^2} -
\big{\langle} \overline{\delta^2}\big{\rangle}\Big{)} \bigg{\rangle}. \label{eq:xi_delta2}
\earr
The first term in Eq.~(\ref{eq:xi_delta2}) would be present even if neglecting the effect of
relative velocities, i.e. setting their distribution
$\mathcal{P}(\bs{v}_0, \bs{v}_x)$ to the product of Dirac functions $\delta_{\rm
  D}(\bs{v}_0) \delta_{\rm D}(\bs{v}_x)$. In terms of our heuristic
derivation in the previous section, this term is of the order of
$(\delta_l^2)^2$. The effect of relative velocities is to
replace it by its average over their distribution, which may
change it by order unity. However, it remains of the order of
$(\xi_{\delta})^2 \ll \xi_{\delta}$ on all scales, and we shall neglect it in this analysis (see Section
\ref{sec:NL} for further discussion). In contrast, the second term in
Eq.~(\ref{eq:xi_delta2}) would vanish if the small-scale power
spectrum were independent of the relative velocity. One could compute
this term including contributions from both $\delta_s$ and $\delta_l$;
however, in practice, $\delta_s \gg \delta_l$ and it is dominated by
the fluctuations of the small-scale variance:
\beq
\xi_{\delta^2}(x) \approx \bigg{\langle}\Big{(}\overline{\delta_s^2}(v_0) - \big{\langle}
\overline{\delta_s^2}\big{\rangle}\Big{)} \Big{(}\overline{\delta_s^2}(v_x) -
\big{\langle} \overline{\delta_s^2}\big{\rangle}\Big{)} \bigg{\rangle},
\eeq
which is precisely the autocorrelation of $(\delta_s^2)_l$ that we
derived with a simple argument leading to Eq.~(\ref{eq:ds2_l}). Since the relative velocities at $0$ and $x$ quickly become uncorrelated for
$x \gtrsim x_{\rm coh}$, this term rapidly vanishes for separations
larger than $x_{\rm coh}$ and as a consequence its Fourier transform
(the power spectrum of $\delta^2$) will have support mostly on large
scales $k_l \leq k_{\rm coh}$, where it may be comparable to the power
spectrum of the linear field.

\subsubsection{Cross-correlation of linear and quadratic terms}

We now consider the cross-correlation function
\beq
\xi_{\delta, \delta^2}(x) \equiv \Big{\langle} \overline{\delta_0
  \delta_x^2} \Big{\rangle}.
\eeq
Using properties of Gaussian random fields at fixed relative
velocities, we get
\beq
\xi_{\delta, \delta^2}(x) = \Big{\langle} \overline{\delta_0}\times 
\overline{\delta_x^2} + 2 \overline{\delta_x}\Big{(}
\overline{\delta_0 \delta_x } - \overline{\delta_0} \times
\overline{\delta_x} \Big{)}\Big{\rangle}. \label{eq:xi_d_d2}
\eeq
Now $\overline{\delta_x^2} = \overline{\delta_{xs}^2}(v_x) +
\overline{\delta_{xl}^2}(v_{0||}^2, v_{x||}^2)$ is an \emph{even}
function of the relative velocities, whereas $\overline{\delta_0}$ has
a linear dependence on $(v_{0||}, v_{x||})$. The first term in
$\xi_{\delta \delta^2}$ therefore vanishes after averaging over
relative velocities. A similar argument shows that $\overline{\delta_x}\Big{(}
\overline{\delta_{0l} \delta_{xl} } - \overline{\delta_{0l}} \times
\overline{\delta_{xl}}\Big{)}$ averages to zero when integrating over
relative velocities. We are therefore only left with
$2 \langle \overline{\delta_x} \times \overline{\delta_{0s}
  \delta_{xs}}\rangle$. From the discussion following
Eq.~(\ref{eq:d0s_dxs}), the correlation function of small-scale
overdensities is also an even function of the relative velocity. This terms
therefore also cancels out upon averaging. In conclusion, we have
shown that the linear overdensity is not correlated with the quadratic
overdensity, even when accounting for fluctuations in relative velocities:
\beq
\Big{\langle} \overline{\delta_0
  \delta_x^2} \Big{\rangle} = 0.
\eeq
Note that this argument applies equally if the fluctuations at the two points are those of different fields (for example
$\delta_{T_{\rm gas}}(\bs{0})$ and $\delta_b^2(\bs{x})$).

\subsubsection{Summary of this section} \label{subsubsec:summary}

To summarize, by modulating the small-scale power spectrum, the
relative velocity leads to large-scale fluctuations of quadratic
quantities, $(i)$ uncorrelated with the fluctuations of linear
quantites, and $(ii)$ with autocorrelation function given by (up to
corrections of relative order $\delta^2 \ll 1$ and
$(\delta_l/\delta_s)^2 \ll 1$):
\beq
\xi_{\delta^2}^{(v_{\rm bc})}(x) =\Big{\langle}\overline{\delta_s^2}(v_0) \overline{\delta_s^2}(v_x) \Big{\rangle}-
\Big{\langle} \overline{\delta_s^2}\Big{\rangle}^2. \label{eq:xidb2-final}
\eeq
In this equation, $\overline{\delta_s^2}(v_{\rm bc})$ is the variance
of the small-scale fluctuation $\delta_s$ given a local value of the
relative velocity, and the averaging $\langle . \rangle$ is to be carried over the
six-dimensional Gaussian probability distribution for $(\bs{v}_0,
\bs{v}_x)$. In Appendix
\ref{app:correlation} we describe the numerical method and analytic approximations
we use to compute this average. 

This result could be obtained with a simpler heuristic
argument, as we discussed in Section \ref{subsec:motiv}; however, here
we have given a detailed derivation which can be generalized to higher-order
statistics if needed.

\subsection{Enhanced large-scale gas temperature fluctuations} \label{sec:TmII}

Whereas the relative velocity has no dynamical effect on the growth of
large-scale overdensities (the non-linear terms in the full fluid equations are
full divergences that integrate to zero), it does lead to additional
large-scale modulations of the gas temperature and ionization
fraction. This can be understood simply from considering the limiting
case of adiabatic cooling: in this case $T_{\rm gas} \propto n_b^{2/3}
= \overline{n}_b^{2/3}(1 + \frac23 \delta_b - \frac19 \delta_b^2
~...)$, and we see that the temperature will get additional
large-scale fluctuations from the modulations of the small-scale
power. The cooling is however non-adiabatic and we need to explicitly
solve for the coupled evolution of the gas temperature and ionization
fraction to second order. We write them in the form
\barr
T_{\rm
  gas} &=& \overline{T}_{\rm gas} \left(1 + \delta_{T_{\rm gas}}^{\rm I} +
  \delta_{T_{\rm gas}}^{\rm II}\right),\\
x_e &=& \overline{x}_e \left(1 + \delta_{x_e}^{\rm I} +
  \delta_{x_e}^{\rm II}\right),
\earr
where we have already written the relevant equations for the
first-order perturbations in Sections \ref{sec:Tgas} and \ref{sec:xe}.

We perturb Eq.~(\ref{eq:first-law}) to second order and obtain the
following equation for $\delta_{T_{\rm gas}}^{\rm II}$:
\barr
&&\dot{\delta}_{T{\rm gas}}^{\rm II} =  \frac23 \dot{\delta}_b
\left(\delta_{T_{\rm gas}}^{\rm I}-  \delta_b \right)\nonumber\\
&&+ \Gamma_{\rm C} \left( \frac{T_{\rm cmb} - \overline{T}_{\rm
      gas}}{\overline{T}_{\rm gas}} \delta_{x_e}^{\rm II} -
  \delta_{x_e}^{\rm I} \delta_{T_{\rm gas}}^{\rm I} - \frac{T_{\rm
      cmb}}{\overline{T}_{\rm gas}} \delta_{T_{\rm gas}}^{\rm II}
\right). ~~~ \label{eq:dT-2}
\earr
This equation has to be solved simultaneously with the second-order
perturbation to the free-electron fraction, whose evolution is
obtained from perturbing Eq.~(\ref{eq:dotxe-final}) to second order. We
define $\delta_{\dot{x}_e}^{\rm II}$ as the part of $\delta
\dot{x}_e/\dot{\overline{x}}_e$ quadratic in the perturbations. The evolution equation for
$\delta_{x_e}^{\rm II}$ is given by
\beq
\dot{\delta}_{x_e}^{\rm II} =
\frac{\dot{\overline{x}}_e}{\overline{x}_e}\left( \delta_{x_e}^{\rm
    II} + \frac{d \log \mathcal{A}_{\rm B}}{d \log T_{\rm gas}}
  \delta_{T_{\rm gas}}^{\rm II}+ \delta_{\dot{x}_e}^{\rm II}
\right). \label{eq:dx-2}
\eeq
We see that we have a coupled linear system for
$(\delta_{T_{\rm gas}}^{\rm II}, \delta_{x_e}^{\rm II})$ sourced by
terms quadratic in the small-scale fluctuations. Note that the full
evolution equation for the large-scale gas temperature and ionization fluctuations
also contains gauge-dependent metric perturbations \cite{Lewis_2007,
  Senatore_2009}. In principle there are also quadratic terms
containing such metric terms. However, only terms quadratic in
\emph{small-scale} perturbations are relevant, and metric terms are
suppressed by $\mathcal{O}(H^2/k_s^2) \ll 1$. We use existing
codes to compute the standard linear large-scale temperature and ionization fluctuations, that properly
account for relativistic corrections. Our correction is uncorrelated
and additive.

\begin{figure}
\includegraphics[width = 90 mm]{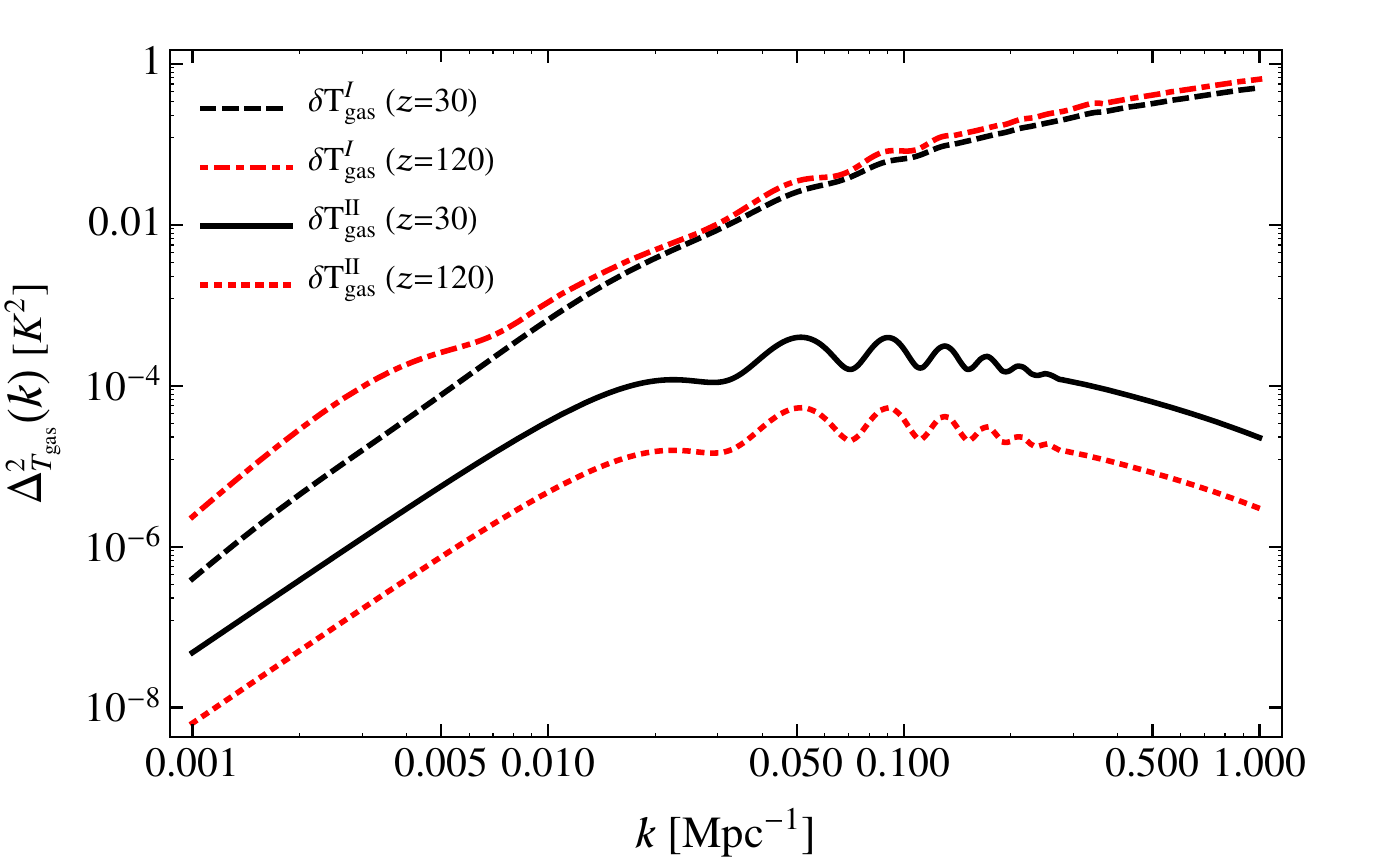}
\caption{Fluctuations of the gas temperature per logarithmic
  $k$-interval, at $z = 30$ and 120. The two upper lines show the standard
  result extracted from \textsc{camb}, and the two lower lines show the
  enhancement resulting from the modulation of small-scale
  fluctuations by the relative velocity.} \label{fig:dTgasII} 
\end{figure}

We average Eqs.~(\ref{eq:dT-2}), (\ref{eq:dx-2}) over a few Mpc
patch. They then become equations for the large-scale fluctuations $\delta_{T_{\rm gas}}^{\rm II}(v_{\rm bc}, z)$ and
$\delta_{x_e}^{\rm II}(v_{\rm bc}, z)$, sourced by the (co)variance of the
quadratic terms, obtained from our small-scale solution
described in Section \ref{sec:small-scale}. For example, the source
term of Eq.~(\ref{eq:dT-2}) is 
\barr
\dot{\delta}_{T_{\rm gas}}^{\rm II}(\textrm{source}) = \frac23 \overline{\theta_b\delta_b} -
 \frac23 \overline{\theta_b \delta_{T_{\rm gas}}^{\rm I}}- \Gamma_{\rm C}
\overline{\delta_{x_e}^{\rm I} \delta_{T_{\rm gas}}^{\rm I}},
\earr
which we compute as a function of relative velocity and redshift by
integrating the small-scale (cross-)power spectra over wavenumbers, for
instance
\beq
\overline{\theta_b\delta_b} = \int \frac{d^3 k_s}{(2 \pi)^3}
P_{\delta_b \theta_b}(\bs{k}_s, \bs{v}_{\rm bc}).
\eeq
After subtracting the average of the sources over relative
velocities, we then solve the coupled system
for $\delta_{T_{\rm gas}}^{\rm II}(v_{\rm bc}, z)$ and
$\delta_{x_e}^{\rm II}(v_{\rm bc}, z)$ with zero initial conditions at
$z_{\rm ini} = 1010$, since at that time the relative velocity has not yet
imprinted large-scale modulations of the small-scale fluctuations. We can
then compute the autocorrelation function of $\delta_{T_{\rm
    gas}}^{\rm II}$ as described in Appendix \ref{app:correlation},
and the resulting power spectrum. We show the latter in
Fig.~\ref{fig:dTgasII}, along with the standard
large-scale temperature fluctuation obtained with \textsc{camb}. We see
that the quadratic correction contributes a $\sim 10\%$ enhancement of
gas temperature fluctuations at $z = 30$ at scales $k \lesssim 0.01$ Mpc$^{-1}$.

\section{21~cm brightness temperature fluctuations during the dark
  ages} \label{sec:21cm}

\subsection{Basic equations}

The subject of 21~cm absorption and its fluctuations during the dark
ages has been treated extensively by multiple authors \cite{Loeb_2004,
  Bharadwaj_2004, Lewis_2007}. We are only concerned with computing $(i)$ corrections
to the small-scale power spectrum and $(ii)$ the enhancement of
large-scale power due to terms quadratic in small-scale fluctuations, which we showed to be uncorrelated with linear terms. We
therefore need not concern ourselves with relativistic corrections on large
scales, treated in detail in LC07. For completeness, and in order to make
all dependencies clear, we briefly summarize the relevant equations below.

\subsubsection{Spin temperature}

Following standard conventions, we define the \emph{spin
  temperature} $T_s$ from the ratio of abundances of neutral hydrogen in the triplet
state $n_1$ and in the singlet state $n_0$ as follows:
\beq
\frac{n_1}{n_0} \equiv 3 \exp\left(- \frac{E_{10}}{T_s}\right) \approx 3
\left(1 - \frac{E_{10}}{T_s}\right),
\eeq
where $E_{10} \approx 0.068$ K is the energy difference between the two states
(corresponding to a transition frequency of 21 cm), and for the second
equality we assumed that $T_s \gg E_{10}$, which is indeed valid at
all times. The spin temperature is determined from a balance between collisional
transitions, which tend to set $T_s \rightarrow T_{\rm gas}$, and radiative
transitions mediated by CMB photons, which tend to set $T_s \rightarrow T_{\rm
  cmb}$. 

The rates of upward and downward collisional transitions are
denoted by $C_{01}$ and $C_{10}$, respectively, and satisfy the detailed
balance relation
\beq
C_{01} = 3 \exp\left(- \frac{E_{10}}{T_{\rm gas}}\right) C_{10}\approx 3
\left(1 - \frac{E_{10}}{T_{\rm gas}}\right) C_{10},
\eeq
where again we used the fact that $T_{\rm gas} \gg E_{10}$. During the dark ages
the Universe is almost fully neutral and collisions with neutral
hydrogen atoms largely dominate the collisional transition rate (see
Fig.~1 of LC07). The coefficient
$C_{10}$ takes the form
\barr
C_{10} = n_{\rm H} \kappa_{10}^{\rm HH}(T_{\rm gas}),
\earr
where the temperature dependence is accurately approximated by the
simple fit $\kappa_{10}^{\rm HH}(T_{\rm gas}) \approx
3.1 \times 10^{-11} T_{\rm gas}^{0.357} \exp(-32/T_{\rm
  gas})$~cm$^3$~s$^{-1}$, with $T_{\rm gas}$ given in Kelvins \cite{Kuhlen_2006}.

We denote by $R_{10}$ and $R_{01}$ the rates of radiative transitions
mediated by CMB photons. The absorption rate $R_{01}$ is related to
the rate of spontaneous and stimulated decays $R_{10}$ through the detailed balance relation:
\beq
R_{01} = 3 \exp\left(- \frac{E_{10}}{T_{\rm cmb}}\right) R_{10} \approx 3
\left(1 - \frac{E_{10}}{T_{\rm cmb}}\right) R_{10}.
\eeq
The latter is given by
\beq
R_{10} = A_{10} \left(1 + \frac{1}{\rme^{E_{10}/T_{\rm cmb}} -
    1}\right) \approx A_{10} \frac{T_{\rm cmb}}{E_{10}},
\eeq
where $A_{10} \approx 2.85 \times 10^{-15}$ s$^{-1}$ is the spontaneous decay rate. At all times during the dark ages the total
transition rate $R_{10} + C_{10}$ surpasses the Hubble rate by several
orders of magnitude. The populations of the hyperfine states can
therefore be obtained to high accuracy by making the steady-state
approximation:
\beq
n_1\left(C_{10} + R_{10}\right) = n_0 \left(C_{01} + R_{01}\right),  
\eeq
which, using the expressions for the transition rates given above and in the
limit $E_{10} \ll T_{\rm gas}, T_{\rm cmb}$, leads to the following
equation for the spin temperature
\beq
T_{s} = T_{\rm cmb} + (T_{\rm gas} - T_{\rm cmb}) \frac{C_{10}}{C_{10}
  + A_{10}\frac{T_{\rm gas}}{E_{10}}}. \label{eq:Ts}
\eeq 

\subsubsection{Brightness temperature}

Following the convention in the field, we define the brightness
temperature $T_{b}$ as the temperature characterizing the \emph{difference}
between the radiation field processed by the 21~cm transition and the
background CMB radiation field. Since $E_{10} \ll T$ we are in the Rayleigh-Jeans
tail of the spectrum. \yacine{In the optically thin limit, and up to
  corrections of the order of its peculiar velocity with respect to
  the CMB \cite{Lewis_2007}, the brightness temperature observed in the gas rest frame is $T_b^{\rm local} = \tau \left(T_s - T_{\rm
      cmb}\right)$, where $\tau$ is the Sobolev optical depth,
  discussed below. The photon phase-space density
  (or $I_{\nu}/\nu^3$ up to multiplicative constants, where $I_{\nu}$ is the
  specific intensity), is a frame-invariant quantity, conserved in
  the absence of emission and absorption. This ensures that the ratio
  $T_b/\nu$ is frame-independent and conserved along the
  photon trajectory. At redshift zero the observed brightness
  temperature is therefore
\beq
T_b = (1 + z)^{-1} \tau \left(T_s - T_{\rm cmb}\right), \label{eq:Tb-local}
\eeq
where again we have neglected corrections of the order of the peculiar
velocity of the gas, as well as the effect of gravitational
potentials along the photon trajectory. The Sobolev optical depth is given by}
\beq
\tau = \frac{3 E_{10}}{32 \pi T_s} x_{\rm HI} n_{\rm H} \lambda_{10}^3
\frac{A_{10}}{H + \partial_{\parallel} v_{\parallel}}, \label{eq:tau}
\eeq
where $\lambda_{10}=21$cm, $x_{\rm HI}$ is the fraction of neutral hydrogen and
$\partial_{\parallel} v_{\parallel}$ is the line-of-sight gradient (in proper space) of the
component of the peculiar velocity along the line of sight. This equation can easily be generalized to arbitrary
optical depth by making the replacement $\tau \rightarrow (1 -
\rme^{-\tau})$; however, the optical depth is at most a few percent during the
dark ages, and we have chosen to keep the lowest-order approximation
in order to have more tractable expressions later on. 

\yacine{In the above derivation} we have assumed that the line is infinitely narrow. In
reality, the line has a finite width due to thermal motions
of the atoms (an additional subtlety being that the spin temperature
is in fact a velocity-dependent function \cite{Hirata_2007}). This
leads to an averaging of fluctuations with radial wavenumber $k_{||}$ larger
than $k_{\rm th} \equiv (1 +z)^{-1} H \sqrt{m_{\rm H}/T_{\rm gas}}$,
which is of the order of the Jeans scale, and is approximately $300$,
400 and 500 Mpc$^{-1}$ at $z = 100, 50$ and 30, respectively. In
practice, observations are made with a finite window function, orders
of magnitude broader than the thermal line width, and the resulting averaging
along the line of sight should dominate any finite line width
effects. 

In closing of this review section, we point out that the term
$\partial_{||} v_{||}$ in the denominator of the optical depth
(\ref{eq:tau}) is often referred to as a ``redshift-distortion''
term. This is a misnomer: although this term is similar to
an actual redshift-space distortion term (see Section
\ref{sec:distortion}), it is very different in nature. Redshift-space
distortions are an \emph{observational} effect, they come from the
inability of an observer to disentangle the intrinsic cosmological redshift of a source
(in a given gauge) from the additional redshifting due to its peculiar
velocity along the line of sight. In contrast, the term $\partial_{||}
v_{||}$ in the optical depth represents a perturbation of the
Hubble expansion rate at the absorber's location, and does not require
any observer (besides the fact that the observer determines the line
of sight). It translates the fact that a photon can resonantly interact with less
atoms the larger their velocity gradient is along the direction of
propagation. See also Ref.~\cite{Mao_2012}.

\subsection{Fluctuations}

\subsubsection{Expansion in density and temperature fluctuations}

The brightness temperature is a function of the local hydrogen density
and gas temperature, and its fluctuations can therefore be expanded
in terms of their perturbations. We neglect fluctuations in $T_{\rm cmb}$ and $x_e$ and only
consider density and temperature fluctuations: 
\barr
n_{\rm H}(z, \bs{x}) &=& \overline{n}_{\rm H}(z)\left(1 + \delta_{\rm
    H}(z, \bs{x})\right),\\
T_{\rm gas}(z, \bs{x}) &=& \overline{T}_{\rm gas}(z)\left(1 + \delta_{T_{\rm
    gas}}(z, \bs{x})\right),
\earr
where we recall that $\delta_{\rm H} = \delta_b$ up to negligible
corrections. We also define the dimensionless small quantity
\beq
\delta_v \equiv \frac{\partial_{\parallel} v_{\parallel}}{H} \equiv \frac{1+z}{H}
\nabla_{\parallel} v_{\parallel}, \label{eq:dv}
\eeq
where $\nabla$ is the comoving gradient. The brightness temperature
depends locally on $\delta_v$ only through $T_b \propto (1 +
\delta_v)^{-1}$, which will simplify the expression for perturbations.

Combining Eqs.~(\ref{eq:Ts}) to (\ref{eq:tau}), we expand the
brightness temperature to second order in the density and temperature
fluctuations.  
\barr
T_b = \overline{T}_b (1 - \delta_v + \delta_v^2)
+ \left(\mathcal{T}_{\rm H}~ \delta_{\rm H} + \mathcal{T}_{T}
~  \delta_{T_{\rm gas}}\right)(1 - \delta_v) \nonumber\\
+  \mathcal{T}_{\rm HH} ~\delta_{\rm H}^2 +\mathcal{T}_{{\rm H}T}~
\delta_{\rm H} \delta_{T_{\rm gas}} + \mathcal{T}_{TT} ~\delta_{\rm
  H}^2  + \mathcal{O}(\delta^3), ~~~\label{eq:expansion}
\earr
where the mean brightness temperature is defined by setting all
perturbations to zero, and all the coefficients $\mathcal{T}_{ij}$ in the expansion are functions of redshift
only. 

\begin{figure}
\includegraphics[width = 87 mm]{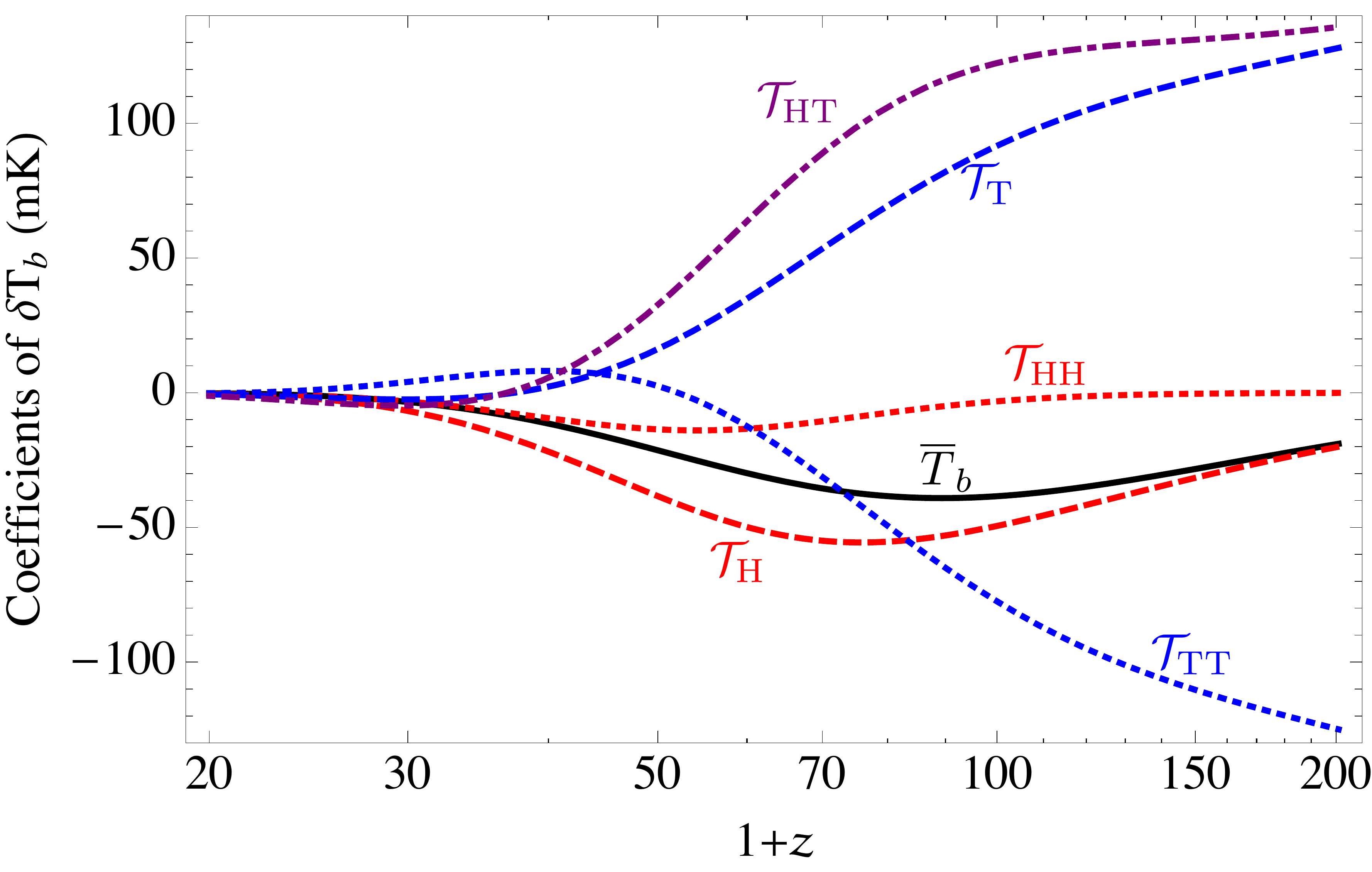}
\caption{Coefficients of the density and temperature fluctuations in
  the expansion of the brightness temperature (\ref{eq:expansion}), as
a function of redshift.} \label{fig:coeffs} 
\end{figure}

We have computed  the relevant coefficients numerically (see e.g. Ref.~\cite{Pillepich_2007} for some explicit
analytic expressions) and show them in
Fig.~\ref{fig:coeffs}. Their qualitative behavior can be easily
understood as follows. 

$\bullet$ For $z \gtrsim 100$, collisions efficiently couple the spin
temperature to the gas temperature, $T_s \approx T_{\rm
  gas}$. Without the velocity gradient term, we therefore have 
\beq
T_b \propto n_{\rm H} \left(1 - \frac{T_{\rm cmb}}{T_{\rm gas}}\right).
\eeq
The dependence on the hydrogen density is linear, so that
$\mathcal{T}_{\rm HH} \rightarrow 0$ and $\mathcal{T}_{\rm H}
\rightarrow \overline{T}_{\rm b}$. The mean brightness temperature
is proportional to $T_{\rm gas} - T_{\rm
  cmb}$, which becomes closer to zero at high redshift due to efficient
Compton heating of the gas by CMB photons. The dependence on $T_{\rm gas}$ in
the denominator implies that $\mathcal{T}_{{\rm H} T} \approx
\mathcal{T}_{T} \approx - \mathcal{T}_{TT}$, and these functions are
not suppressed as $\overline{T}_b$ as they do not have a factor of
$(T_{\rm gas} - T_{\rm cmb})$: they instead increase at high redshift
proportionally to the optical depth $\tau \propto (1 + z)^{3/2}$

$\bullet$ For $z \lesssim 50$ collisions become very inefficient and
$T_s \approx T_{\rm cmb}$, with a small difference proportional to the
collision coefficient: $T_s - T_{\rm cmb} \propto n_{\rm H}
\kappa_{10}(T_{\rm gas}$). This implies that the dependence of the
brightness temperature on $n_{\rm
  H}$ is approximately quadratic so that $\mathcal{T}_{\rm H} \approx
2 \overline{T}_b \approx 2 \mathcal{T}_{\rm HH}$. As time progresses
the optical depth gets smaller and all coefficients are rapidly damped.

\subsubsection{Redshift-space distortions} \label{sec:distortion}

In what follows we shall assume that the observer's peculiar velocity
with respect to the CMB can be accurately determined from independent
observations, and subtracted. 

Let us consider a parcel of absorbing material at redshift $z$,
i.e. at comoving radial position 
\beq
r(z) = \int_0^{z} \frac{dz'}{H(z')}. \label{eq:r(z)}
\eeq
If the parcel is moving along our line of sight with respect to its local comoving frame with a peculiar
velocity $v_{\parallel}$ (where $v_{\parallel} > 0$ if the gas is moving away from
us), then the \emph{observed} wavelength of the redshifted 21~cm line
is, to first order in $v_{||}$, 
\beq
\lambda_{\rm obs} = \lambda_{10} (1 + v_{\parallel})(1 + z).
\eeq
Therefore the \emph{observed} redshift, which is the only measurable
quantity, is given by
\beq
1+ z_{\rm obs} \equiv \frac{\lambda_{\rm obs}}{\lambda_{10}} = (1 +z) (1 + v_{\parallel}).
\eeq
From this measured redshift, one would infer a radial comoving
distance $r(z_{\rm obs})$, which is related to the actual position $r(z)$ by
\beq
r(z) \approx r(z_{\rm obs}) - \frac{1 +
z_{\rm obs}}{H(z_{\rm obs})}~ v_{\parallel}. \label{eq:rz-rzobs}
\eeq
The brightness temperature observed at a given
wavelength $\lambda_{\rm obs}$ arises from absorption at $r(z)$:
$T_b^{\rm obs} = T_b(r(z))$. Using Eq.~(\ref{eq:rz-rzobs}), and to
linear order in $v_{||}$, this is related to $r(z_{\rm obs})$ through
\barr
T_b^{\rm obs} =\left[T_b - \frac{1
  + z}{H}~ v_{\parallel}\nabla_{\parallel} (\delta T_b)\right]_{r_{\rm obs}},
\earr
where the gradient is with respect to comoving distance along the
line of sight (at fixed redshift\footnote{Note that throughout we have neglected terms of relative order $a
H/k$, such as, for instance, the term $v_{||} (1+z) \partial
T_b/\partial z$. We also do not account for metric perturbations along
the photon trajectory, which are pure large-scale terms.}), and only
acts on the perturbation $\delta T_b$. This equation and the resulting
Fourier transform are equivalent to
Eqs.~(51) and (56) of Ref.~\cite{Mao_2012}, in the optically thin
limit, and to lowest order in $v_{||}$.

The perturbation to the observed brightness temperature is therefore:
\beq
\delta T_b^{\rm obs} = \delta T_b(1 + \delta_v) - \frac{1+z}{H} \nabla_{\parallel}(v_{\parallel}
\delta T_b), \label{eq:Tb-obs}
\eeq
where we have simply used the definition (\ref{eq:dv}) of $\delta_v$
and rewritten $\nabla_{\parallel}(v_{\parallel} ~\delta T_b) = (\nabla_{\parallel} v_{\parallel})
\delta T_b+ v_{\parallel} \nabla_{\parallel} \delta T_b$.

The last term in Eq.~(\ref{eq:Tb-obs}) is the total derivative of a
quadratic term and does not fluctuate on large scales. Indeed, when
approximating the spatial average by a statistical average, we have, for any two scalar quantities $\delta_1, \delta_2$,
\barr
\langle \nabla (\delta_1 \delta_2) \rangle &=& \langle \delta_1 \nabla
\delta_2 +  (\nabla
\delta_1) \delta_2 \rangle \nonumber\\
&=& \int \frac{d^3k}{(2 \pi)^3}\langle\delta_1^* i \bs{k}
  \delta_2 +(i \bs{k} \delta_1)^* \delta_2\rangle = 0.
\earr
Using Eq.~(\ref{eq:expansion}) we therefore have, to second order in all fluctuations,
\barr
\delta T_b^{\rm obs} &=&  \mathcal{T}_{\rm H} ~\delta_{\rm H} + \mathcal{T}_{T}
 ~ \delta_{T_{\rm gas}} - \overline{T}_b ~ \delta_v
  \nonumber\\
&+&  \mathcal{T}_{\rm HH} \Delta(\delta_{\rm H}^2) +
\mathcal{T}_{TT} \Delta( \delta_{T_{\rm
 gas}}^2)  + \mathcal{T}_{{\rm H}T} 
\Delta ( \delta_{\rm H} \delta_{T_{\rm gas}} ) ,\nonumber
\earr
where $\Delta (\delta_i \delta_j )$ is the \emph{fluctuation} of the
quadratic term $\delta_i \delta_j$ about its mean value\footnote{The mean of the quadratic terms
should be formally included in $\overline{T}_b$, even though we do not add
these terms in practice as they are completely negligible.}. We see that quadratic terms involving $\delta_v$ very conveniently cancel out
but emphasize that this is only valid in the optically thin limit; there are additional corrections of order $\tau$ that do
contain such terms and that we are neglecting for simplicity. 

Following LC07, we define the ``monopole source'' as:
\beq
\delta_s \equiv \frac{\mathcal{T}_{\rm H} \delta_{\rm H} +
  \mathcal{T}_{\rm T} \delta_{T_{\rm gas}}^{\rm I}}{\overline{T}_b}.
\eeq
We also define $\delta T_b^{\rm II}$ as the total contribution
of quadratic terms (and remind the reader that $\delta_{T_{\rm gas}} =
\delta_{T_{\rm gas}}^{\rm I} +  \delta_{T_{\rm gas}}^{\rm II}$
effectively contains quadratic terms itself): 
\barr
\delta T_{\rm b}^{\rm II} &\equiv&  \mathcal{T}_{\rm HH} \Delta(\delta_{\rm H}^2) +
\mathcal{T}_{TT} \Delta( \delta_{T_{\rm
 gas}}^2) \nonumber\\
 &+& \mathcal{T}_{{\rm H}T} 
\Delta ( \delta_{\rm H} \delta_{T_{\rm gas}} ) +\mathcal{T}_{T}  ~ \delta_{T_{\rm gas}}^{\rm II}.
\earr
Finally, we bear in mind that our expression does not account for relativistic
corrections on large scales, of order $\sim \overline{T}_b \phi,
\overline{T}_b v$, which we denote by $\delta T_b^{\rm rel}$. 

Our final expression for the observed brightness temperature is
therefore
\beq
\delta T_b^{\rm obs} = \overline{T}_b (\delta_s - \delta_v) + \delta
T_b^{\rm II} + \delta T_b^{\rm rel}. \label{eq:dTb-final}
\eeq

\subsection{Angular power spectrum}

We define $P_0(k)$ as the power spectrum of the terms independent of
the direction of the line of sight, i.e. $\overline{T}_b
\delta_s + \delta T_b^{\rm II} +\delta T_b^{\rm rel}$. In Fourier space,
$\delta_v = (\hat{n} \cdot \hat{k})^2 \theta_b/H$, and we define $P_v(k)$ as the power
spectrum of $\theta_b/H$. Finally, we define $P_{0 v}(k)$ as the cross-power
spectrum of the two. 

The \emph{angular} power spectrum of 21~cm brightness temperature fluctuations
from observed redshift $z \equiv z_{\rm obs} \equiv \nu_{21}/\nu_{\rm obs} -1$
is then given by \cite{Bharadwaj_2004, Lewis_2007}
\barr
C_{\ell}(z) &=& \rme^{- 2 \tau_{\rm reion}} \Big{[}4 \pi \int \frac{d^3 k}{(2 \pi)^3} P_0(k,
z) \alpha_{\ell}(k, z)^2\nonumber\\
&+& 8 \pi \int \frac{d^3 k}{(2 \pi)^3} P_{0v}(k,
z) \alpha_{\ell}(k, z)\beta_{\ell}(k, z)\nonumber\\
&+& 4 \pi \int \frac{d^3 k}{(2 \pi)^3} P_{v}(k,
z) \beta_{\ell}(k, z)^2 \Big{]}\label{eq:Cl}
\earr
where
\barr
\alpha_{\ell}(k, z) &\equiv& \int dr' j_{\ell}(k r') W_{z}(r'), \label{eq:alpha_l}\\
\beta_{\ell}(k, z) &\equiv& \int dr' j_{\ell}''(k r') W_{z}(r'),
\earr
and $W_z(r')$ is a window function centered at the radial comoving
distance $r(z)$ accounting for the finite spectral resolution $\Delta
\nu$. The term $\rme^{- 2 \tau_{\rm reion}}$ accounts for Thomson scattering of photons out of the line
of sight by free electrons after reionization.
In Eq.~(\ref{eq:Cl}) we have neglected the variation of the
various power spectra across the redshift interval $\Delta z$
corresponding to the width of the window function. Since the power
spectra vary on a redshift scale $\Delta z \sim z$, this amounts to
neglecting terms of order $(\Delta \nu/\nu)^2$ provided $\int r'
W_z(r') dr' = r(z)$.

For $z \gg 1$ and for our fiducial cosmology, $r(z) \approx r(\infty)
= 14.9$ Gpc. During matter domination, the
change in comoving separation corresponding to a frequency
width $\Delta \nu/\nu = \Delta z/(1 + z)$ is therefore
\barr
\frac{\Delta r}{r} &\approx& \frac{c \Delta z}{r(\infty) H_0 \Omega_m^{1/2} (1+z)^{3/2}}
\approx \frac{0.57}{\sqrt{1 + z}} \frac{\Delta \nu}{\nu} \\
&\approx& 4 \times 10^{-4} \frac{\Delta \nu}{0.1 ~\rm MHz} \sqrt{\frac{1+z}{101}},
\earr
where $\nu = 1420$ MHz$/(1+z)$ is the observed frequency of the 21~cm transition at redshift $z$.
We may use the Limber approximation for $ \ell \gg (\Delta
r/r)^{-1}$, that is for
\beq
\ell \gg 2500 ~\frac{0.1 ~\rm MHz}{\Delta \nu} \sqrt{\frac{101}{1+z}}.
\eeq
In this regime, the velocity terms are suppressed (see Appendix \ref{app:Cl-Limber}), and the Limber approximation gives
\cite{Lewis_2007}
\beq
\frac{\ell^2 C_{\ell}}{2 \pi} \approx \frac{\pi r(z)}{\ell}
\frac{k^3 P_0(k)}{2 \pi^2}\big{|}_{k = \ell/r} \int dr' W_z(r')^2.
\eeq
For scales $\ell \lesssim r/\Delta r$, we compute the angular power
spectrum numerically. We first generate the spherical Bessel function
up to $\ell = 10^4$ with sufficient resolution in both $\ell$ and
$k$ using a modified version of \textsc{cmbfast} \cite{ZS1996}. We then use a trapezoidal integration scheme to integrate the stored Bessel functions over a Gaussian window function with varying width as
prescribed in Eq.~\eqref{eq:alpha_l}. We checked for convergence and
determined that 200 steps in $r$ are sufficient. In addition we have
checked our code for consistency with analytical expressions for a
top-hat window function. We also found good agreement with the monopole spectrum generated with \textsc{camb} sources.

\subsubsection{Corrections to the small-scale angular
power spectrum} \label{sec:limber}

We first consider the small-scale angular power spectrum, $\ell
\gtrsim 10^5$ corresponding to $k$ greater than a few Mpc$^{-1}$. At
these scales we only need to consider the terms linear in the baryon density and temperature fluctuations (see
Eq.~(\ref{eq:Xs}) and associated discussion). For definiteness, we shall assume a window function $\Delta \nu = 0.1$
MHz and use the Limber approximation, in which the velocity term
$\delta_v$ cancels. The only relevant term is
therefore the ```monopole'' term, which must be averaged over relative velocities.

\begin{figure}
\includegraphics[width = 87mm]{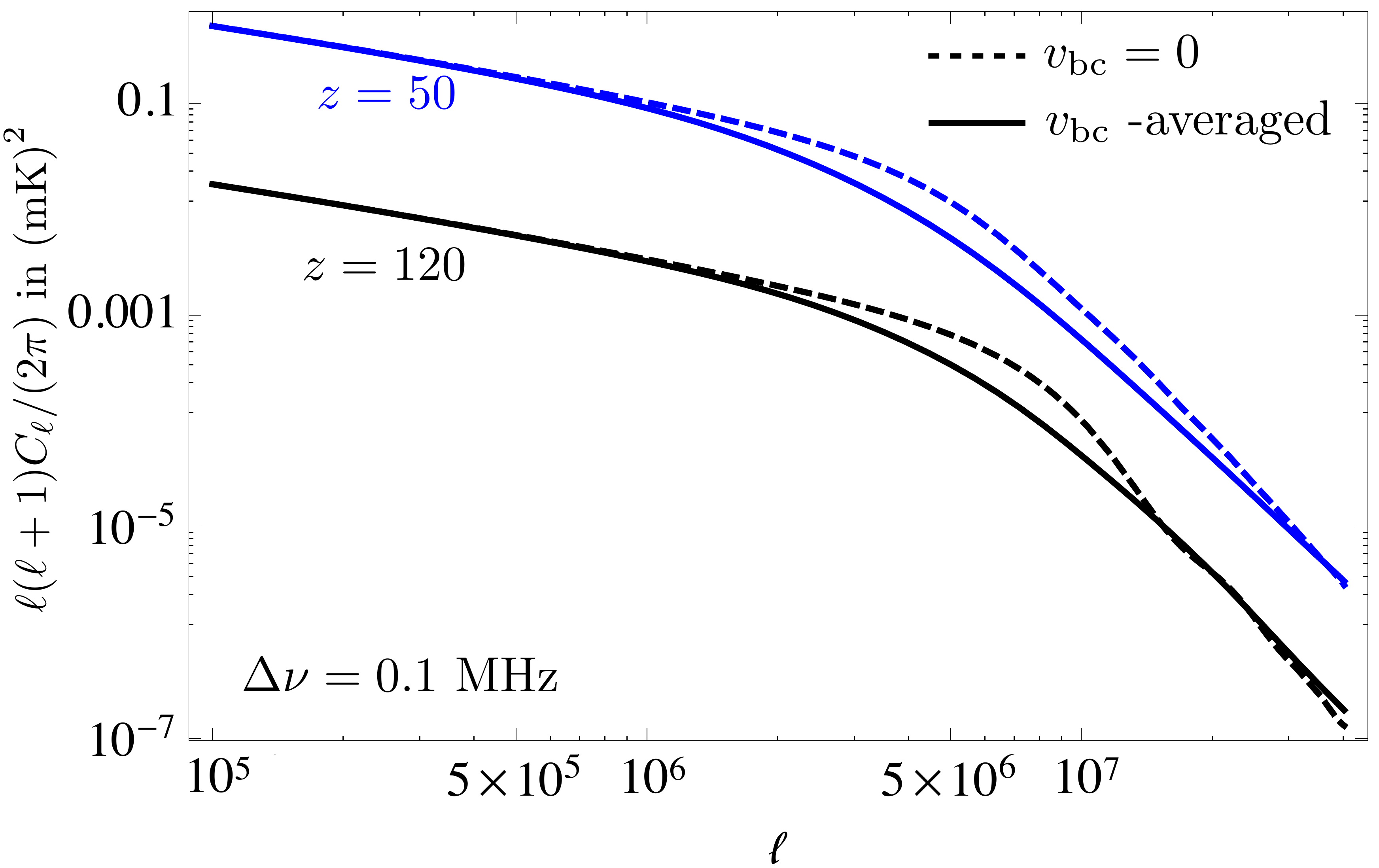}
\caption{Small-scale angular power spectrum of 21~cm brightness
  temperature fluctuations at redshifts 120 and 50, neglecting the
  effect of relative velocities (dashed lines), and averaging over
  relative velocities (sold lines). The relative change is more than
  50\% at $\ell \approx 5 \times 10^6$.}\label{fig:Cl-small} 
\end{figure}

We show the resulting small-scale power spectrum in
Fig.~\ref{fig:Cl-small} and compare it to the case without
relative velocities. We see that the relative velocities lead to power being suppressed by as much as
$\sim 50\%$ at the ``knee'' corresponding to the Jeans scale, $\ell \approx 5
\times 10^6$. Fluctuations can be enhanced for $\ell \gtrsim 2 \times
10^7$, due to the resonant excitation of acoustic waves which we
described in Section \ref{sec:deltab-ss}. 

Even though the relative velocity affects the small-scale angular
power spectrum at order unity, observations of the highly-redshifted
21~cm radiation with an angular resolution $\Delta \theta \lesssim
10^{-5}$ steradian would be
extremely challenging, if not merely impossible. We now turn to the
still challenging but more accessible large angular scales.

\subsubsection{Corrections to the large-scale angular power spectrum}

\begin{figure}[h]
\includegraphics[width = 88 mm]{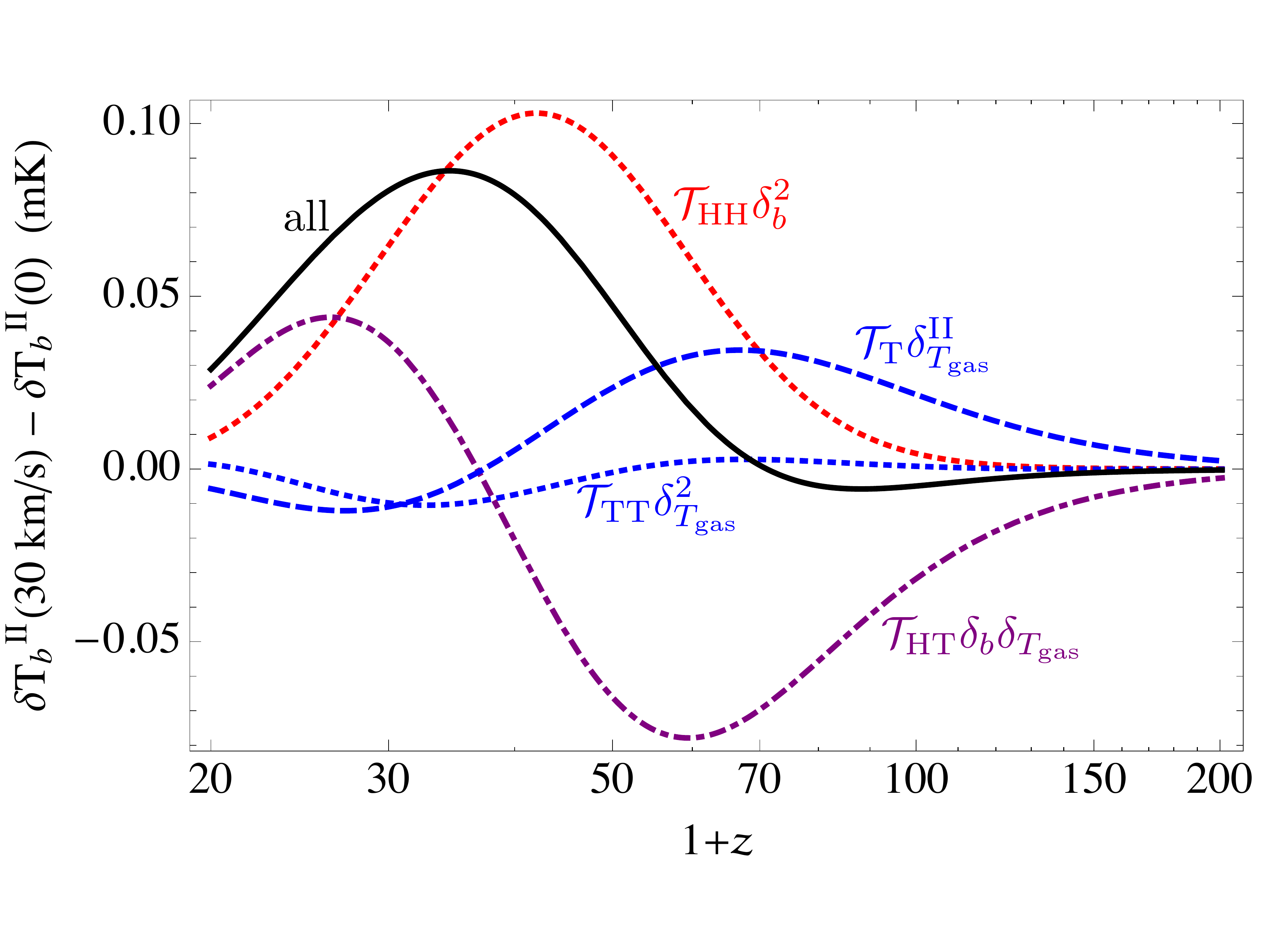}
\caption{Characteristic amplitude of the quadratic part of
  21~cm brightness temperature fluctuations, $\delta T_b^{\rm
    II}(v_{\rm bc} = 30$ km/s$) - \delta T_b^{\rm II}(0$ km/s), as a
  function of redshift. The colored lines show the contributions of
  the different terms, and the black solid line is the sum of them.} \label{fig:dTbII_z} 
\end{figure}

\begin{figure}
\includegraphics[width = 87 mm]{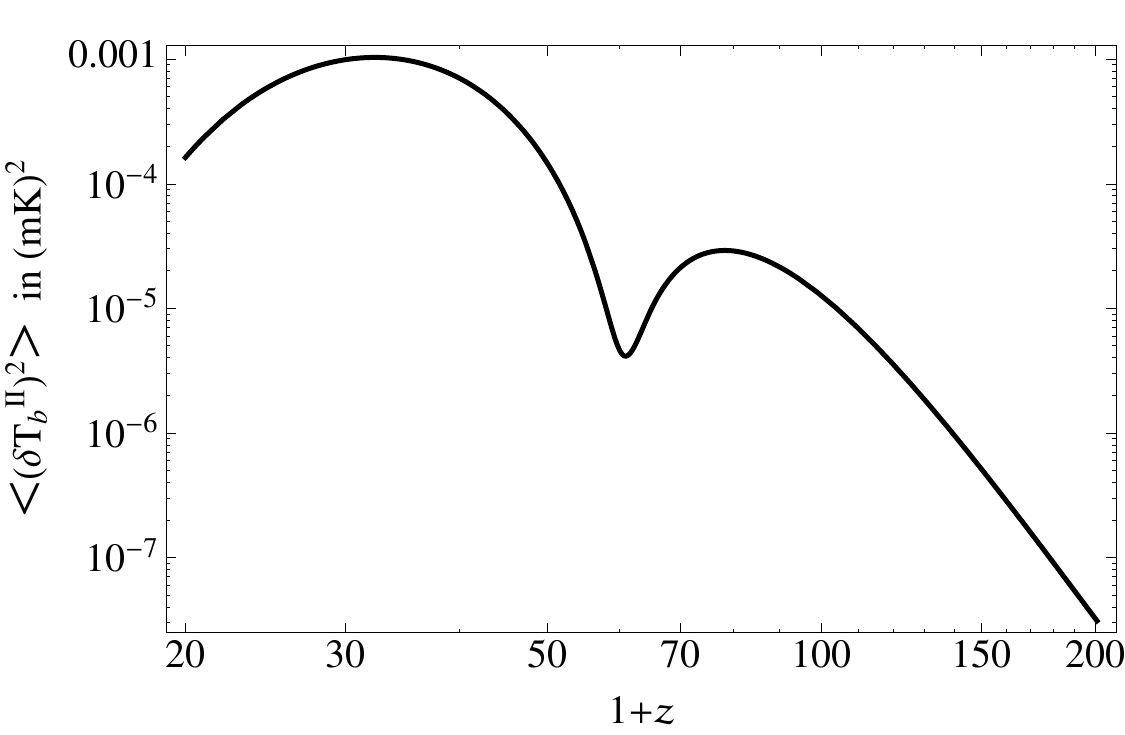}
\caption{Variance of the additional large-scale fluctuation of the 21~cm brightness
  temperature.} \label{fig:dTbII_rms} 
\end{figure}
On large angular scales all terms in Eq.~(\ref{eq:dTb-final}) are
relevant. All terms but the quadratic term were already computed by LC07, and we use the code \textsc{camb}
\emph{sources} to compute them. As we
showed earlier, the quadratic terms are uncorrelated with linear terms
and we therefore only need to compute the power spectrum of $\delta
T_b^{\rm II}$, and add it to the LC07 result.

Figure \ref{fig:dTbII_z} illustrates the
redshift dependence of the different terms contributing to $\delta
T_b^{\rm II}$. We see that they are all of comparable amplitude and
happen to nearly cancel out at $z \gtrsim 60$. Figure.~\ref{fig:dTbII_rms} shows the variance of the total
additional large-scale contribution $\delta T_b^{\rm II}$ as a
function of redshift. Because of the near-cancellation of the
different terms at $z \gtrsim 60$, the fluctuations of the quadratic
term peak around $z \approx 30$, at a lower redshift than the
fluctuations of the overall 21~cm signal.

\begin{figure}
\includegraphics[width = 90 mm]{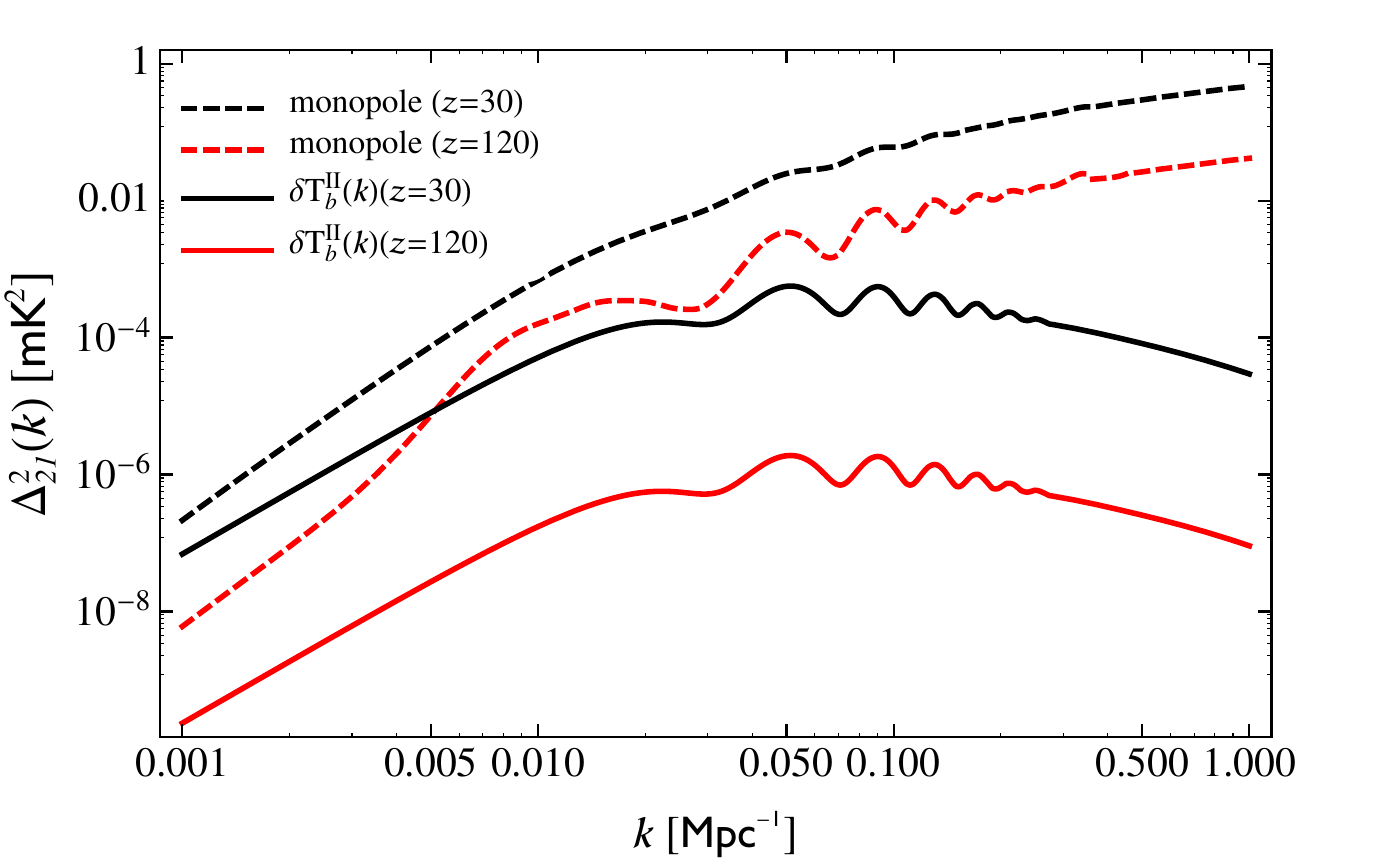}
\caption{Variance of fluctuations
  per logarithmic $k$-interval for the quadratic correction (solid)
  and the standard monopole term (dashed) at $z=30$ (top) and $z=120$ (bottom). The
  correction is of order tens of percent at large scales and low redshift.}
   \label{fig:monopole}
\end{figure}

Figure \ref{fig:monopole} shows the power spectrum of $\delta T_b^{\rm
  II}$ compared to the large-scale monopole fluctuations. We see that
at $z = 30$ the quadratic terms have fluctuations greater than $\sim
10\%$ of those of the monopole term for $k \lesssim 0.01$ Mpc$^{-1}$.

\begin{figure*}
\includegraphics[width = 89 mm]{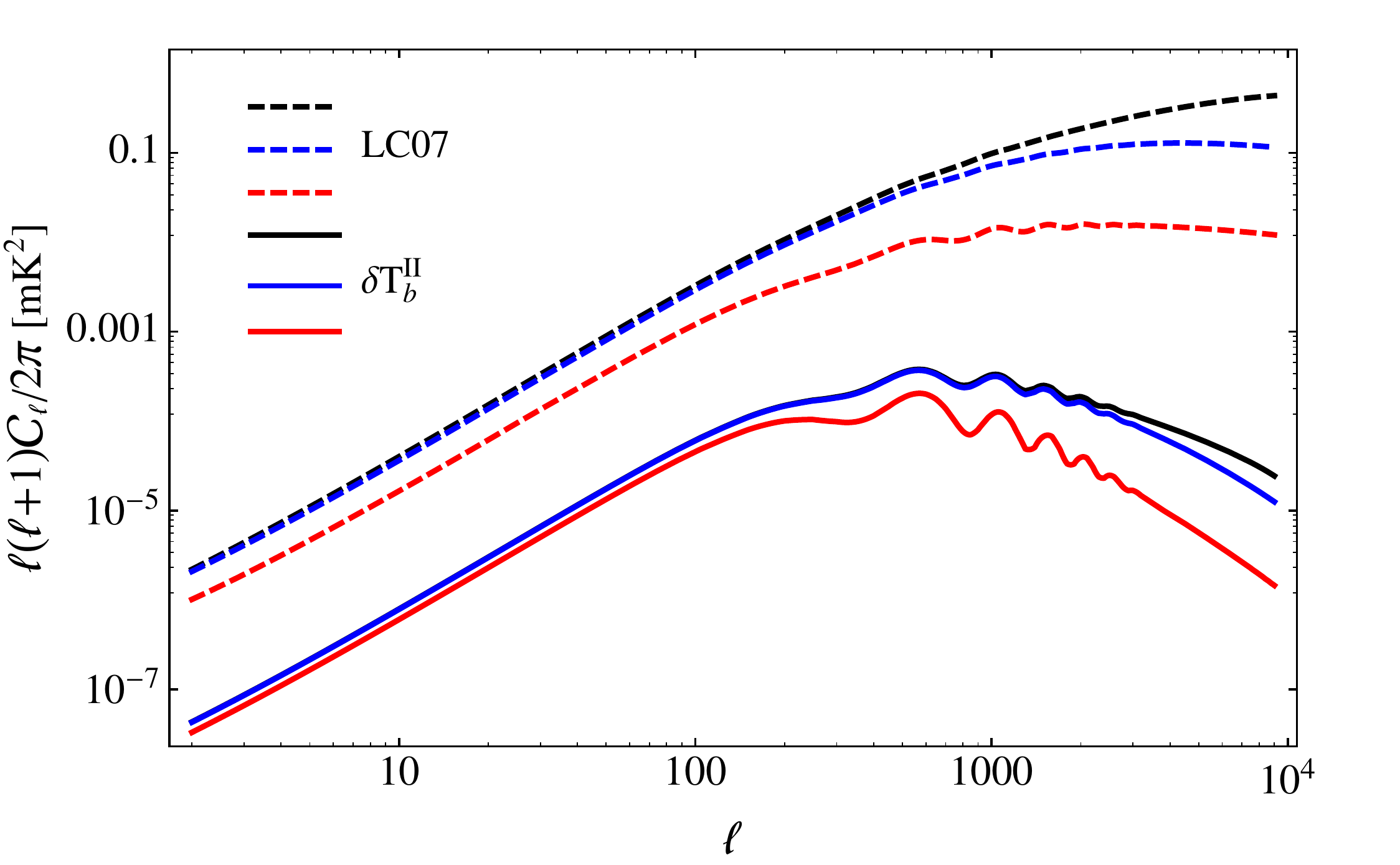}
\includegraphics[width = 89 mm]{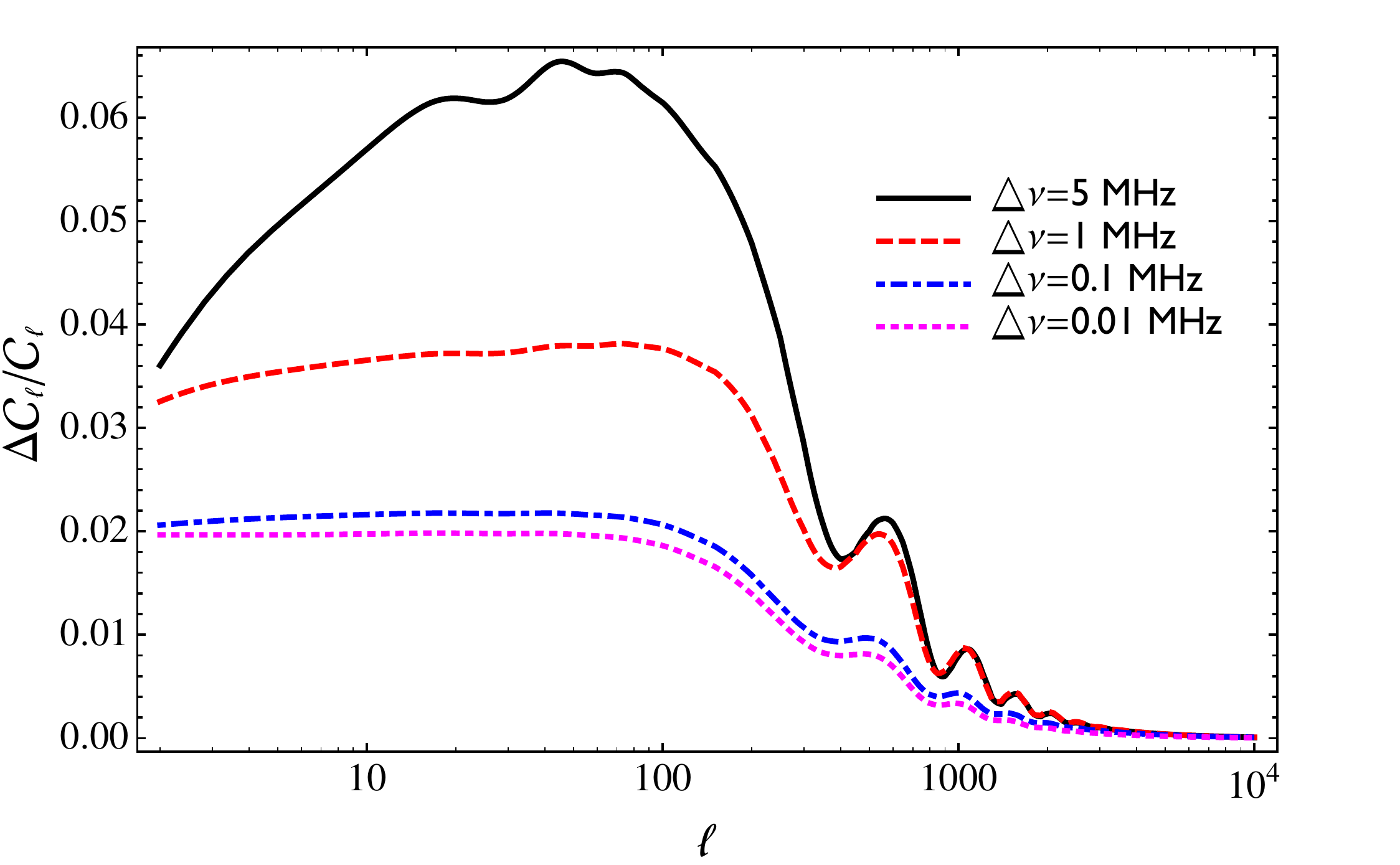}
\caption{\emph{Left}: Computed large-scale power spectrum (LC07,
  including relativistic corrections) and its correction due to the
  relative velocity between baryons and cold dark matter at redshift
  30 and through 3 different windows $\Delta\nu=0.01,0.1$ and $1$ Mhz
  (top to bottom). \emph{Right} The relative contribution of the
  correction at redshift 30. Applying a bigger window transfers more
  power from large scales, leading to a larger relative
  contribution.}\label{fig:TbII_final}
\end{figure*}

Figure \ref{fig:TbII_final} is our main result: it shows the
large-scale angular power spectrum $C_{\ell}$ of the quadratic terms, compared
with the standard $C_{\ell}$. Because the monopole fluctuation is a rapidly
increasing function of $k$, its large-scale angular fluctuations are
actually dominated by small-scale power \cite{Lewis_2007}. As a
consequence, the correction to the \emph{angular} power spectrum is
smaller than one would expect from comparing the Fourier-space
fluctuations. We still find that quadratic terms enhance the
large-scale power spectrum by a few percent at $z = 30$ and for $\ell$
up to a few hundred. The relative increase is larger when using a
larger window function (see right panel of Fig.~\ref{fig:TbII_final});
however in that case the absolute power is also decreased. We note
that with the standard cosmological scenario considered, the correction to the large-scale power spectrum is maximal around $z \approx 30$, due to the near-cancellation of various terms at
higher redshitfs. One should keep in mind that at these redshifts the radiation from the first stars may
alread have a significant impact on the 21 cm signal, depending on the
model considered \cite{Fialkov_2013b}.

Finally, we point out that we have only considered a standard
cosmology here, and simply extrapolated the small-scale power spectrum
from its known shape at much larger scales. Any unusual feature in the
small-scale power spectrum (due, fore example to a running of the
spectral index, or to a warm dark matter \cite{Loeb_2004}) would also
have some effect on large angular scales through the relative velocity
effect. This effect therefore potentially allows to measure
small-scale physics through observations of large angular scales, an
aspect which we shall explore in future works.

\subsection{Comment on other non-linear terms} \label{sec:NL}

In this paper we are considering quadratic terms only insofar as they
are \yacine{significantly} modulated on large scales by the relative velocity. We are
neglecting the term $2\langle \delta_0 \delta_x\rangle^2$ in the
autocorrelation function of $\delta^2$, as well as terms of similar
order that would result from the correlation of linear terms with cubic terms,
$\langle \delta_0 \delta_x^3 \rangle = 3 \langle \delta^2\rangle
\langle \delta_0 \delta_x\rangle$. This neglect is formally justified,
since our correction to the simple linear analysis at large scales is of relative
order $(\delta_s^2/\delta_l )^2 \sim 1$, whereas other non-linear terms are
formally corrections of order $\delta^2 \ll 1$. In practice, however,
our large-scale correction \yacine{is numerically of the order of tens
  of percent, and is the largest at $z \sim 30$. By then
the variance of the density fluctuation is already several
percent, and the neglected non-linear terms could therefore be of
comparable magnitude as the one we have accounted for, even though they are \emph{formally} of a different order.}
\yacine{To our knowledge, the effect of higher-order terms in the brightness
  temperature expansion has not been investigated yet (beside Ref.~\cite{Shaw_2008}, where the non-linear velocity
  gradient terms are considered, see also Ref.~\cite{Mao_2012})}. 
Including the other non-linear terms consistently would also require
accounting for the non-linear growth of overdensities. This would
significantly complicate the analysis, and we defer it to a future work.

\section{Conclusions} \label{sec:conclusions}

We have revisited the theoretical prediction for the 21~cm intensity
fluctuations during the dark ages, accounting for the relative velocity
between baryons and CDM recently discussed by Tseliakhovich and
Hirata \cite{Tsel_2010}. We have focused on isolating the consequences
of this effect and for the sake of simplicity have made several
assumptions regarding other effects which can be important at the few-percent
level. Some of these effects are treated elsewhere in the literature and we
list them here for completeness. First, we have computed the signal to lowest
order in the small optical depth and neglected fluctuations of the residual free
electron fraction, which lead to a few percent correction
\cite{Lewis_2007}. This can be straightforwardly accounted for in our
computation, and we have not done so simply for the sake of conciseness. Secondly, we have neglected the thermal
broadening of the 21~cm line and have assumed it can be described by a single,
velocity-independent spin temperature, effects which can be important
at the percent-level \cite{Hirata_2007}. Finally, we have used linear perturbation theory to follow
the growth of density perturbations, and neglected non-linear
corrections which affect the small-scale power spectrum at the several percent level at $z
\lesssim 50$. Computing these corrections accurately
is technically challenging and has only been done approximately so far
\cite{Lewis_2007}. We have also neglected higher-order terms in the
expansion of the brightness temperature, which could lead to
corrections at the several percent level at low redshift. To our
knowledge, these corrections have not yet been explored. Last but
not least, we have neglected the impact that early-formed stars may
have on the signal at $ z \approx 30$.

Our findings are as follows. The relative velocity between baryons and CDM leads to a
suppression of baryonic density and temperature fluctuations on scales $k
\gtrsim 30$ Mpc$^{-1}$ by several tens of percent, which result in a
similar suppression of the 21~cm fluctuations on angular scales $\ell
\gtrsim 5 \times 10^5$. Less intuitively, we find an
\emph{enhancement} of the 21~cm fluctuations in two scale
regimes. First, on scales much smaller than the Jeans scale, we find
that the streaming of cold dark matter perturbations relative to
baryonic ones lead to a resonant amplification of acoustic waves. This
translates to an enhancement of the 21~cm power spectrum for angular
scales $\ell \gtrsim 5 \times 10^7$. Most importantly (and as
anticipated by TH10), the large-scale fluctuations of the relative velocity field are
imprinted on the 21~cm signal, at scales $k \sim 0.005-1$ Mpc$^{-1}$,
corresponding to angular scales $\ell \lesssim 10^4$. This enhancement
is due to the combination of two facts. On the one hand, the 21~cm
brightness temperature depends non-linearly on the underlying baryonic fluctuations. On
the other hand, the large-scale modulation by the relative velocity of the \emph{square} of
small-scale perturbations is comparable to the linear large-scale fluctuations at
$z \lesssim 100$.

One of the prime appeals of 21~cm fluctuations from the dark ages is to
access the small-scale power spectrum at $k \gtrsim$ few Mpc$^{-1}$,
currently unaccessible to other probes \cite{Loeb_2004,
  Tegmark_2009}. If observed directly, these Fourier modes correspond
to multipoles $\ell$ of several tens of thousands at least, i.e.~an
angular resolution better than $10^{-4}$ radians. Reaching this
resolution at the highly redshifted frequency of the 21~cm transition
would be highly challenging, requiring very large baselines. Our results show that
detection prospects are in fact more optimistic (though still
challenging): the relative
velocity imprints the characteristic amplitude of the small-scale
density power spectrum (around $k \sim 100$ Mpc$^{-1}$) on large
angular fluctuations of the 21~cm signal, around $\ell \lesssim 1000$. Note that the relative velocity perturbations have support
on scales which are well measured by current cosmological probes, and
can therefore be computed exactly. Any deviation from the standard
cosmological model on small scales, such as warm dark matter or a
running of the primordial power spectrum, would therefore not only
affect the small angular scales of 21~cm fluctuations, but also the
regime $\ell \lesssim 1000$. The relative velocity should also
significantly change the effect that dark matter annihilations would
have on the 21~cm signal fluctuations \cite{Natarajan_2009}. We plan
to investigate these issues in future work.

Another extension to the work presented here is to include effects of
primordial non-Gaussianity; similarly to the relative velocity,
non-Gaussianities modulate the small-scale power spectrum on large
scales in the squeezed limit. It is interesting to know how these
effects compare, both as a function of scale as well as amplitude, and
whether the relative velocity may hamper or help detection of
primordial non-gaussianities with 21 cm fluctuations.

Finally, the analytical results presented here also encourage to look for
semi-analytical modeling of the low redshift Universe. So far, this
has predominantly been a numerical effort, but it is not unlikely that
some of the physics at late times can be modeled analytically. We
shall tackle this problem in future work.

\section*{Acknowledgements}
We would like to thank Simone Ferraro, Anastasia Fialkov, Daniel Grin, Chris Hirata, Antony Lewis,
Avi Loeb and Matias Zaldarriaga for useful discussions and comments
on this work.

Y.~A.-H.~was supported by the Frank and Peggy Taplin fellowship at the
Institute for Advanced Study. P.~D.~M.~was supported by
the Netherlands Organization for Scientific Research
(NWO), through a Rubicon fellowship and the John Templeton Foundation
grant number 37426. S.~H.~was funded by the Princeton Undergraduate
Summer Research Program.

\appendix
\begin{widetext}

\section{Autocorrelation of functions of the relative velocity} \label{app:correlation}

In Section \ref{sec:21cm} we had to compute the autocorrelation function of the form $\langle F(v_0) F(v_x) \rangle$ of terms quadratic in
small-scale fluctuations which depend on the magnitude of the local
relative velocity (for which we have dropped the subscript bc). In
this Appendix we describe our numerical method and derive analytical
approximations for the two limiting cases of weak and strong correlation. 

 This autocorrelation takes the following integral form: 
\barr
\langle F(v_0) F(v_x) \rangle \equiv 
\int d^3 \bs{u}_{\bs{0}}~ d^3\bs{u}_{\bs x} P(\bs{u}_{\bs 0},
\bs{u}_{\bs x}) F(\sigma_{1d} u_0) F(\sigma_{1d} u_x),~~~~\label{eq:corr-integral}
\earr
where $P(\bs{u}_{\bs 0}, \bs{u}_{\bs x})$ is the six-dimensional joint Gaussian
probability distribution for the normalized relative velocities
$\bs{u}_0 \equiv \bs{v}_0/\sigma_{1d}, \bs{u}_x \equiv
\bs{v}_x/\sigma_{1d}$, at two points
separated by comoving distance $\bs{x}$:
\beq
P(\bs{u}_{\bs 0}, \bs{u}_{\bs x}) = \frac1{(2 \pi)^3
  \sqrt{1 - c_{||}^2}(1 - c_{\bot}^2)} \exp\left[- \frac12
  \frac{u_{0 ||}^2 + u_{x ||}^2 - 2 c_{||}u_{0||} u_{x
      ||}}{1-  c_{||}^2} - \frac12
  \frac{\bs{u}_{0\bot}^2 + \bs{u}_{x \bot}^2 - 2 c_{\bot}
    \bs{u}_{0\bot} \cdot \bs{u}_{x \bot}}{1 - c_{\bot}^2} \right],
\eeq
where $\bs{u}_{||} = \bs{u} \cdot \hat{x}$, $\bs{u}_{ \bot} =
\bs{u} - u_{||}\hat{x}$, and the dimensionless correlation
coefficients $c_{||}(x), c_{\bot}(x)$ were defined in
Eq.~(\ref{eq:cpara-cperp}).

\subsection{General case}

When the correlation coefficients are neither small nor very close to
unity, we have to compute the integral (\ref{eq:corr-integral})
numerically. Using spherical polar coordinates with $\hat{x}$ as the
polar axis, one of the angular integrals
is trivial, and the other can be performed analytically, so that the
remaining integral is only four-dimensional, and takes the form
\cite{Dalal_2010}: 
\beq
\langle F(v_0) F(v_x) \rangle = \iint_0^{\infty} d u_0 du_x F(\sigma_{1d} u_0)
F(\sigma_{1d} u_x)
\mathcal{P}(u_0, u_x), \label{eq:corr-2d}
\eeq
where the joint probability distribution for the normalized magnitudes
is given by
\barr
\mathcal{P}(u_0, u_x) \equiv \frac{u_0^2 u_x^2}{2 \pi
  \sqrt{1 - c_{||}^2}(1 - c_{\bot}^2)}\iint_{-1}^1 d \mu_0 d \mu_x \exp\left[-\frac12
  \frac{u_{0 ||}^2 + u_{x ||}^2  - 2 c_{||} u_{0 ||} u_{x ||}}{1 -
    c_{||}^2} -\frac12 \frac{u_{0 \bot}^2  + u_{x \bot}^2}{1 -
    c_{\bot}^2}\right] \mathcal{I}_0\left[\frac{c_{\bot} u_{0 \bot}
    u_{x \bot}}{1 - c_{\bot}^2} \right],~~~~~
\earr
where $u_{0||} \equiv u_0 \mu_0, u_{0\bot} \equiv u_0 \sqrt{1 -
  \mu_0^2}$ and similarly for $u_{x||}, u_{x\bot}$, and
$\mathcal{I}_0$ is the zero-th order modified Bessel function of the
first kind. 

In order to speed up computations, we first pre-compute the redshift-independent
distribution $\mathcal{P}(u_0, u_x)$ as a function of $u_0, u_x$ and
the magnitude $x$ of the separation vector. We can then quickly compute the
remaining two-dimensional integral for any given specific function
$F$, in particular for the same physical quantity at different redshifts.

\subsection{Small separation, strong correlation limit}

When $x \rightarrow 0$, $c_{\parallel}, c_{\bot} \rightarrow 1$ and the joint
probability distribution $P(\bs{u}_0,
\bs{u}_x)$ becomes sharply peaked around $\bs{u}_x = \bs{u}_0$,
which makes direct numerical integration difficult. In this section we
derive an asymptotic expression valid in this regime. We start by rewriting 
\beq
P(\bs{u}_0, \bs{u}_x) = P(\bs{u_0})\prod_i P(u_x^i | u_0^i),
\eeq
where $P(\bs{u}_0)$ is an isotropic three-dimensional Gaussian distribution with unit
variance per axis and $P(u_x^i|u_0^i)$ is a one-dimensional Gaussian
distribution with mean $c_i u_0^i$ and variance $1 - c_i^2$, with $c_1
= c_{||}$ and $c_2 = c_3 = c_{\bot}$. We now
Taylor-expand $\tilde{F}(u_x) \equiv F(\sigma_{1d} u_x)$ around $\bs{u}_0$. In order to get a correct
expression at order $\mathcal{O}(1 - c_i)$ we need to carry the expansion to second order in $\Delta^i \equiv u_x^i -
u_0^i$. Dropping the tilde on $F$, we have:
\beq
F(u_x) \approx F(u_0) + \sum_i \Delta^i \partial_i F + \frac12\sum_{ij}\Delta^i \Delta^j \partial_i \partial_j F 
+ \mathcal{O}(\Delta^3).
\eeq
We integrate this expression over the constrained distribution of
$u_x^i$ and obtain, to order $1 - c_i$:
\barr
\langle \Delta^i \rangle &=& - (1 - c_i) u_0^i,\\
\langle \Delta^i \Delta^j \rangle &=& \delta^{ij}(1 - c_i^2) + (1 -
c_i)(1-c_j) u_0^i u_0^j \approx 2 \delta^{ij}(1 - c_i).
\earr
We therefore obtain
\barr
\langle F(u_0) F(u_x) \rangle &\approx& \langle F(u_0)^2 \rangle 
+ \sum_i (1 - c_i)\left[ \langle F \partial_i^2 F\rangle- \langle
  u_0^i F  \partial_i F \rangle\right] + \mathcal{O}(1 - c_i)^2,\label{eq:average1}
\earr
where the argument $u_0$ is implicit everywhere. We now recall that
the Gaussian probability distribution $P(\bs{u}_0)$
satisfies the differential equation $\partial_i P = - u_0^i P$, which, after
integration by parts, leads to the identity $\langle u_0^i G \rangle =
\langle \partial_i G \rangle$ for any function $G$. This allows us to simplify equation
(\ref{eq:average1}):
\barr
\langle F(u_0) F(u_x) \rangle \approx \langle F^2 \rangle -
\sum_i (1 - c_i) \langle (\partial_i F)^2 \rangle + \mathcal{O}(1 - c_i)^2.
\earr
From the isotropy of $F$ and $P$ we have
$\langle (\partial_i F)^2 \rangle = \frac13 \langle (\nabla F)^2
\rangle = \frac 13 \langle (F')^2 \rangle$. We therefore arrive at the
following expression, valid in the small-separation limit:
\beq
\langle F(u_0) F(u_x) \rangle \approx \langle F^2 \rangle - (1 - \overline{c})\langle (F')^2 \rangle, 
\eeq
where $\overline{c} \equiv \frac13 c_{||} + \frac23 c_{\bot}$ is the
spherically-averaged correlation coefficient. 

It is in principle straightforward to carry on this expansion to
higher order in $(1 - c_i)$. However, the resulting coefficients
depend on higher-order derivatives of $F$, which is itself a
numerically evaluated function, and whose numerical higher-order derivatives are
less and less accurate. We have therefore chosen to stop at the first
order given here. In practice we use this
expansion for $x \leq 3$ Mpc, for which $1 - \overline{c} \leq 0.03$,
and switch to numerical integration beyond that value.

\subsection{Large separation, weak correlation limit}

In the other limiting regime, $x \rightarrow \infty$, $c_i \rightarrow
0$, the autocorrelation of the mean-subtracted function $F$ becomes
vanishingly small. Direct numerical integration cannot properly
capture the near-vanishing of the integral, and here also we may use a
series expansion. We expand the probability distribution $P(\bs{u}_0,
\bs{u}_x)$ to second order in $c_i \ll 1$:
\beq
P(\bs{u}_0, \bs{u}_x) =  \frac{\exp\left[- \frac12 \sum_i
  \frac{u_{0 i}^2 + u_{x i}^2}{1-  c_{i}^2} \right]}{(2 \pi)^3
  \sqrt{1 - c_{||}^2}(1 - c_{\bot}^2)} \left( 1 + \sum_i
  \frac{c_i u_{0i} u_{x i}}{1 - c_i^2}  + \frac12 \sum_{ij} c_i c_j
  u_{0i} u_{xi} u_{0j} u_{xj} + \mathcal{O}(c_i^3)\right).
\eeq
Since the function $F$ only depends on the magnitude of $\bs{u}$, it
is an even function of the $u_i$. Therefore upon integration against
$F(u_0) F(u_x)$, only the term $c_i^2 u_{0i}^2 u_{xi}^2$ survives, and
to lowest order we get
\beq
\langle F(u_0) F(u_x) \rangle \approx \frac12 \sum_i c_i^2 \langle
u_{i}^2 F(u)\rangle^2 = \frac1{18} \langle
u^2 F(u)\rangle^2 \sum_i c_i^2,
\eeq
where the radial averaging is to be carried with an isotropic
Gaussian distribution, and we recall that $\langle F\rangle = 0$. In
practice, we use this approximation for $\sum_i c_i^2 \leq 10^{-4}$.

As an example, we show the autocorrelation function of $\delta_b^2$ and the resulting power
spectrum obtained by Fourier transforming it in Fig.~\ref{fig:P_db2}, where we compare it to the power
spectrum of the linear overdensity. We see that the power spectrum of
$\delta_b^2$ can be comparable to that of $\delta_b$ on very large
scales ($k \lesssim 0.01$ Mpc$^{-1}$) and at low redshifts. For $z =
30$, the ratio of power spectra is greater than 10 percent for $k
\lesssim 0.1$ Mpc$^{-1}$. Even at $z
= 120$, the ratio is still of order a percent or more on scales $k \lesssim
0.01$ Mpc$^{-1}$.

\begin{figure*}
\includegraphics[width = 180 mm]{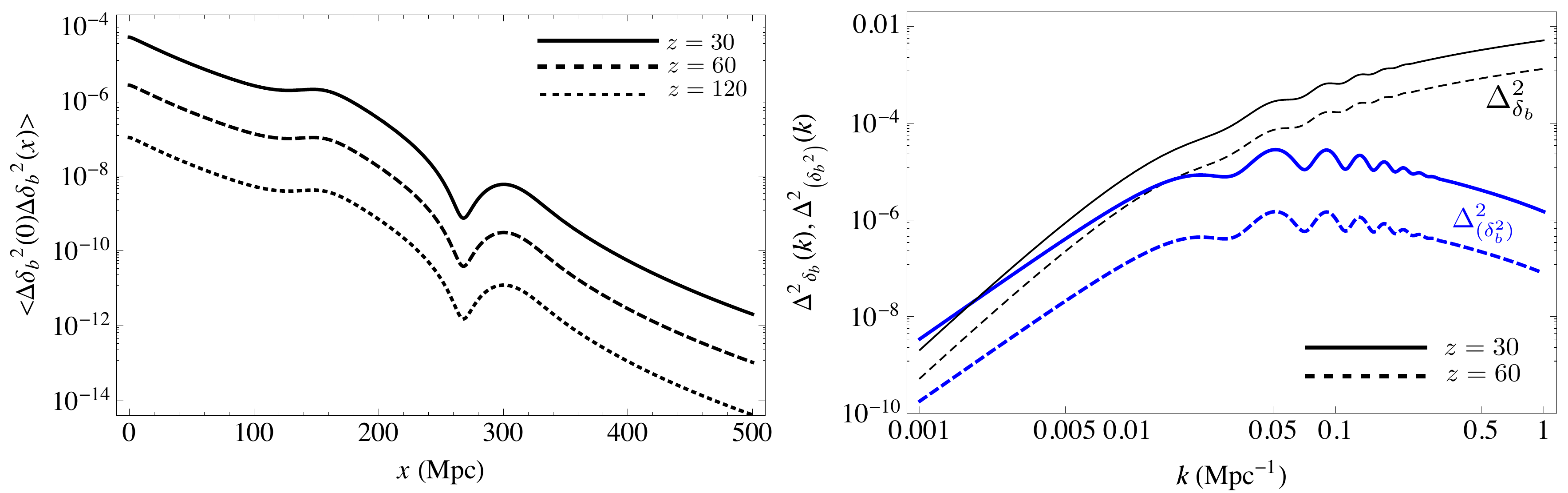}
\caption{\emph{Left}: Autocorrelation function of the fluctuations of $\delta_b^2$
  due to the modulation of small-scale power by the relative velocity
  of baryons and CDM. \emph{Right}: Variance of fluctuations of the baryon
  overdensity (thin black lines) and of its
square (thick blue lines) per logarithmic $k$-interval, at $z = 60$ and 30. The large-scale power
spectrum of $\delta_b$ is computed with \textsc{camb} in the
synchronous gauge. We only show scales inside the horizon for
which the overdensity is not strongly dependent on the chosen gauge.} \label{fig:P_db2} 
\end{figure*}

\section{Analytic expressions for the angular power spectrum for $\Delta^2(k)
\propto k$.}\label{app:Cl-Limber}

In this section we give analytic expressions for the angular power
spectrum, valid for all $\ell \gg 1$ and all widths of observational
window function $\Delta \equiv \Delta r /r \ll 1$, if the underlying
three-dimensional power spectra grow as $\Delta^2(k) \propto k$. \yacine{The
suppression factor $\rme^{- 2 \tau_{\rm reion}}$ is implicit everywhere.}

The angular power spectrum at redshift $z$ takes the form $C_{\ell}(z)
\equiv C_{\ell}^0 + C_{\ell}^{0 v} + C_{\ell}^v$, where the three
components are given in Eq.~(\ref{eq:Cl}). In this section we shall
derive analytic expressions in the case where $\Delta_0^2(k) \equiv
k^3 P_0(k)/(2 \pi^2) \propto k$, and similarly for $\Delta_{0v}^2$ and $\Delta_v^2$.

With this assumption on the scale dependence, the first term is
\beq
C_{\ell}^0 = 4 \pi \Delta_0^2(\ell/r_z) \frac{r_z}{\ell} \int dr_1 dr_2 W_{z}(r_1) W_{z}(r_2)\frac1{r_1}
\int d x j_{\ell}(x) j_{\ell}\left((r_2/r_1) x \right). \label{eq:ClT}
\eeq
This integral involves the function
\barr
F_{\ell}(R) \equiv \int d x j_{\ell}(x) j_{\ell}(Rx).
\earr
Using the differential equation satisfied by the spherical Bessel
functions, we obtain the following differential equation for $F_{\ell}(R)$:
\beq
R^2 F_{\ell}'' + 2 R F_{\ell}' - \ell(\ell + 1) F_{\ell} = - R^2 \int
dx ~x^2 j_{\ell}(x) j_{\ell}(R x) = -\frac{\pi}2 \delta(R-1),
\eeq
where in the second equality we have used the orthogonality
relation for the spherical Bessel functions. The homogeneous solutions
of this equation are $F_{\ell}(R) \propto R^{\ell}$ and $F_{\ell}(R)
\propto R^{-(\ell +1)}$. Integrating the ODE with initial condition
$F_{\ell}(0) = 0$, requiring continuity of $F_{\ell}$ at $R = 1$ and the jump
condition for its derivative $F_{\ell}'(1^+) - F_{\ell}'(1^-)
= - \pi/2$, we arrive at
\barr
F_{\ell}(R) \  = \begin{dcases}
     \frac{\pi R^{\ell}}{2(2 \ell +1)}& \text{if } R\leq 1\\
    \frac{\pi R^{-(\ell+1)}}{2(2 \ell +1)}             & \text{if } R\geq 1\\
 \end{dcases} \approx \frac{\pi}{4 \ell}\rme^{- \ell|R-1|} \ \
 \text{if}  \ |R -1 |
 \ll 1 \ \text{and} \  \ \ell \gg 1,
\earr
where the limit is valid for either
sign of $R-1$. We rewrite Eq.~(\ref{eq:ClT}) with $r_1 = r_z(1 + \epsilon_1)$ and $r_2 = r_z(1 +
\epsilon_2)$. For a top-hat window function the outer integral becomes,
to lowest order in $\epsilon_1, \epsilon_2$ 
\beq
C_{\ell}^0
\approx \frac{\pi^2}{\ell^2}\Delta_0^2(\ell/r_z) \frac1{\Delta^2}\iint_{- \Delta/2}^{\Delta/2} d \epsilon_1
d \epsilon_2 ~\rme^{- \ell
|\epsilon_2 - \epsilon_1|} = \frac{\pi^2}{\ell^2}\Delta_0^2(\ell/r_z)
\frac{2(\ell \Delta -  1 +  \rme^{- \ell \Delta})}{(\ell \Delta)^2}. \label{eq:CellT}
\eeq
We therefore obtain the following general expression and asymptotic limits:
\barr
\frac{\ell^2}{2 \pi} C_{\ell}^0  \approx \frac{\pi}{2}
\Delta_0^2(\ell/r_z) \frac{2(\ell \Delta -  1 +  \rme^{- \ell
    \Delta})}{(\ell \Delta)^2} \approx \begin{dcases} 
\frac{\pi}{2}\Delta_0^2(\ell/r_z)& \text{if}  \ \ \ell \Delta \ll 1,\\
\frac{\pi}{\ell \Delta}\Delta_0^2(\ell/r_z) & \text{if} \ \ \ell
\Delta \gg 1.
\end{dcases} 
\earr
Next we consider the cross term. We need to compute the function
\beq
G_{\ell}(R) \equiv \int d x j_{\ell}(x) j_{\ell}''(Rx) =
\frac{d^2}{dR^2} \int \frac{d x}{x^2} j_{\ell}(x) j_{\ell}(Rx) \equiv \frac{d^2}{dR^2}H_{\ell}(R),
\eeq
where the second equality is valid for $R \neq 1$ and the last one
defines the function $H_{\ell}$. Using again the differential equation
satisfied by $j_{\ell}$, we obtain the following equation for $H_{\ell}(R)$:
\beq
R^2 H_{\ell}'' + 2 R H_{\ell}' - \ell(\ell+1) H_{\ell} = - R^2 F_{\ell},
\eeq
from which we get the following equation for $G_{\ell} = H_{\ell}''$:
\beq
R^2 G_{\ell}'' + 6 R G_{\ell}'  + \left(6 - \ell(\ell+1)\right) G_{\ell} = - \frac{d^2}{dR^2}(R^2 F_{\ell}).
\eeq
One can obtain an explicit solution given the boundary conditions
$G_{\ell}(0) = 
G_{\ell}(\infty) = 0$ and requiring that $G_{\ell}$ is continuous at
$R = 1$. In the limit $|R -1| \ll 1, \ell \gg 1$ of interest, we obtain
\beq
G_{\ell}(R) \approx -\frac{\pi}{8
  \ell} \rme^{- \ell |R-1|}( 1 - \ell |R - 1|),
\eeq
and as a consequence, 
\barr
C_{\ell}^{0 v} \approx - \frac{\pi^2}{\ell^2}\Delta_{0 v}^2(\ell
/r_z) \frac1{\Delta^2} \iint_{- \Delta/2}^{\Delta/2} d\epsilon_1 d\epsilon_2  \rme^{- \ell |\epsilon_2 -
  \epsilon_1|}( 1 - \ell |\epsilon_2 - \epsilon_1|) = - \frac{\pi^2}{\ell^2}\Delta_{0 v}^2(\ell
/r_z) \frac{2(1 - \rme^{- \ell \Delta}(1 + \ell \Delta))}{(\ell \Delta)^2}.
\earr
We therefore arrive at the following general expression and
corresponding asymptotic regimes for the cross term:
\barr
\frac{\ell^2}{2 \pi} C_{\ell}^{0v}  \approx  - \frac{\pi}{2}\Delta_{0 v}^2(\ell
/r_z) \frac{2(1 - \rme^{- \ell \Delta}(1 + \ell \Delta))}{(\ell \Delta)^2} \approx \begin{dcases} 
- \frac{\pi}{2}\Delta_{0v}^2(\ell/r_z)& \text{if}  \ \ \ell \Delta \ll 1,\\
- \frac{\pi}{(\ell \Delta)^2}\Delta_{0v}^2(\ell/r_z) & \text{if} \ \ \ell
\Delta \gg 1,
\end{dcases}
\earr
We compute the power spectrum of the velocity term with similar
techniques, and arrive at
\barr
\frac{\ell^2}{2 \pi} C_{\ell}^{v}  \approx  \frac{\pi}{8}\Delta_{v}^2(\ell
/r_z)\frac{1 - \rme^{- \ell \Delta}(1 + \ell \Delta - \ell^2
  \Delta^2)}{(\Delta \ell)^2} \approx \begin{dcases} 
\frac{3 \pi}{16}\Delta_{v}^2(\ell/r_z)& \text{if}  \ \ \ell \Delta \ll 1,\\
\frac{\pi}{8 (\ell \Delta)^2}\Delta_{v}^2(\ell/r_z) & \text{if} \ \ \ell
\Delta \gg 1,
\end{dcases}
\earr
To conclude, we find, for power spectra scaling as $\Delta^2(k) \propto
k$ (i.e.~for equal power per linear $k$-interval), that, in the narrow
window regime, we get
\beq
\frac{\ell^2}{2 \pi}C_{\ell} \approx \frac{\pi}{2}\Delta_0^2(\ell
r_z) - \frac{\pi}{2} \Delta_{0v}^2(\ell
r_z) + \frac{3\pi}{16} \Delta_{v}^2(\ell
r_z) , \ \ \text{for} \ \ell \Delta \ll 1,
\eeq
which agrees with equation (41) of LC07. In the
large-window function regime, the terms involving velocities along the
line of sight are
suppressed by $1/(\ell \Delta r/r)^2$, whereas the ``monopole'' term
is only suppressed by $1/(\ell \Delta r /r)$ and therefore dominates
the angular power spectrum:
\beq
\frac{\ell^2}{2 \pi}C_{\ell} \approx \frac{\pi}{\ell \Delta}
\Delta_{0}^2(\ell/r_z),\ \ \text{for} \ \ell \Delta \gg 1,
\eeq
in agreement with equation (43) of LC07. This appendix moreover
provides explicit forms for the transition regime valid for $\Delta^2(k) \propto
k$.

\end{widetext}

\bibliography{vbc_21cm.bib}

\begin{thebibliography}{47}%
\makeatletter
\providecommand \@ifxundefined [1]{%
 \@ifx{#1\undefined}
}%
\providecommand \@ifnum [1]{%
 \ifnum #1\expandafter \@firstoftwo
 \else \expandafter \@secondoftwo
 \fi
}%
\providecommand \@ifx [1]{%
 \ifx #1\expandafter \@firstoftwo
 \else \expandafter \@secondoftwo
 \fi
}%
\providecommand \natexlab [1]{#1}%
\providecommand \enquote  [1]{``#1''}%
\providecommand \bibnamefont  [1]{#1}%
\providecommand \bibfnamefont [1]{#1}%
\providecommand \citenamefont [1]{#1}%
\providecommand \href@noop [0]{\@secondoftwo}%
\providecommand \href [0]{\begingroup \@sanitize@url \@href}%
\providecommand \@href[1]{\@@startlink{#1}\@@href}%
\providecommand \@@href[1]{\endgroup#1\@@endlink}%
\providecommand \@sanitize@url [0]{\catcode `\\12\catcode `\$12\catcode
  `\&12\catcode `\#12\catcode `\^12\catcode `\_12\catcode `\%12\relax}%
\providecommand \@@startlink[1]{}%
\providecommand \@@endlink[0]{}%
\providecommand \url  [0]{\begingroup\@sanitize@url \@url }%
\providecommand \@url [1]{\endgroup\@href {#1}{\urlprefix }}%
\providecommand \urlprefix  [0]{URL }%
\providecommand \Eprint [0]{\href }%
\providecommand \doibase [0]{http://dx.doi.org/}%
\providecommand \selectlanguage [0]{\@gobble}%
\providecommand \bibinfo  [0]{\@secondoftwo}%
\providecommand \bibfield  [0]{\@secondoftwo}%
\providecommand \translation [1]{[#1]}%
\providecommand \BibitemOpen [0]{}%
\providecommand \bibitemStop [0]{}%
\providecommand \bibitemNoStop [0]{.\EOS\space}%
\providecommand \EOS [0]{\spacefactor3000\relax}%
\providecommand \BibitemShut  [1]{\csname bibitem#1\endcsname}%
\let\auto@bib@innerbib\@empty
\bibitem [{\citenamefont {{Furlanetto}}\ \emph {et~al.}(2006)\citenamefont
  {{Furlanetto}}, \citenamefont {{Oh}},\ and\ \citenamefont
  {{Briggs}}}]{Fuetal2006}%
  \BibitemOpen
  \bibfield  {author} {\bibinfo {author} {\bibfnamefont {S.~R.}\ \bibnamefont
  {{Furlanetto}}}, \bibinfo {author} {\bibfnamefont {S.~P.}\ \bibnamefont
  {{Oh}}}, \ and\ \bibinfo {author} {\bibfnamefont {F.~H.}\ \bibnamefont
  {{Briggs}}},\ }\href {\doibase 10.1016/j.physrep.2006.08.002} {\bibfield
  {journal} {\bibinfo  {journal} {\physrep}\ }\textbf {\bibinfo {volume}
  {433}},\ \bibinfo {pages} {181} (\bibinfo {year} {2006})},\ \Eprint
  {http://arxiv.org/abs/arXiv:astro-ph/0608032} {arXiv:astro-ph/0608032}
  \BibitemShut {NoStop}%
\bibitem [{\citenamefont {{Pritchard}}\ and\ \citenamefont
  {{Loeb}}(2008)}]{Pritchard_2008}%
  \BibitemOpen
  \bibfield  {author} {\bibinfo {author} {\bibfnamefont {J.~R.}\ \bibnamefont
  {{Pritchard}}}\ and\ \bibinfo {author} {\bibfnamefont {A.}~\bibnamefont
  {{Loeb}}},\ }\href {\doibase 10.1103/PhysRevD.78.103511} {\bibfield
  {journal} {\bibinfo  {journal} {\prd}\ }\textbf {\bibinfo {volume} {78}},\
  \bibinfo {eid} {103511} (\bibinfo {year} {2008})},\ \Eprint
  {http://arxiv.org/abs/0802.2102} {arXiv:0802.2102} \BibitemShut {NoStop}%
\bibitem [{\citenamefont {{Furlanetto}}\ \emph {et~al.}(2009)\citenamefont
  {{Furlanetto}} \emph {et~al.}}]{Fur2009proc}%
  \BibitemOpen
  \bibfield  {author} {\bibinfo {author} {\bibfnamefont {S.~R.}\ \bibnamefont
  {{Furlanetto}}} \emph {et~al.},\ }in\ \href@noop {} {\emph {\bibinfo
  {booktitle} {astro2010: The Astronomy and Astrophysics Decadal Survey}}},\
  \bibinfo {series} {ArXiv Astrophysics e-prints}, Vol.\ \bibinfo {volume}
  {2010}\ (\bibinfo {year} {2009})\ p.~\bibinfo {pages} {82},\ \Eprint
  {http://arxiv.org/abs/0902.3259} {arXiv:0902.3259 [astro-ph.CO]} \BibitemShut
  {NoStop}%
\bibitem [{\citenamefont {{Loeb}}\ and\ \citenamefont
  {{Zaldarriaga}}(2004)}]{Loeb_2004}%
  \BibitemOpen
  \bibfield  {author} {\bibinfo {author} {\bibfnamefont {A.}~\bibnamefont
  {{Loeb}}}\ and\ \bibinfo {author} {\bibfnamefont {M.}~\bibnamefont
  {{Zaldarriaga}}},\ }\href {\doibase 10.1103/PhysRevLett.92.211301} {\bibfield
   {journal} {\bibinfo  {journal} {Physical Review Letters}\ }\textbf {\bibinfo
  {volume} {92}},\ \bibinfo {eid} {211301} (\bibinfo {year} {2004})},\ \Eprint
  {http://arxiv.org/abs/arXiv:astro-ph/0312134} {arXiv:astro-ph/0312134}
  \BibitemShut {NoStop}%
\bibitem [{\citenamefont {Mao}\ \emph {et~al.}(2008)\citenamefont {Mao},
  \citenamefont {Tegmark}, \citenamefont {McQuinn}, \citenamefont
  {Zaldarriaga},\ and\ \citenamefont {Zahn}}]{Mao:2008ug}%
  \BibitemOpen
  \bibfield  {author} {\bibinfo {author} {\bibfnamefont {Y.}~\bibnamefont
  {Mao}}, \bibinfo {author} {\bibfnamefont {M.}~\bibnamefont {Tegmark}},
  \bibinfo {author} {\bibfnamefont {M.}~\bibnamefont {McQuinn}}, \bibinfo
  {author} {\bibfnamefont {M.}~\bibnamefont {Zaldarriaga}}, \ and\ \bibinfo
  {author} {\bibfnamefont {O.}~\bibnamefont {Zahn}},\ }\href {\doibase
  10.1103/PhysRevD.78.023529} {\bibfield  {journal} {\bibinfo  {journal}
  {Phys.Rev.}\ }\textbf {\bibinfo {volume} {D78}},\ \bibinfo {pages} {023529}
  (\bibinfo {year} {2008})},\ \Eprint {http://arxiv.org/abs/0802.1710}
  {arXiv:0802.1710 [astro-ph]} \BibitemShut {NoStop}%
\bibitem [{\citenamefont {{Lewis}}\ and\ \citenamefont
  {{Challinor}}(2007)}]{Lewis_2007}%
  \BibitemOpen
  \bibfield  {author} {\bibinfo {author} {\bibfnamefont {A.}~\bibnamefont
  {{Lewis}}}\ and\ \bibinfo {author} {\bibfnamefont {A.}~\bibnamefont
  {{Challinor}}},\ }\href {\doibase 10.1103/PhysRevD.76.083005} {\bibfield
  {journal} {\bibinfo  {journal} {\prd}\ }\textbf {\bibinfo {volume} {76}},\
  \bibinfo {eid} {083005} (\bibinfo {year} {2007})},\ \Eprint
  {http://arxiv.org/abs/arXiv:astro-ph/0702600} {arXiv:astro-ph/0702600 [LC07]}
  \BibitemShut {NoStop}%
\bibitem [{\citenamefont {{Bharadwaj}}\ and\ \citenamefont
  {{Ali}}(2004)}]{Bharadwaj_2004}%
  \BibitemOpen
  \bibfield  {author} {\bibinfo {author} {\bibfnamefont {S.}~\bibnamefont
  {{Bharadwaj}}}\ and\ \bibinfo {author} {\bibfnamefont {S.~S.}\ \bibnamefont
  {{Ali}}},\ }\href {\doibase 10.1111/j.1365-2966.2004.07907.x} {\bibfield
  {journal} {\bibinfo  {journal} {\mnras}\ }\textbf {\bibinfo {volume} {352}},\
  \bibinfo {pages} {142} (\bibinfo {year} {2004})},\ \Eprint
  {http://arxiv.org/abs/arXiv:astro-ph/0401206} {arXiv:astro-ph/0401206}
  \BibitemShut {NoStop}%
\bibitem [{\citenamefont {{Tseliakhovich}}\ and\ \citenamefont
  {{Hirata}}(2010)}]{Tsel_2010}%
  \BibitemOpen
  \bibfield  {author} {\bibinfo {author} {\bibfnamefont {D.}~\bibnamefont
  {{Tseliakhovich}}}\ and\ \bibinfo {author} {\bibfnamefont {C.}~\bibnamefont
  {{Hirata}}},\ }\href {\doibase 10.1103/PhysRevD.82.083520} {\bibfield
  {journal} {\bibinfo  {journal} {\prd}\ }\textbf {\bibinfo {volume} {82}},\
  \bibinfo {eid} {083520} (\bibinfo {year} {2010})}\BibitemShut {NoStop}%
\bibitem [{\citenamefont {{Tseliakhovich}}\ \emph {et~al.}(2011)\citenamefont
  {{Tseliakhovich}}, \citenamefont {{Barkana}},\ and\ \citenamefont
  {{Hirata}}}]{Tsel_2011}%
  \BibitemOpen
  \bibfield  {author} {\bibinfo {author} {\bibfnamefont {D.}~\bibnamefont
  {{Tseliakhovich}}}, \bibinfo {author} {\bibfnamefont {R.}~\bibnamefont
  {{Barkana}}}, \ and\ \bibinfo {author} {\bibfnamefont {C.~M.}\ \bibnamefont
  {{Hirata}}},\ }\href {\doibase 10.1111/j.1365-2966.2011.19541.x} {\bibfield
  {journal} {\bibinfo  {journal} {\mnras}\ }\textbf {\bibinfo {volume} {418}},\
  \bibinfo {pages} {906} (\bibinfo {year} {2011})}\BibitemShut {NoStop}%
\bibitem [{\citenamefont {{Bittner}}\ and\ \citenamefont
  {{Loeb}}(2011)}]{Bittner_2011}%
  \BibitemOpen
  \bibfield  {author} {\bibinfo {author} {\bibfnamefont {J.~M.}\ \bibnamefont
  {{Bittner}}}\ and\ \bibinfo {author} {\bibfnamefont {A.}~\bibnamefont
  {{Loeb}}},\ }\href@noop {} {\bibfield  {journal} {\bibinfo  {journal} {ArXiv
  e-prints}\ } (\bibinfo {year} {2011})},\ \Eprint
  {http://arxiv.org/abs/1110.4659} {arXiv:1110.4659 [astro-ph.CO]} \BibitemShut
  {NoStop}%
\bibitem [{\citenamefont {{Fialkov}}\ \emph {et~al.}(2012)\citenamefont
  {{Fialkov}}, \citenamefont {{Barkana}}, \citenamefont {{Tseliakhovich}},\
  and\ \citenamefont {{Hirata}}}]{Fialkov_2012}%
  \BibitemOpen
  \bibfield  {author} {\bibinfo {author} {\bibfnamefont {A.}~\bibnamefont
  {{Fialkov}}}, \bibinfo {author} {\bibfnamefont {R.}~\bibnamefont
  {{Barkana}}}, \bibinfo {author} {\bibfnamefont {D.}~\bibnamefont
  {{Tseliakhovich}}}, \ and\ \bibinfo {author} {\bibfnamefont {C.~M.}\
  \bibnamefont {{Hirata}}},\ }\href {\doibase 10.1111/j.1365-2966.2012.21318.x}
  {\bibfield  {journal} {\bibinfo  {journal} {\mnras}\ }\textbf {\bibinfo
  {volume} {424}},\ \bibinfo {pages} {1335} (\bibinfo {year} {2012})},\ \Eprint
  {http://arxiv.org/abs/1110.2111} {arXiv:1110.2111 [astro-ph.CO]} \BibitemShut
  {NoStop}%
\bibitem [{\citenamefont {{Visbal}}\ \emph {et~al.}(2012)\citenamefont
  {{Visbal}}, \citenamefont {{Barkana}}, \citenamefont {{Fialkov}},
  \citenamefont {{Tseliakhovich}},\ and\ \citenamefont
  {{Hirata}}}]{Visbal_2012}%
  \BibitemOpen
  \bibfield  {author} {\bibinfo {author} {\bibfnamefont {E.}~\bibnamefont
  {{Visbal}}}, \bibinfo {author} {\bibfnamefont {R.}~\bibnamefont {{Barkana}}},
  \bibinfo {author} {\bibfnamefont {A.}~\bibnamefont {{Fialkov}}}, \bibinfo
  {author} {\bibfnamefont {D.}~\bibnamefont {{Tseliakhovich}}}, \ and\ \bibinfo
  {author} {\bibfnamefont {C.~M.}\ \bibnamefont {{Hirata}}},\ }\href {\doibase
  10.1038/nature11177} {\bibfield  {journal} {\bibinfo  {journal} {\nat}\
  }\textbf {\bibinfo {volume} {487}},\ \bibinfo {pages} {70} (\bibinfo {year}
  {2012})},\ \Eprint {http://arxiv.org/abs/1201.1005} {arXiv:1201.1005
  [astro-ph.CO]} \BibitemShut {NoStop}%
\bibitem [{\citenamefont {{Fialkov}}\ \emph
  {et~al.}(2013{\natexlab{a}})\citenamefont {{Fialkov}}, \citenamefont
  {{Barkana}}, \citenamefont {{Visbal}}, \citenamefont {{Tseliakhovich}},\ and\
  \citenamefont {{Hirata}}}]{Fialkov_2013}%
  \BibitemOpen
  \bibfield  {author} {\bibinfo {author} {\bibfnamefont {A.}~\bibnamefont
  {{Fialkov}}}, \bibinfo {author} {\bibfnamefont {R.}~\bibnamefont
  {{Barkana}}}, \bibinfo {author} {\bibfnamefont {E.}~\bibnamefont {{Visbal}}},
  \bibinfo {author} {\bibfnamefont {D.}~\bibnamefont {{Tseliakhovich}}}, \ and\
  \bibinfo {author} {\bibfnamefont {C.~M.}\ \bibnamefont {{Hirata}}},\ }\href
  {\doibase 10.1093/mnras/stt650} {\bibfield  {journal} {\bibinfo  {journal}
  {\mnras}\ }\textbf {\bibinfo {volume} {432}},\ \bibinfo {pages} {2909}
  (\bibinfo {year} {2013}{\natexlab{a}})},\ \Eprint
  {http://arxiv.org/abs/1212.0513} {arXiv:1212.0513 [astro-ph.CO]} \BibitemShut
  {NoStop}%
\bibitem [{\citenamefont {{Fialkov}}\ \emph
  {et~al.}(2013{\natexlab{b}})\citenamefont {{Fialkov}}, \citenamefont
  {{Barkana}}, \citenamefont {{Pinhas}},\ and\ \citenamefont
  {{Visbal}}}]{Fialkov_2013b}%
  \BibitemOpen
  \bibfield  {author} {\bibinfo {author} {\bibfnamefont {A.}~\bibnamefont
  {{Fialkov}}}, \bibinfo {author} {\bibfnamefont {R.}~\bibnamefont
  {{Barkana}}}, \bibinfo {author} {\bibfnamefont {A.}~\bibnamefont {{Pinhas}}},
  \ and\ \bibinfo {author} {\bibfnamefont {E.}~\bibnamefont {{Visbal}}},\
  }\href@noop {} {\bibfield  {journal} {\bibinfo  {journal} {ArXiv e-prints}\ }
  (\bibinfo {year} {2013}{\natexlab{b}})},\ \Eprint
  {http://arxiv.org/abs/1306.2354} {arXiv:1306.2354 [astro-ph.CO]} \BibitemShut
  {NoStop}%
\bibitem [{\citenamefont {{Wouthuysen}}(1952)}]{Wouthuysen_1952}%
  \BibitemOpen
  \bibfield  {author} {\bibinfo {author} {\bibfnamefont {S.~A.}\ \bibnamefont
  {{Wouthuysen}}},\ }\href {\doibase 10.1086/106661} {\bibfield  {journal}
  {\bibinfo  {journal} {\aj}\ }\textbf {\bibinfo {volume} {57}},\ \bibinfo
  {pages} {31} (\bibinfo {year} {1952})}\BibitemShut {NoStop}%
\bibitem [{\citenamefont {{Field}}(1958)}]{Field_1958}%
  \BibitemOpen
  \bibfield  {author} {\bibinfo {author} {\bibfnamefont {G.~B.}\ \bibnamefont
  {{Field}}},\ }\href {\doibase 10.1109/JRPROC.1958.286741} {\bibfield
  {journal} {\bibinfo  {journal} {Proceedings of the IRE}\ }\textbf {\bibinfo
  {volume} {46}},\ \bibinfo {pages} {240} (\bibinfo {year} {1958})}\BibitemShut
  {NoStop}%
\bibitem [{\citenamefont {{Hirata}}(2006)}]{Hirata_2006}%
  \BibitemOpen
  \bibfield  {author} {\bibinfo {author} {\bibfnamefont {C.~M.}\ \bibnamefont
  {{Hirata}}},\ }\href {\doibase 10.1111/j.1365-2966.2005.09949.x} {\bibfield
  {journal} {\bibinfo  {journal} {\mnras}\ }\textbf {\bibinfo {volume} {367}},\
  \bibinfo {pages} {259} (\bibinfo {year} {2006})},\ \Eprint
  {http://arxiv.org/abs/arXiv:astro-ph/0507102} {arXiv:astro-ph/0507102}
  \BibitemShut {NoStop}%
\bibitem [{\citenamefont {{Carilli}}\ \emph {et~al.}(2007)\citenamefont
  {{Carilli}}, \citenamefont {{Hewitt}},\ and\ \citenamefont
  {{Loeb}}}]{Carilli_2007}%
  \BibitemOpen
  \bibfield  {author} {\bibinfo {author} {\bibfnamefont {C.~L.}\ \bibnamefont
  {{Carilli}}}, \bibinfo {author} {\bibfnamefont {J.~N.}\ \bibnamefont
  {{Hewitt}}}, \ and\ \bibinfo {author} {\bibfnamefont {A.}~\bibnamefont
  {{Loeb}}},\ }\href@noop {} {\bibfield  {journal} {\bibinfo  {journal} {ArXiv
  Astrophysics e-prints}\ } (\bibinfo {year} {2007})},\ \Eprint
  {http://arxiv.org/abs/arXiv:astro-ph/0702070} {arXiv:astro-ph/0702070}
  \BibitemShut {NoStop}%
\bibitem [{\citenamefont {{Planck Collaboration}}\ \emph
  {et~al.}(2013)\citenamefont {{Planck Collaboration}}, \citenamefont {{Ade}},
  \citenamefont {{Aghanim}}, \citenamefont {{Armitage-Caplan}}, \citenamefont
  {{Arnaud}}, \citenamefont {{Ashdown}}, \citenamefont {{Atrio-Barandela}},
  \citenamefont {{Aumont}}, \citenamefont {{Baccigalupi}}, \citenamefont
  {{Banday}},\ and\ \citenamefont {et~al.}}]{Planck_params}%
  \BibitemOpen
  \bibfield  {author} {\bibinfo {author} {\bibnamefont {{Planck
  Collaboration}}}, \bibinfo {author} {\bibfnamefont {P.~A.~R.}\ \bibnamefont
  {{Ade}}}, \bibinfo {author} {\bibfnamefont {N.}~\bibnamefont {{Aghanim}}},
  \bibinfo {author} {\bibfnamefont {C.}~\bibnamefont {{Armitage-Caplan}}},
  \bibinfo {author} {\bibfnamefont {M.}~\bibnamefont {{Arnaud}}}, \bibinfo
  {author} {\bibfnamefont {M.}~\bibnamefont {{Ashdown}}}, \bibinfo {author}
  {\bibfnamefont {F.}~\bibnamefont {{Atrio-Barandela}}}, \bibinfo {author}
  {\bibfnamefont {J.}~\bibnamefont {{Aumont}}}, \bibinfo {author}
  {\bibfnamefont {C.}~\bibnamefont {{Baccigalupi}}}, \bibinfo {author}
  {\bibfnamefont {A.~J.}\ \bibnamefont {{Banday}}}, \ and\ \bibinfo {author}
  {\bibnamefont {et~al.}},\ }\href@noop {} {\bibfield  {journal} {\bibinfo
  {journal} {ArXiv e-prints}\ } (\bibinfo {year} {2013})},\ \Eprint
  {http://arxiv.org/abs/1303.5076} {arXiv:1303.5076 [astro-ph.CO]} \BibitemShut
  {NoStop}%
\bibitem [{\citenamefont {{Eisenstein}}\ and\ \citenamefont
  {{Hu}}(1998)}]{Eisenstein_1998}%
  \BibitemOpen
  \bibfield  {author} {\bibinfo {author} {\bibfnamefont {D.~J.}\ \bibnamefont
  {{Eisenstein}}}\ and\ \bibinfo {author} {\bibfnamefont {W.}~\bibnamefont
  {{Hu}}},\ }\href {\doibase 10.1086/305424} {\bibfield  {journal} {\bibinfo
  {journal} {\apj}\ }\textbf {\bibinfo {volume} {496}},\ \bibinfo {pages} {605}
  (\bibinfo {year} {1998})},\ \Eprint
  {http://arxiv.org/abs/arXiv:astro-ph/9709112} {arXiv:astro-ph/9709112}
  \BibitemShut {NoStop}%
\bibitem [{\citenamefont {{Dalal}}\ \emph {et~al.}(2010)\citenamefont
  {{Dalal}}, \citenamefont {{Pen}},\ and\ \citenamefont
  {{Seljak}}}]{Dalal_2010}%
  \BibitemOpen
  \bibfield  {author} {\bibinfo {author} {\bibfnamefont {N.}~\bibnamefont
  {{Dalal}}}, \bibinfo {author} {\bibfnamefont {U.-L.}\ \bibnamefont {{Pen}}},
  \ and\ \bibinfo {author} {\bibfnamefont {U.}~\bibnamefont {{Seljak}}},\
  }\href {\doibase 10.1088/1475-7516/2010/11/007} {\bibfield  {journal}
  {\bibinfo  {journal} {\jcap}\ }\textbf {\bibinfo {volume} {11}},\ \bibinfo
  {eid} {007} (\bibinfo {year} {2010})},\ \Eprint
  {http://arxiv.org/abs/1009.4704} {arXiv:1009.4704 [astro-ph.CO]} \BibitemShut
  {NoStop}%
\bibitem [{\citenamefont {Lewis}\ \emph {et~al.}(2000)\citenamefont {Lewis},
  \citenamefont {Challinor},\ and\ \citenamefont {Lasenby}}]{Lewis:1999bs}%
  \BibitemOpen
  \bibfield  {author} {\bibinfo {author} {\bibfnamefont {A.}~\bibnamefont
  {Lewis}}, \bibinfo {author} {\bibfnamefont {A.}~\bibnamefont {Challinor}}, \
  and\ \bibinfo {author} {\bibfnamefont {A.}~\bibnamefont {Lasenby}},\ }\href
  {\doibase 10.1086/309179} {\bibfield  {journal} {\bibinfo  {journal}
  {Astrophys.J.}\ }\textbf {\bibinfo {volume} {538}},\ \bibinfo {pages} {473}
  (\bibinfo {year} {2000})},\ \Eprint {http://arxiv.org/abs/astro-ph/9911177}
  {arXiv:astro-ph/9911177 [astro-ph]} \BibitemShut {NoStop}%
\bibitem [{\citenamefont {{Bernardeau}}\ \emph {et~al.}(2012)\citenamefont
  {{Bernardeau}}, \citenamefont {{van de Rijt}},\ and\ \citenamefont
  {{Vernizzi}}}]{Bernardeau_2012}%
  \BibitemOpen
  \bibfield  {author} {\bibinfo {author} {\bibfnamefont {F.}~\bibnamefont
  {{Bernardeau}}}, \bibinfo {author} {\bibfnamefont {N.}~\bibnamefont {{van de
  Rijt}}}, \ and\ \bibinfo {author} {\bibfnamefont {F.}~\bibnamefont
  {{Vernizzi}}},\ }\href {\doibase 10.1103/PhysRevD.85.063509} {\bibfield
  {journal} {\bibinfo  {journal} {\prd}\ }\textbf {\bibinfo {volume} {85}},\
  \bibinfo {eid} {063509} (\bibinfo {year} {2012})},\ \Eprint
  {http://arxiv.org/abs/1109.3400} {arXiv:1109.3400 [astro-ph.CO]} \BibitemShut
  {NoStop}%
\bibitem [{\citenamefont {{Bernardeau}}\ \emph {et~al.}(2013)\citenamefont
  {{Bernardeau}}, \citenamefont {{Van de Rijt}},\ and\ \citenamefont
  {{Vernizzi}}}]{Bernardeau_2013}%
  \BibitemOpen
  \bibfield  {author} {\bibinfo {author} {\bibfnamefont {F.}~\bibnamefont
  {{Bernardeau}}}, \bibinfo {author} {\bibfnamefont {N.}~\bibnamefont {{Van de
  Rijt}}}, \ and\ \bibinfo {author} {\bibfnamefont {F.}~\bibnamefont
  {{Vernizzi}}},\ }\href {\doibase 10.1103/PhysRevD.87.043530} {\bibfield
  {journal} {\bibinfo  {journal} {\prd}\ }\textbf {\bibinfo {volume} {87}},\
  \bibinfo {eid} {043530} (\bibinfo {year} {2013})},\ \Eprint
  {http://arxiv.org/abs/1209.3662} {arXiv:1209.3662 [astro-ph.CO]} \BibitemShut
  {NoStop}%
\bibitem [{\citenamefont {{Ma}}\ and\ \citenamefont
  {{Bertschinger}}(1995)}]{Ma_1995}%
  \BibitemOpen
  \bibfield  {author} {\bibinfo {author} {\bibfnamefont {C.-P.}\ \bibnamefont
  {{Ma}}}\ and\ \bibinfo {author} {\bibfnamefont {E.}~\bibnamefont
  {{Bertschinger}}},\ }\href {\doibase 10.1086/176550} {\bibfield  {journal}
  {\bibinfo  {journal} {\apj}\ }\textbf {\bibinfo {volume} {455}},\ \bibinfo
  {pages} {7} (\bibinfo {year} {1995})},\ \Eprint
  {http://arxiv.org/abs/arXiv:astro-ph/9506072} {arXiv:astro-ph/9506072}
  \BibitemShut {NoStop}%
\bibitem [{\citenamefont {{Naoz}}\ and\ \citenamefont
  {{Barkana}}(2005)}]{Naoz_2005}%
  \BibitemOpen
  \bibfield  {author} {\bibinfo {author} {\bibfnamefont {S.}~\bibnamefont
  {{Naoz}}}\ and\ \bibinfo {author} {\bibfnamefont {R.}~\bibnamefont
  {{Barkana}}},\ }\href {\doibase 10.1111/j.1365-2966.2005.09385.x} {\bibfield
  {journal} {\bibinfo  {journal} {\mnras}\ }\textbf {\bibinfo {volume} {362}},\
  \bibinfo {pages} {1047} (\bibinfo {year} {2005})},\ \Eprint
  {http://arxiv.org/abs/arXiv:astro-ph/0503196} {arXiv:astro-ph/0503196}
  \BibitemShut {NoStop}%
\bibitem [{\citenamefont {{Switzer}}\ and\ \citenamefont
  {{Hirata}}(2008)}]{Switzer_2008}%
  \BibitemOpen
  \bibfield  {author} {\bibinfo {author} {\bibfnamefont {E.~R.}\ \bibnamefont
  {{Switzer}}}\ and\ \bibinfo {author} {\bibfnamefont {C.~M.}\ \bibnamefont
  {{Hirata}}},\ }\href {\doibase 10.1103/PhysRevD.77.083006} {\bibfield
  {journal} {\bibinfo  {journal} {\prd}\ }\textbf {\bibinfo {volume} {77}},\
  \bibinfo {eid} {083006} (\bibinfo {year} {2008})},\ \Eprint
  {http://arxiv.org/abs/arXiv:astro-ph/0702143} {arXiv:astro-ph/0702143}
  \BibitemShut {NoStop}%
\bibitem [{\citenamefont {{Seager}}\ \emph {et~al.}(2000)\citenamefont
  {{Seager}}, \citenamefont {{Sasselov}},\ and\ \citenamefont
  {{Scott}}}]{Seager_2000}%
  \BibitemOpen
  \bibfield  {author} {\bibinfo {author} {\bibfnamefont {S.}~\bibnamefont
  {{Seager}}}, \bibinfo {author} {\bibfnamefont {D.~D.}\ \bibnamefont
  {{Sasselov}}}, \ and\ \bibinfo {author} {\bibfnamefont {D.}~\bibnamefont
  {{Scott}}},\ }\href {\doibase 10.1086/313388} {\bibfield  {journal} {\bibinfo
   {journal} {\apjs}\ }\textbf {\bibinfo {volume} {128}},\ \bibinfo {pages}
  {407} (\bibinfo {year} {2000})},\ \Eprint
  {http://arxiv.org/abs/arXiv:astro-ph/9912182} {arXiv:astro-ph/9912182}
  \BibitemShut {NoStop}%
\bibitem [{\citenamefont {{Senatore}}\ \emph {et~al.}(2009)\citenamefont
  {{Senatore}}, \citenamefont {{Tassev}},\ and\ \citenamefont
  {{Zaldarriaga}}}]{Senatore_2009}%
  \BibitemOpen
  \bibfield  {author} {\bibinfo {author} {\bibfnamefont {L.}~\bibnamefont
  {{Senatore}}}, \bibinfo {author} {\bibfnamefont {S.}~\bibnamefont
  {{Tassev}}}, \ and\ \bibinfo {author} {\bibfnamefont {M.}~\bibnamefont
  {{Zaldarriaga}}},\ }\href {\doibase 10.1088/1475-7516/2009/08/031} {\bibfield
   {journal} {\bibinfo  {journal} {\jcap}\ }\textbf {\bibinfo {volume} {8}},\
  \bibinfo {eid} {031} (\bibinfo {year} {2009})},\ \Eprint
  {http://arxiv.org/abs/0812.3652} {arXiv:0812.3652} \BibitemShut {NoStop}%
\bibitem [{\citenamefont {{Pillepich}}\ \emph {et~al.}(2007)\citenamefont
  {{Pillepich}}, \citenamefont {{Porciani}},\ and\ \citenamefont
  {{Matarrese}}}]{Pillepich_2007}%
  \BibitemOpen
  \bibfield  {author} {\bibinfo {author} {\bibfnamefont {A.}~\bibnamefont
  {{Pillepich}}}, \bibinfo {author} {\bibfnamefont {C.}~\bibnamefont
  {{Porciani}}}, \ and\ \bibinfo {author} {\bibfnamefont {S.}~\bibnamefont
  {{Matarrese}}},\ }\href {\doibase 10.1086/517963} {\bibfield  {journal}
  {\bibinfo  {journal} {\apj}\ }\textbf {\bibinfo {volume} {662}},\ \bibinfo
  {pages} {1} (\bibinfo {year} {2007})},\ \Eprint
  {http://arxiv.org/abs/astro-ph/0611126} {astro-ph/0611126} \BibitemShut
  {NoStop}%
\bibitem [{\citenamefont {{Chen}}\ and\ \citenamefont
  {{Kamionkowski}}(2004)}]{Chen_2004}%
  \BibitemOpen
  \bibfield  {author} {\bibinfo {author} {\bibfnamefont {X.}~\bibnamefont
  {{Chen}}}\ and\ \bibinfo {author} {\bibfnamefont {M.}~\bibnamefont
  {{Kamionkowski}}},\ }\href {\doibase 10.1103/PhysRevD.70.043502} {\bibfield
  {journal} {\bibinfo  {journal} {\prd}\ }\textbf {\bibinfo {volume} {70}},\
  \bibinfo {eid} {043502} (\bibinfo {year} {2004})},\ \Eprint
  {http://arxiv.org/abs/astro-ph/0310473} {astro-ph/0310473} \BibitemShut
  {NoStop}%
\bibitem [{\citenamefont {{Giesen}}\ \emph {et~al.}(2012)\citenamefont
  {{Giesen}}, \citenamefont {{Lesgourgues}}, \citenamefont {{Audren}},\ and\
  \citenamefont {{Ali-Ha{\"i}moud}}}]{Giesen_2012}%
  \BibitemOpen
  \bibfield  {author} {\bibinfo {author} {\bibfnamefont {G.}~\bibnamefont
  {{Giesen}}}, \bibinfo {author} {\bibfnamefont {J.}~\bibnamefont
  {{Lesgourgues}}}, \bibinfo {author} {\bibfnamefont {B.}~\bibnamefont
  {{Audren}}}, \ and\ \bibinfo {author} {\bibfnamefont {Y.}~\bibnamefont
  {{Ali-Ha{\"i}moud}}},\ }\href {\doibase 10.1088/1475-7516/2012/12/008}
  {\bibfield  {journal} {\bibinfo  {journal} {\jcap}\ }\textbf {\bibinfo
  {volume} {12}},\ \bibinfo {eid} {008} (\bibinfo {year} {2012})},\ \Eprint
  {http://arxiv.org/abs/1209.0247} {arXiv:1209.0247 [astro-ph.CO]} \BibitemShut
  {NoStop}%
\bibitem [{\citenamefont {{Dvorkin}}\ \emph {et~al.}(2013)\citenamefont
  {{Dvorkin}}, \citenamefont {{Blum}},\ and\ \citenamefont
  {{Zaldarriaga}}}]{Dvorkin_2013}%
  \BibitemOpen
  \bibfield  {author} {\bibinfo {author} {\bibfnamefont {C.}~\bibnamefont
  {{Dvorkin}}}, \bibinfo {author} {\bibfnamefont {K.}~\bibnamefont {{Blum}}}, \
  and\ \bibinfo {author} {\bibfnamefont {M.}~\bibnamefont {{Zaldarriaga}}},\
  }\href {\doibase 10.1103/PhysRevD.87.103522} {\bibfield  {journal} {\bibinfo
  {journal} {\prd}\ }\textbf {\bibinfo {volume} {87}},\ \bibinfo {eid} {103522}
  (\bibinfo {year} {2013})},\ \Eprint {http://arxiv.org/abs/1302.4753}
  {arXiv:1302.4753 [astro-ph.CO]} \BibitemShut {NoStop}%
\bibitem [{\citenamefont {{Ali-Ha{\"i}moud}}\ and\ \citenamefont
  {{Hirata}}(2011)}]{hyrec}%
  \BibitemOpen
  \bibfield  {author} {\bibinfo {author} {\bibfnamefont {Y.}~\bibnamefont
  {{Ali-Ha{\"i}moud}}}\ and\ \bibinfo {author} {\bibfnamefont {C.~M.}\
  \bibnamefont {{Hirata}}},\ }\href {\doibase 10.1103/PhysRevD.83.043513}
  {\bibfield  {journal} {\bibinfo  {journal} {\prd}\ }\textbf {\bibinfo
  {volume} {83}},\ \bibinfo {eid} {043513} (\bibinfo {year} {2011})},\ \Eprint
  {http://arxiv.org/abs/1011.3758} {arXiv:1011.3758 [astro-ph.CO]} \BibitemShut
  {NoStop}%
\bibitem [{\citenamefont {{Hirata}}\ and\ \citenamefont
  {{Forbes}}(2009)}]{Hirata_2009}%
  \BibitemOpen
  \bibfield  {author} {\bibinfo {author} {\bibfnamefont {C.~M.}\ \bibnamefont
  {{Hirata}}}\ and\ \bibinfo {author} {\bibfnamefont {J.}~\bibnamefont
  {{Forbes}}},\ }\href {\doibase 10.1103/PhysRevD.80.023001} {\bibfield
  {journal} {\bibinfo  {journal} {\prd}\ }\textbf {\bibinfo {volume} {80}},\
  \bibinfo {eid} {023001} (\bibinfo {year} {2009})},\ \Eprint
  {http://arxiv.org/abs/0903.4925} {arXiv:0903.4925 [astro-ph.CO]} \BibitemShut
  {NoStop}%
\bibitem [{\citenamefont {{Chluba}}\ and\ \citenamefont
  {{Sunyaev}}(2010)}]{Chluba_2010}%
  \BibitemOpen
  \bibfield  {author} {\bibinfo {author} {\bibfnamefont {J.}~\bibnamefont
  {{Chluba}}}\ and\ \bibinfo {author} {\bibfnamefont {R.~A.}\ \bibnamefont
  {{Sunyaev}}},\ }\href {\doibase 10.1051/0004-6361/200912263} {\bibfield
  {journal} {\bibinfo  {journal} {\aap}\ }\textbf {\bibinfo {volume} {512}},\
  \bibinfo {eid} {A53} (\bibinfo {year} {2010})},\ \Eprint
  {http://arxiv.org/abs/0904.0460} {arXiv:0904.0460 [astro-ph.CO]} \BibitemShut
  {NoStop}%
\bibitem [{\citenamefont {{Peebles}}(1968)}]{Peebles_1968}%
  \BibitemOpen
  \bibfield  {author} {\bibinfo {author} {\bibfnamefont {P.~J.~E.}\
  \bibnamefont {{Peebles}}},\ }\href {\doibase 10.1086/149628} {\bibfield
  {journal} {\bibinfo  {journal} {\apj}\ }\textbf {\bibinfo {volume} {153}},\
  \bibinfo {pages} {1} (\bibinfo {year} {1968})}\BibitemShut {NoStop}%
\bibitem [{\citenamefont {{Zel'dovich}}\ \emph {et~al.}(1969)\citenamefont
  {{Zel'dovich}}, \citenamefont {{Kurt}},\ and\ \citenamefont
  {{Syunyaev}}}]{Zeldovich_1969}%
  \BibitemOpen
  \bibfield  {author} {\bibinfo {author} {\bibfnamefont {Y.~B.}\ \bibnamefont
  {{Zel'dovich}}}, \bibinfo {author} {\bibfnamefont {V.~G.}\ \bibnamefont
  {{Kurt}}}, \ and\ \bibinfo {author} {\bibfnamefont {R.~A.}\ \bibnamefont
  {{Syunyaev}}},\ }\href@noop {} {\bibfield  {journal} {\bibinfo  {journal}
  {Soviet Journal of Experimental and Theoretical Physics}\ }\textbf {\bibinfo
  {volume} {28}},\ \bibinfo {pages} {146} (\bibinfo {year} {1969})}\BibitemShut
  {NoStop}%
\bibitem [{\citenamefont {{Pequignot}}\ \emph {et~al.}(1991)\citenamefont
  {{Pequignot}}, \citenamefont {{Petitjean}},\ and\ \citenamefont
  {{Boisson}}}]{Pequignot_1991}%
  \BibitemOpen
  \bibfield  {author} {\bibinfo {author} {\bibfnamefont {D.}~\bibnamefont
  {{Pequignot}}}, \bibinfo {author} {\bibfnamefont {P.}~\bibnamefont
  {{Petitjean}}}, \ and\ \bibinfo {author} {\bibfnamefont {C.}~\bibnamefont
  {{Boisson}}},\ }\href@noop {} {\bibfield  {journal} {\bibinfo  {journal}
  {\aap}\ }\textbf {\bibinfo {volume} {251}},\ \bibinfo {pages} {680} (\bibinfo
  {year} {1991})}\BibitemShut {NoStop}%
\bibitem [{\citenamefont {{Ali-Ha{\"i}moud}}\ and\ \citenamefont
  {{Hirata}}(2010)}]{emla}%
  \BibitemOpen
  \bibfield  {author} {\bibinfo {author} {\bibfnamefont {Y.}~\bibnamefont
  {{Ali-Ha{\"i}moud}}}\ and\ \bibinfo {author} {\bibfnamefont {C.~M.}\
  \bibnamefont {{Hirata}}},\ }\href {\doibase 10.1103/PhysRevD.82.063521}
  {\bibfield  {journal} {\bibinfo  {journal} {\prd}\ }\textbf {\bibinfo
  {volume} {82}},\ \bibinfo {eid} {063521} (\bibinfo {year} {2010})},\ \Eprint
  {http://arxiv.org/abs/1006.1355} {arXiv:1006.1355 [astro-ph.CO]} \BibitemShut
  {NoStop}%
\bibitem [{\citenamefont {{Kuhlen}}\ \emph {et~al.}(2006)\citenamefont
  {{Kuhlen}}, \citenamefont {{Madau}},\ and\ \citenamefont
  {{Montgomery}}}]{Kuhlen_2006}%
  \BibitemOpen
  \bibfield  {author} {\bibinfo {author} {\bibfnamefont {M.}~\bibnamefont
  {{Kuhlen}}}, \bibinfo {author} {\bibfnamefont {P.}~\bibnamefont {{Madau}}}, \
  and\ \bibinfo {author} {\bibfnamefont {R.}~\bibnamefont {{Montgomery}}},\
  }\href {\doibase 10.1086/500548} {\bibfield  {journal} {\bibinfo  {journal}
  {\apjl}\ }\textbf {\bibinfo {volume} {637}},\ \bibinfo {pages} {L1} (\bibinfo
  {year} {2006})},\ \Eprint {http://arxiv.org/abs/arXiv:astro-ph/0510814}
  {arXiv:astro-ph/0510814} \BibitemShut {NoStop}%
\bibitem [{\citenamefont {{Hirata}}\ and\ \citenamefont
  {{Sigurdson}}(2007)}]{Hirata_2007}%
  \BibitemOpen
  \bibfield  {author} {\bibinfo {author} {\bibfnamefont {C.~M.}\ \bibnamefont
  {{Hirata}}}\ and\ \bibinfo {author} {\bibfnamefont {K.}~\bibnamefont
  {{Sigurdson}}},\ }\href {\doibase 10.1111/j.1365-2966.2006.11321.x}
  {\bibfield  {journal} {\bibinfo  {journal} {\mnras}\ }\textbf {\bibinfo
  {volume} {375}},\ \bibinfo {pages} {1241} (\bibinfo {year} {2007})},\ \Eprint
  {http://arxiv.org/abs/arXiv:astro-ph/0605071} {arXiv:astro-ph/0605071}
  \BibitemShut {NoStop}%
\bibitem [{\citenamefont {{Mao}}\ \emph {et~al.}(2012)\citenamefont {{Mao}},
  \citenamefont {{Shapiro}}, \citenamefont {{Mellema}}, \citenamefont
  {{Iliev}}, \citenamefont {{Koda}},\ and\ \citenamefont {{Ahn}}}]{Mao_2012}%
  \BibitemOpen
  \bibfield  {author} {\bibinfo {author} {\bibfnamefont {Y.}~\bibnamefont
  {{Mao}}}, \bibinfo {author} {\bibfnamefont {P.~R.}\ \bibnamefont
  {{Shapiro}}}, \bibinfo {author} {\bibfnamefont {G.}~\bibnamefont
  {{Mellema}}}, \bibinfo {author} {\bibfnamefont {I.~T.}\ \bibnamefont
  {{Iliev}}}, \bibinfo {author} {\bibfnamefont {J.}~\bibnamefont {{Koda}}}, \
  and\ \bibinfo {author} {\bibfnamefont {K.}~\bibnamefont {{Ahn}}},\ }\href
  {\doibase 10.1111/j.1365-2966.2012.20471.x} {\bibfield  {journal} {\bibinfo
  {journal} {\mnras}\ }\textbf {\bibinfo {volume} {422}},\ \bibinfo {pages}
  {926} (\bibinfo {year} {2012})},\ \Eprint {http://arxiv.org/abs/1104.2094}
  {arXiv:1104.2094 [astro-ph.CO]} \BibitemShut {NoStop}%
\bibitem [{\citenamefont {{Seljak}}\ and\ \citenamefont
  {{Zaldarriaga}}(1996)}]{ZS1996}%
  \BibitemOpen
  \bibfield  {author} {\bibinfo {author} {\bibfnamefont {U.}~\bibnamefont
  {{Seljak}}}\ and\ \bibinfo {author} {\bibfnamefont {M.}~\bibnamefont
  {{Zaldarriaga}}},\ }\href {\doibase 10.1086/177793} {\bibfield  {journal}
  {\bibinfo  {journal} {\apj}\ }\textbf {\bibinfo {volume} {469}},\ \bibinfo
  {pages} {437} (\bibinfo {year} {1996})},\ \Eprint
  {http://arxiv.org/abs/astro-ph/9603033} {astro-ph/9603033} \BibitemShut
  {NoStop}%
\bibitem [{\citenamefont {{Shaw}}\ and\ \citenamefont
  {{Lewis}}(2008)}]{Shaw_2008}%
  \BibitemOpen
  \bibfield  {author} {\bibinfo {author} {\bibfnamefont {J.~R.}\ \bibnamefont
  {{Shaw}}}\ and\ \bibinfo {author} {\bibfnamefont {A.}~\bibnamefont
  {{Lewis}}},\ }\href {\doibase 10.1103/PhysRevD.78.103512} {\bibfield
  {journal} {\bibinfo  {journal} {\prd}\ }\textbf {\bibinfo {volume} {78}},\
  \bibinfo {eid} {103512} (\bibinfo {year} {2008})},\ \Eprint
  {http://arxiv.org/abs/0808.1724} {arXiv:0808.1724} \BibitemShut {NoStop}%
\bibitem [{\citenamefont {{Tegmark}}\ and\ \citenamefont
  {{Zaldarriaga}}(2009)}]{Tegmark_2009}%
  \BibitemOpen
  \bibfield  {author} {\bibinfo {author} {\bibfnamefont {M.}~\bibnamefont
  {{Tegmark}}}\ and\ \bibinfo {author} {\bibfnamefont {M.}~\bibnamefont
  {{Zaldarriaga}}},\ }\href {\doibase 10.1103/PhysRevD.79.083530} {\bibfield
  {journal} {\bibinfo  {journal} {\prd}\ }\textbf {\bibinfo {volume} {79}},\
  \bibinfo {eid} {083530} (\bibinfo {year} {2009})},\ \Eprint
  {http://arxiv.org/abs/0805.4414} {arXiv:0805.4414} \BibitemShut {NoStop}%
\bibitem [{\citenamefont {{Natarajan}}\ and\ \citenamefont
  {{Schwarz}}(2009)}]{Natarajan_2009}%
  \BibitemOpen
  \bibfield  {author} {\bibinfo {author} {\bibfnamefont {A.}~\bibnamefont
  {{Natarajan}}}\ and\ \bibinfo {author} {\bibfnamefont {D.~J.}\ \bibnamefont
  {{Schwarz}}},\ }\href {\doibase 10.1103/PhysRevD.80.043529} {\bibfield
  {journal} {\bibinfo  {journal} {\prd}\ }\textbf {\bibinfo {volume} {80}},\
  \bibinfo {eid} {043529} (\bibinfo {year} {2009})},\ \Eprint
  {http://arxiv.org/abs/0903.4485} {arXiv:0903.4485 [astro-ph.CO]} \BibitemShut
  {NoStop}%
\end{thebibliography}%

\end{document}